\documentclass{emulateapj}
\usepackage{natbib}
\slugcomment{{\sc Accepted to the ApJ:} August 16, 2008} 
\usepackage{lscape}
\def\CIVdblt{{\rm C~}\kern 0.1em{\sc iv}~$\lambda\lambda 1548, 1550$}
\def\MgIIdblt{{\rm Mg~}\kern 0.1em{\sc ii}~$\lambda\lambda 2796, 2803$}
\def\NVdblt{{\rm N}\kern 0.1em{\sc v}~$\lambda\lambda 1238, 1242$}  
\def\OVIdblt{{\rm O}\kern 0.1em{\sc vi}~$\lambda\lambda 1031, 1037$}
\def\SiIVdblt{{\rm Si~}\kern 0.1em{\sc iv}~$\lambda\lambda1394, 1403$}
\def\AlIIIdblt{{\rm Al~}\kern 0.1em{\sc iii}~$\lambda\lambda1855,1863$}
\def\FeIIdblt{{\rm Fe~}\kern 0.1em{\sc ii}~$\lambda\lambda 2383, 2600$}

\def\AlII{\hbox{{\rm Al~}\kern 0.1em{\sc ii}}}
\def\AlIII{\hbox{{\rm Al~}\kern 0.1em{\sc iii}}}
\def\CaI{\hbox{{\rm Ca}\kern 0.1em{\sc i}}}
\def\CaII{\hbox{{\rm Ca}\kern 0.1em{\sc ii}}}
\def\CrII{\hbox{{\rm Cr}\kern 0.1em{\sc ii}}}
\def\CII{\hbox{{\rm C~}\kern 0.1em{\sc ii}}}
\def\CIII{\hbox{{\rm C~}\kern 0.1em{\sc iii}}}
\def\CIV{\hbox{{\rm C~}\kern 0.1em{\sc iv}}}
\def\CV{\hbox{{\rm C}\kern 0.1em{\sc v}}}
\def\H{\hbox{{\rm H}}}
\def\HI{\hbox{{\rm H~}\kern 0.1em{\sc i}}}
\def\HII{\hbox{{\rm H~}\kern 0.1em{\sc ii}}}
\def\Lya{\hbox{{\rm Ly}\kern 0.1em$\alpha$}}
\def\Lyb{\hbox{{\rm Ly}\kern 0.1em$\beta$}}
\def\Lyg{\hbox{{\rm Ly}\kern 0.1em$\gamma$}}
\def\Lyfive{\hbox{{\rm Ly}\kern 0.1em$5$}}
\def\Lysix{\hbox{{\rm Ly}\kern 0.1em$6$}}
\def\Lyseven{\hbox{{\rm Ly}\kern 0.1em$7$}}
\def\Lyeight{\hbox{{\rm Ly}\kern 0.1em$8$}}
\def\Lynine{\hbox{{\rm Ly}\kern 0.1em$9$}}
\def\Lyten{\hbox{{\rm Ly}\kern 0.1em$10$}}
\def\HeI{\hbox{{\rm He}\kern 0.1em{\sc i}}}
\def\HeII{\hbox{{\rm He}\kern 0.1em{\sc ii}}}
\def\FeI{\hbox{{\rm Fe~}\kern 0.1em{\sc i}}}
\def\FeII{\hbox{{\rm Fe~}\kern 0.1em{\sc ii}}}
\def\FeIII{\hbox{{\rm Fe~}\kern 0.1em{\sc iii}}}
\def\MnII{\hbox{{\rm Mn}\kern 0.1em{\sc ii}}}
\def\MgI{\hbox{{\rm Mg~}\kern 0.1em{\sc i}}}
\def\MgII{\hbox{{\rm Mg~}\kern 0.1em{\sc ii}}}
\def\MgIII{\hbox{{\rm Mg~}\kern 0.1em{\sc iii}}}
\def\MgIV{\hbox{{\rm Mg~}\kern 0.1em{\sc iv}}}
\def\NaI{\hbox{{\rm Na}\kern 0.1em{\sc i}}}
\def\NV{\hbox{{\rm N}\kern 0.1em{\sc v}}}
\def\NII{\hbox{{\rm N}\kern 0.1em{\sc ii}}}
\def\NIII{\hbox{{\rm N}\kern 0.1em{\sc iii}}}
\def\OVI{\hbox{{\rm O}\kern 0.1em{\sc vi}}}
\def\OII{\hbox{[{\rm O}\kern 0.1em{\sc ii}]}}
\def\SiII{\hbox{{\rm Si~}\kern 0.1em{\sc ii}}}
\def\SiIII{\hbox{{\rm Si~}\kern 0.1em{\sc iii}}}
\def\SiIV{\hbox{{\rm Si~}\kern 0.1em{\sc iv}}}
\def\SII{\hbox{{\rm S}\kern 0.1em{\sc ii}}}
\def\SIII{\hbox{{\rm S}\kern 0.1em{\sc iii}}}
\def\SIV{\hbox{{\rm S}\kern 0.1em{\sc iv}}}
\def\TiII{\hbox{{\rm Ti}\kern 0.1em{\sc ii}}}
\def\ZnII{\hbox{{\rm Zn}\kern 0.1em{\sc ii}}}
\newcommand{\kms}{\hbox{km~s$^{-1}$}}
\newcommand{\cmsq}{\hbox{cm$^{-2}$}}
\newcommand{\cc}{\hbox{cm$^{-3}$}}
\def\kms{\hbox{km~s$^{-1}$}}      
\def\cmsq{\hbox{cm$^{-2}$}}
\def\cc{\hbox{cm$^{-3}$}}
\def\etal{et~al.\ }

\begin{document}

\title{The Chemical and Ionization Conditions in Weak {\MgII} Absorbers}
\author{Anand~Narayanan\altaffilmark{1}, Jane~C.~Charlton\altaffilmark{1}, Toru~Misawa\altaffilmark{1}, Rebecca~E.~Green\altaffilmark{1}, \\ and Tae-Sun Kim\altaffilmark{2}}

\altaffiltext{1}{Department of Astronomy and Astrophysics, The Pennsylvania State University, University Park, PA 16802. Email: anand, charlton, misawa, reg5021@astro.psu.edu}

\altaffiltext{2}{Astrophysiakalisches Institut Potsdam, An der Sternwarte 16, 14482 Potsdam, Germany, {\it tkim@aip.de}}

\begin{abstract} 
We present an analysis of the chemical and ionization conditions in a sample of $100$ weak {\MgII} absorbers identified in the VLT/UVES archive of quasar spectra. In addition to {\MgII},
we present equivalent width and column density measurements of other low ionization species such as {\MgI}, {\FeII}, {\AlII}, {\CII}, {\SiII} and also {\AlIII}. We find that the column densities of {\CII}  and {\SiII} are strongly correlated with the column density of {\MgII}, with minimal scatter in the relationships. The column densities of {\FeII} exhibit an appreciable scatter when compared with the column density of {\MgII}, with some fraction of clouds having $N(\FeII) \sim N(\MgII)$, in which case the density is constrained to n$_{\H} > 0.05$~{\cc}. Other clouds in which $N(\FeII) << N(\MgII)$ have much lower densities. From ionization models, we infer that the metallicity in a significant fraction of weak {\MgII} clouds is constrained to values of solar or higher, if they are sub-Lyman limit systems. Based on the observed constraints, we hypothesize that weak {\MgII} absorbers are predominantly tracing two different astrophysical processes/structures.  A significant population of weak {\MgII} clouds, those in which $N(\FeII) << N(\MgII)$, identified at both low ($z \sim 1$) and high ($z \sim 2$) redshift, are likely to be tracing gas in the extended halos of galaxies, analogous to the Galactic high velocity clouds. These absorbers might correspond to $\alpha$-enhanced interstellar gas expelled from star-forming galaxies, in correlated supernova events. The $N(\MgII)$ and $N(\FeII)/N(\MgII)$ in such clouds are also closely comparable to those measured for the high velocity components in strong {\MgII} systems. An evolution is found in $N(\FeII)/N(\MgII)$ from $z = 2.4$ to $z = 0.4$, with an absence of weak {\MgII} clouds with $N(\FeII) \sim N(\MgII)$ at high-$z$. The $N(\FeII) \sim N(\MgII)$ clouds, which are prevalent at lower redshifts ($z < 1.5$), must be tracing Type Ia enriched gas in small, high metallicity pockets in dwarf galaxies, tidal debris, or other intergalactic structures.  

\end{abstract}

\section{INTRODUCTION}
\label{sec:1}
The {\HI} gas directly associated with galaxies that intercept the
line of sight to background quasars appears as optically thick Lyman
Limit systems in the quasar spectrum. The prominent metal lines 
associated with these intervening absorbers are typically observed to be
kinematically broad ($\Delta v \sim 100 - 400$~{\kms}), strong, and often saturated
\citep[e.g.,][]{ss92,cwcvogt01} . Studying the
properties of a large population of such strong {\MgII} absorbers is a
technique used for constraining the evolution of metals in the
interstellar media, gaseous halos and coronae of galaxies over a large
history of the universe \citep{lanzetta87,cwc96}. Apparently distinct from these strong {\MgII} absorbers are 
the population of quasar absorption line systems
in which the low ionization metal lines are weak.  These systems are
separated from the strong ones based on the standard definition of the
rest-frame equivalent width of $\MgII~\lambda 2796$~{\AA} line being
$W_r(2796) < 0.3$~{\AA}. This is not a firm criterion for division,
but has been followed as a convention on the following basis. The
survey of \citet{ss92}, which identified a large
population of strong {\MgII} absorbers used a sample of intermediate
resolution spectra ($\Delta \lambda \sim 5$~{\AA}) which had an
equivalent width threshold of $\sim 0.3$~{\AA}. Later surveys of
higher sensitivity and spectral resolution found that the equivalent
width distribution of {\MgII} systems at $z \sim 1$ increases steeply
for $W_r(2796) < 0.3$~{\AA} such that $\sim 67\%$ of all {\MgII}
absorbers (down to $0.02$~{\AA}) from that epoch are in fact weak
\citep{weak1, anand07}. It later became clear that such an empirical 
basis for the classification of {\MgII}
systems into {\it strong} and {\it weak} does bear some physical
significance in that the two classes might be tracing two or more different
populations of objects \citep{weak1, weak2, charlton03}.

The class of weak {\MgII} quasar absorption systems have several
remarkable properties that are unique. To begin with, unlike the
strong systems, the weak {\MgII} systems are optically thin in neutral
hydrogen and produce metal lines that are narrow [$b$(Mg) $\sim
4$~{\kms}] and often unsaturated \citep{weak1}. If weak
{\MgII} absorbers are sub-Lyman limit systems with $10^{15.8} < N(\HI)
< 10^{16.8}$~{\cmsq}, they would account for a significant fraction
($> 25\%$) of the high column density regime of the {\Lya} forest
\citep{weak2}. Surveys of quasar fields to identify host
galaxies have not often found weak {\MgII} systems at close impact
parameters (physical distance, D $< 30$~Kpc) of luminous star forming galaxies ($L >
0.05$~$L_*$) \citep{cwc98, cwc05, milni06}. This is a surprising result particularly in
light of the fact that, in a substantial number of weak systems, the
metallicity of the low ionization gas where the {\MgII} absorption
arises is constrained to values greater than 0.1Z$_\odot$.  In some
cases the best constraints require metallicities that are between
Z$_\odot$ and 10Z$_\odot$ \citep{weak2, charlton03, misawa07}. 
Thus, even though they have {\HI} column
densities that are $\sim 4$ orders of magnitude smaller than DLAs,
weak {\MgII} absorbers are produced in gas clouds with metallicities
that are 0.5 - 2 dex higher than the average metallicity of DLA
absorbers.

The astrophysical systems associated with weak {\MgII} absorbers have
not been identified yet. Several possibilities exist, which partly
account for the observed statistical and physical properties of weak
{\MgII} systems. Examples include extragalactic high velocity clouds
\citep{anand07}, dwarf galaxies \citep{lynch06}, gas clouds expelled 
in  super winds from dwarfs \citep[e.g.][]{zonak04, stocke04, keeney06} 
and/or massive starburst galaxies and metal enriched gas in intergalactic star clusters
\citep{weak2}. 
Recently, a number of authors have reported the detection
of several high metallicity ($> 0.1$~Z$_\odot$) gas clouds that are
residing in the intergalactic medium, both in the local
universe \citep{aracil06, tripp06} and at high redshift [$z > 2$, \citet{simcoe04, 
schaye07}]. The observed column densities of {\MgII}, {\CII}, {\SiII} and {\FeII}
in these gas clouds, and the metallicities inferred for them are comparable to several weak {\MgII} absorbers studied so far \citep{weak2,lynch07,misawa07}. 

Photoionization models of specific weak {\MgII} systems have shown
that they possess a two phase structure. The low ionization gas,
traced by such species as {\MgII}, {\FeII}, {\CII}, {\SiII}, etc. has
a gas density of n$_{\H} > 0.1$~{\cc} that is roughly 2 - 3 orders of
magnitude higher than the density of the associated high ionization
gas traced by {\CIV} lines \citep{charlton03, lynch07, misawa07}. 
The column density measured for
weak {\MgII} lines is typically $N(\MgII) \sim 10^{12} - 10^{13}$
{\cmsq}. For such relatively small values, the high number density of
ions derived from the photoionization modeling constraints the
thickness of the low ionization gas to $\sim 10$~pc. The weak {\MgII}
population occupies a significant volume of the universe with a
cross-section similar to the absorption cross section of luminous
($L_*$) galaxies \citep{weak1, anand07}. Thus, given their small thickness, 
if their gas clouds had a
spherical geometry then they would be a million times more numerous
than luminous galaxies at $z \sim 1$ in order to reproduce the
observed cross-section on the sky. However, an analysis comparing the
relative incidence of high and low ionization gas in a sample of weak
{\MgII} and {\CIV} systems at low redshift ($z < 0.3$) favors a
filamentary or sheet-like configuration for the absorber's physical
geometry \citep{milni06}, instead of millions of individual, {\it spherical} low 
ionization {\MgII} clouds of $\sim 10$~pc size, embedded in a $\sim $~kpc 
higher ionization halo, traced by {\CIV}.

In addition, \citet{lynch06} and \citet{anand07} discovered an evolution
in the redshift number density ($dN/dz$) of weak {\MgII}
absorbers over the interval $0.4 < z < 2.4$. The $dN/dz$ was found
to peak at $z = 1.2$, and subsequently decline (from a no-evolution trend,) towards 
higher redshift. The equivalent width distribution 
function was also found to be different between low and high redshift.
At $z \sim 1$, the equivalent width distribution of weak {\MgII} is
significantly higher than an extrapolation of the exponential
distribution for strong {\MgII} absorbers. In contrast, at $z \sim 2$, 
the equivalent width distribution of weak {\MgII} clouds
is only slightly in excess of an extrapolation of the strong {\MgII}
distribution. In the context of these observed changes in the 
absorber statistics between $z \sim 1$ and $z \sim 2$, it becomes
important to investigate if the changes are reflective of
an underlying change in the absorbers' physical or chemical
properties. Such an investigation may yield valuable clues
into the physical nature of the kind of astrophysical processes/structures
that produce weak {\MgII} absorption at low and high redshifts. The large
sample size considered here, provides the scope for such
an analysis.

In this paper we present constraints on the chemical abundances,
metallicity and
ionization conditions for a sample of 100 weak {\MgII} absorbers
identified in spectra that were extracted from the
VLT/UVES archive. We compare the observed line properties between the
various chemical elements in order to constrain the metallicity,
density and line-of-sight thickness of the low ionization
gas. Previous studies have focused on individual weak {\MgII} systems,
whereas here we have attempted to derive the range of properties for a
large ensemble of weak {\MgII} systems, which have only recently been
discovered (Narayanan {\etal} 2007). In \S~\ref{sec:2} and
\S~\ref{sec:3}, we explain the measurement of line parameters and the
comparison between the prominent metal lines in these systems. In
\S~\ref{sec:4}, we present the Cloudy photoionization constraints on
the densities and metallicities of these absorbers. An observed trend
in the {\FeII} to {\MgII} ratio with redshift is presented in \S~\ref{sec:5}, and a comparison
of weak {\MgII} clouds with the high velocity subsystems in a sample of
strong {\MgII} systems in \S~\ref{sec:6}. We conclude the paper with a summary
of the significant results (in \S~\ref{sec:7}) and a detailed discussion (\S~\ref{sec:8}) on the 
nature of the gaseous structures selected by weak {\MgII} absorption.

\section{THE SAMPLE OF WEAK {\MgII} ABSORBERS}
\label{sec:2}

The sample of weak {\MgII} systems presented in this study were
identified in $51$ quasar spectra extracted from the ESO archive. The
quasars were observed using the UVES high resolution echelle
spectrograph \citep{dekker00} on VLT at $R \sim 45,000$ (FWHM~$= 6.6$~{\kms}). The
detailed information on each quasar observation, such as the exposure time, emission
redshift of the quasar, wavelength coverage, program ID and PI of the observation
are listed in Table 1 of \cite{anand07}. The
reduction and wavelength calibration of the spectra were carried out
using the ESO-provided MIDAS pipeline. When multiple exposures of the
same target were available, they were co-added to enhance the $S/N$
ratio of the spectrum, after applying vacuum-heliocentric velocity
corrections. The final combined spectrum was continuum fitted using
IRAF\footnote{IRAF is distributed by the National Optical Astronomy Observatories (NOAO), which are operated by AURA, Inc., under cooperative agreement with NSF}, and subsequently normalized. The detailed reduction procedure can
be found in \S~2.1 of \citet{anand07}. The redshift path
length searched in each quasar spectrum for weak {\MgIIdblt} lines
excluded regions blueward of the {\Lya} emission to avoid
contamination from the Ly-$\alpha$ forest.

The 100 {\MgII} systems that we analyze here are taken from 
\citet{anand07} which described a survey for weak {\MgII} systems at
$0.4 < z < 2.4$. In addition to magnesium, we use the information from
ions of four other elements, viz. iron, carbon, silicon and aluminum,
in each system to estimate the chemical and ionization
conditions in the gas. Specifically, the lines that we consider
are the following; $\MgI$~$\lambda$~2853~{\AA}; {\MgIIdblt}~{\AA}; {\FeIIdblt}~{\AA};
{\AlIIIdblt}~{\AA}; $\AlII$~ $\lambda$~1671~{\AA}; $\CII$~$\lambda$~1335~{\AA}; 
and $\SiII$~$\lambda$~1260~{\AA}\footnote{The wavelengths are vacuum wavelengths rounded to the nearest natural number}. The coverage
of the individual lines vary depending on the redshift of the system
and the wavelength coverage of the spectrum in which the line is identified.
The system plots are presented in Figure~\ref{fig:1} ({\it Note: The full set of system plots will  be available in the online version of the journal. Here we provide only a few as examples}). 

Most weak {\MgII} absorbers also have associated high ionization gas in a 
separate phase, traced by {\CIV} lines \citep[e.g.][]{weak1}. 
The relative incidence of high and low ionization
phase is a useful constraint on the physical geometry of the absorber population
\citep{milni06}. The UVES spectra offer simultaneous coverage of {\CIV} and {\MgII} over the redshift interval $1 < z < 2.4$. Within this interval, in almost all cases {\CIV} is detected. However, the {\CIV} absorption profile is in many cases offset in velocity with {\MgII}, indicating the presence of a separate phase for the high ionization gas. The {\CIVdblt} profiles are shown in the various system plots of Figure~\ref{fig:1} ({\it Note: The full set of system plots will  be available in the online version of the journal. Here we provide only a few as examples}). In this paper, our focus is on determining the ionization conditions and metallicity in the low ionization  gas, and hence we defer the detailed analysis of the high ionization {\CIV} phase and its association with the low ionization gas to a forthcoming paper.

\subsection{Measurement of Equivalent Widths}
\label{sec:2.1}

For each system within the redshift interval of $0.4 < z < 2.4$,
besides {\MgIIdblt}~{\AA}, only {\MgI}~$\lambda 2853$~{\AA} and
{\FeIIdblt} lines have wavelength long enough to be in the regions of
the spectrum typically uncontaminated by {\HI} lines in the
forest. The other prominent metal lines that we have measured have
rest-frame wavelengths, $\lambda < 2000$~{\AA}. As a consequence they
are susceptible to blending with {\Lya} forest lines, particularly
since the redshift of the intervening absorber is often much less than
the emission redshift of the quasar. In instances where line blending
is apparent, we quote an upper limit on the measurement of the
rest-frame equivalent width. For doublet/multiplet lines such as {\FeII} and
{\AlIII} we have measured the equivalent width of the stronger member
of the doublet. We also quote a $3\sigma$ upper limit when a line is
not detected at the $3\sigma$ level. Table 1 lists the rest-frame
equivalent width measured for the various lines in each system.

\subsection{Measurement of Column Densities}
\label{sec:2.2}

Absorption lines were fit with a Voigt function to estimate the column
density. An initial model for the line profile was derived using the
automated profile fitter AUTOVP \citep{dave97}. The AUTOVP
routine generated its model profile by performing a Voigt profile
decomposition of the absorption feature and subsequently minimizing
the $\chi^2$ by adjusting the velocity ($v$), column density ($N$) and
Doppler parameter ($b$) for all the components in the model. The
output of AUTOVP was then refined using a maximum likelihood least
square fitter, MINFIT, which returns a best-fit model with a minimum
number of Voigt profile components based on an F-test \citep{cwc03}. 
To retain a component, requires an improvement in the model fit at an
80\% significance.  MINFIT derives the
model absorption profile after convolving with a Gaussian kernel of
FWHM = 6.6~{\kms}, corresponding to the UVES spectral resolution 
of $R=45,000$. The column density and Doppler parameter with 
their $1 \sigma$ errors are obtained for this final model.

Voigt profile fits were applied to the following lines associated with each system : {\MgIIdblt};
{\MgI}~$\lambda 2853$~{\AA}; {\FeIIdblt}; {\AlIIIdblt};
{\AlII}~$\lambda 1671$~{\AA}; {\CII}~$\lambda 1335$~{\AA}; and {\SiII}
~$\lambda 1260 $~{\AA}. In
the situation where a line is not detected at the $3\sigma$ level, we
quote an upper limit on the column density determined from the
$3\sigma$ limit on the equivalent width. Our sample consists of
only relatively high-$S/N$ spectra. The $3\sigma$ limits are hence low, so that we can
assume linear part of the curve-of-growth for estimating the
corresponding upper limit in column density. To get robust constraints
on the fit parameters, we use both members of the doublet while
fitting profiles for {\MgII} and {\AlIII}, and both of the
strong members of the multiplet in the case of {\FeII}, viz. {\FeIIdblt}. Weaker 
members of the {\FeII} multiplet were rarely detected at the 3-$\sigma$ level. 
By simultaneous fitting of members of the doublet/multiplet, it 
is possible to recover the true column density, even if the
stronger member of the doublet/multiplet is saturated \citep[see Sec 4.4.2][]{cwcthesis}. 
Thus for example, in the case of {\MgII}, by using both members of the doublet, it is possible to 
recover the true column density for values of $N(\MgII)$ up to 
$10^{14}$~{\cmsq} \citep[see Figure 4.3][]{cwcthesis}. For lines that are not doublets,
Voigt profile fits are unique only when the lines are unsaturated. In our
sample, this would a problem only for the strongest of {\CII}~$\lambda 1335$~{\AA}
and {\SiII}~$\lambda 1260 $~{\AA} lines. 

Table 2 lists the line parameters ($v$, $N$ and $b$) thus measured for the various lines in
each system. As mentioned earlier, the lines with rest-wavelength
$\lambda < 2000$~{\AA} are often found within the region of the
spectrum that is contaminated by the {\Lya} forest.  For 
{\AlIII}, we found that blending with {\HI} lines of the forest could
be identified by comparing the profile shapes of the individual
members of the doublets. For the rest of the transitions, their
profiles were compared to {\MgII} to rule out possible contamination.
Figure~\ref{fig:1} shows the line profiles of the various low ionization 
transitions and {\AlIII} associated with each system in our 
sample. For each line, the positions of the individual clouds, 
determined from Voigt profile fitting, are labeled. 

\section{RESULTS FROM MEASUREMENT OF METAL LINES}
\label{sec:3}

\subsection{The Population of Single and Multiple Clouds}
\label{sec:3.1_01}

From comparing the frequency distribution of the number of clouds per
system between strong and weak absorbers, \citet{weak2} discovered
that unlike strong absorbers, weak {\MgII} systems have a
non-Poissonian frequency distribution. Approximately two-thirds of the
weak systems in their sample of 30 at $z \sim 1$ had absorption in a
single cloud, isolated in redshift. The clouds were narrow ($ b \sim
4$~{\kms}) indicating a small temperature and velocity dispersion in
the gas. These systems were consequently called {\it single cloud}
weak {\MgII} absorbers referring to the low ionization gas in a single
narrow component, unresolved at $R=45,000$ (FWHM$=
6.6$~{\kms}). The other set of weak absorbers were called {\it
multiple cloud} weak {\MgII} systems as they had the low ionization
absorption in multiple clouds that are resolved at $R=45,000$
and kinematically broad ($\Delta v > 30$~{\kms}) compared to 
single clouds. 

The incidence of the number of low ionization clouds in any given weak
{\MgII} system is important for considering the physical geometry of
the absorbing structure \citep{ellison04, milni06}.
Figure~\ref{fig:2} shows the distribution of the number of Voigt profile
components per system in our sample. In nine systems
\footnote{$z=0.599512$ in Q2217-2818, $z=1.091866$ in Q0042-2930,
$z=1.153704$ in Q1151+068, $z=1.330502$ in Q1157+014, $z=1.395635$ in
Q0011+0055, $z=1.405367$ in Q2347-4342, $z=1.491972$ in Q0551-3637,
$z=1.755704$ in Q2243-6031 and $z=1.796233$ in Q2347-4342}, we found the
{\MgII} line profile to have a slight asymmetry, sometimes yielding
two components in the Voigt profile model. We have classified these as
single cloud systems, and plotted them in the $N_c = 1$ bin, since the
low ionization gas is predominantly still in a single gas cloud with
an internal velocity dispersion is less than 6.6~{\kms}.  A similar
asymmetry in single cloud line profiles was also noticed by \citet{weak1}
in HIRES/Keck high resolution spectra. However, their
formal fitting procedure, with the lower $S/N$ data, did not
statistically favor a two component fit.  The occasional asymmetry in
the line profile is likely due to a contribution to the low ionization
absorption from a slightly offset higher ionization gas
cloud. Photoionization models have succeeded in reproducing the
observed asymmetry in the line profile using a single low ionization
phase and separate high ionization phases (e.g., see the ionization
models for $z=1.405367$ and $z=1.796237$ systems in \citet{lynch07}
and the $z=1.755704$ system in \citet{misawa07}. We have therefore
chosen to classify the above nine systems as single cloud systems.

Taking this into account, in our larger sample of weak systems we find
that the single cloud absorbers account for 48\% of the total
population, which is much smaller than their observed fraction in the
HIRES sample described in \citet{weak1} and \citet{weak2}, 
but is consistent within 1$\sigma$. Within the redshift
interval of $0.4 \leq z \leq 1.4$, identical to the redshift path
length covered by the \citet{weak1} sample, we find that
only 45\% (34/76) of the weak absorbers are single cloud systems,
indicating that our results are not affected by any evolutionary
effect in which a larger fraction of $z > 1.4$ systems have multiple
clouds. This is further confirmed in Figure~\ref{fig:3} where we illustrate the
distribution of single and multiple cloud absorbers as a function of
redshift. We find weak absorbers showing absorption in both single and
multiple clouds at all redshifts within $0.4 < z < 2.4$. A preference is
not evident for a certain type of weak absorber (i.e. single or multiple
cloud) towards either low or high redshift.

\subsection{Equivalent Width of {\MgII}}
\label{sec:3.1_02}

Figure~\ref{fig:4} shows the distribution of rest-frame equivalent width of
{\MgII}$\lambda 2796$ as a function of system redshift. The
strength of the low ionization phase as traced by {\MgII} demonstrates
considerable scatter within the interval $0.4 < z < 2.4$. A
Spearmann-Kendall test supports the null hypothesis that the
equivalent width is statistically uncorrelated (Spearman's $\rho =
0.04$) with the redshift of the absorber. In strong {\MgII} absorbers, 
the low ionization absorption is never confined to a single cloud. The
line profiles are often kinematically complex with the absorption
spread in several clouds separated in velocity. From the statistical
analysis of 23 strong {\MgII} systems along 18 quasar lines of sight,
\citet{cwc03} found an average of $\sim 8$ clouds per
system, with the absorption profile of one system resolved into as
many as 19 different components. In addition, in that sample of strong
{\MgII} systems, a very strong correlation was found between the
number of clouds and the rest-frame equivalent width $W_r(2796)$. In
the bottom panel of Figure~\ref{fig:2}, we illustrate that such a strong
correlation ($>9\sigma$) also exists for weak {\MgII} systems, where
most of the weaker systems ($W_r(2796) < 0.1$~{\AA}) have absorption only in
one or two clouds. Both \citet{weak1} and \citet{anand07} found that the equivalent 
width distribution of weak systems at
$z \sim 1$ rises rapidly towards smaller equivalent widths. This
observation is also reflected in the bottom panel of Figure~\ref{fig:2} where we
find 67\% of our sample of weak absorbers to be at $W_r(2796) <
0.15$~{\AA}.

\subsection{Comparison of Rest-Frame Equivalent Width}
\label{sec:3.2}

In Figure~\ref{fig:5}, we present the rest-frame equivalent width of the various
metal lines compared to the equivalent width of {\MgII}~$\lambda 2796$
line for both single and multiple clouds. The difference in the number
of data points in each plot is attributed to the spectral coverage for
the various transitions. A Spearman-Kendall nonparametric
correlation test shows that the rest-frame equivalent width of {\MgI},
{\FeII}, {\CII}, {\AlII}, and {\AlIII} are all correlated with
the rest equivalent width of {\MgII} at a greater than 98\% confidence level. Owing to fewer data
points, {\SiII} exhibits a correlation with {\MgII} of lesser
significance ($ > 3 \sigma$). The Spearman and Kendall 
correlation tests were carried out using the ASURV astrostatistics
package which takes into consideration measurements that are upper
limits \citep{edf85, lavalley92}. 

We note that the {\CII} equivalent width has a strong linear
relationship with {\MgII}, with little scatter.  The {\SiII} may have
a similar relationship, but it is hard to demonstrate with the smaller
number of data points.  All other transitions, though we have shown
their equivalent widths to be correlated with {\MgII}, show a much
larger scatter in this relationship. 

For {\FeII}, there is more than a factor of ten spread in the ratio
$W_r(2383)/W_r(2796)$ at $0.2 < W_r(2796) < 0.3$~{\AA}.  At small
$W_r(2796)$, many of the $W_r(2383)$ measurements are upper limits,
but a spread of more than a factor of two can still be demonstrated at
$W_r(2796) \sim 0.05$~{\AA}.  A similarly large spread is also found
for the relationships between {\MgI} and {\MgII}, {\AlII} and {\MgII}, and
{\AlIII} and {\MgII}.

The large scatter in the observed ratios between the various
transitions can be brought about by a number of factors, and can be
exploited to diagnose the physical conditions of the absorber.  For a
given strength of {\MgII}, the spread in the strength of the other
transitions can be due to variations in abundance patterns or to
differences in the density/ionization parameters of the gas clouds (addressed
in \S~\ref{sec:4.2},~\ref{sec:4.6} and ~\ref{sec:5}).
There is also the possibility that absorption from two different ions
arises in separate phases, in which case the scatter between their
equivalent widths could be quite large.  If we are to distinguish
between these different factors, it is important not to use the observed
equivalent widths since they average together the contributions from
different clouds.  The physical conditions are better probed through
comparison of cloud-by-cloud column densities.

\subsection{Comparison of Column Densities}
\label{sec:3.3}

In Figure~\ref{fig:6}, we compare the {\MgII} column density measured for
each cloud with the corresponding column densities in {\MgI}, {\FeII},
{\SiII}, {\CII}, {\AlII} and {\AlIII}. In multiple cloud systems, the
comparison is between each component of {\MgII} and the corresponding
component in the other transition. Non-detections at the $3\sigma$
level are given as upper limits. To test for likely dependence between
the measured quantities, we apply Spearman and Kendall's
non-parametric correlation tests. We find that the column densities of
all ionization species except for {\SiII} are correlated with the
column density of {\MgII} at greater than $7\sigma$
significance. The statistical measure of the correlation is smaller for {\SiII}
($3\sigma$ significance) because of fewer data, 40\% of which are
censored points. The strongest correlation is observed between
$N(\CII)$ and $N(\MgII)$ (rank correlation coefficient, $\rho =
0.762$), in spite of being limited by fewer data points. Such a strong
positive correlation in the equivalent width and column density of
{\CII} with {\MgII} justifies the use of {\CII} lines, in conjunction with
other low ionization lines such as {\SiII}, to select weak
absorbers at redshifts $z > 2.5$ where it becomes more difficult to
use {\MgIIdblt} lines because of the redshifting of these lines
into the near-infrared regime.

Among the various ions, the column densities of {\FeII}, {\CII} and {\SiII}
display the least scatter with the column density of {\MgII}. The
correlation between these ions and {\MgII} can be formalized as:

\begin{center}
log $N(\FeII) = (1.15 \pm 0.14)$ log $N(\MgII) - 2.36$, ($\sigma = 0.37$) 

log $N(\CII) = (0.82 \pm 0.10)$ log $N(\MgII) + 3.42$, ($\sigma = 0.19$)

log $N(\SiII) = (1.02 \pm 0.25)$ log $N(\MgII) - 0.26$, ($\sigma = 0.28$)
\end{center}

\noindent The best-fit slope, the y-intercept and the corresponding $1 \sigma$
uncertainties of these regression lines were calculated using the
survival analysis package ASURV Rev 1.2 \citep{edf85, lavalley92}
which implements the methods
presented in \citet{isobe86}. The $\sigma$ values
are the standard deviation of the respective fits.

In Figure~\ref{fig:6}, we also plot the Solar composition of Fe, C, Si and Al
with respect to Mg, for reference. The abundance of carbon
[log (C/Mg)$_\odot$ = 0.970] is taken from \citet{allende01, allende02}, of
silicon [log (Si/Mg)$_\odot$ = -0.030], iron [log (Fe/Mg)$_\odot$ = -0.069] and
magnesium from \citet{holweger01}, and for aluminum [log (Al/Mg)$_\odot$ = -1.110] from \citet{grevesse98}. The observed ratio of column densities between the various
ions and {\MgII} when compared with the respective solar abundance ratios, can indicate whether the ionization fractions of {\CII}, {\SiII}, {\AlII}, {\AlIII} and
{\MgI} are comparable with that of {\MgII} in the low 
ionization gas. We note that this is the case for the observed {\SiII} to {\MgII} and
{\AlII} to {\MgII} column density ratios, which closely follow the respective
solar abundance ratios. 
On the other hand, the observed {\CII} to {\MgII} ratios 
are above the solar abundance ratio, while the {\FeII} to {\MgII} ratios are below. 
A number of factors such as differences in ionization parameter, differences in the elemental abundances and/or contributions from different gas phases can combine to produce these observed trends, which are discussed in the next section. 

In addition, differential depletion of elements onto dust can lead to deviations from Solar composition. The presence of dust has not been directly measured in weak {\MgII} systems. However, it has been found that dust extinction is significant only in stronger {\MgII} absorbers \citep[$W_r(\MgII) > 1.5$~{\AA};][]{khare05, york06}. For low column density absorbers such as the weak {\MgII} systems, interstellar dust may not be a substantial component influencing metallicity estimates derived from gas phase abundances. In addition, CLOUDY models incorporating varying amounts of dust levels find dust having a negligible effect on the density of the absorbing gas as well \citep{weak2}.

\section{CHEMICAL AND IONIZATION PROPERTIES OF WEAK {\MgII} ABSORBERS}
\label{sec:4}

The physical conditions in the low ionization gas clouds are
constrained using the standard photoionization code Cloudy
\citep[ver.07.02.01,][]{ferland98}. The primary objective is to
derive limits on the metallicity, density, and line-of-sight thickness
for the gas phase where the bulk of the {\MgII} absorption arises. For
this purpose, the observed column densities of the other prominent low
and intermediate ionization transitions - namely {\MgI}, {\FeII},
{\SiII}, {\AlII}, {\CII}, and {\AlIII}, and their ratios to {\MgII} - 
are used.

The ionization fraction for a given element is controlled by the
density in the gas cloud as well as by the strength of the incident
ionizing radiation. Weak {\MgII} systems are not known to reside at
small impact parameters (d~$< 30$~kpc) from luminous star-forming
galaxies ($L > 0.05L_*$, \citet{weak1}). Hence the ionization balance in them is
likely dictated by the intensity of the extragalactic background
radiation (EBR). We choose the \citet{hm96} model for
the EBR which incorporates ionizing photons from quasars and
star-forming galaxies after propagation through a thick IGM.  A 10\%
escape fraction from galaxies is used for ionizing photons with
$\lambda \leq  912$~{\AA}.

To determine the overall properties for our sample of weak {\MgII}
absorbers, we generate a grid of Cloudy models for a range of
ionization parameters ($-8.0 <$ log $U < -1.0$) and neutral hydrogen
column densities ($14.0 <$ log $N(\HI) < 19.0$). The weak {\MgII}
systems in our sample span the redshift range $0.4 < z <
2.4$. Therefore we consider two separate Cloudy grids modeled using
the integrated ionizing photon density ($h\nu \geq 1$~Ryd) at $z = 1$
and $z = 2$.  The difference of $\sim 0.5$~dex in the intensity
of the extragalactic background radiation field between these two redshifts
does not critically affect the output of the Cloudy models. Nonetheless, to have
a more tenable comparison between the data and the models, we plot 
the $z < 1.5$ and $z \geq 1.5$ systems on the $z=1$ and the $z=2$ grids 
respectively. We adopt a solar abundance pattern for the Cloudy models, but 
discuss the effects of abundance variations. In the following sections, we discuss the
constraints that the various ions provide towards the chemical and
ionization conditions in the absorbing gas.

\subsection {Constraints from {\MgII}}
\label{sec:4.1}

In our sample, the {\MgII} column densities of the individual clouds
in weak {\MgII} absorbers fall within the range $10^{11.0} < N(\MgII)
< 10^{13.3}$~{\cmsq}. Our search for weak {\MgII} systems along the 81
quasar lines of sight is 86\% complete down to the equivalent width
threshold of $W_r (2796) = 0.02$~{\AA}, corresponding to $N(\MgII)
\sim 10^{11.8}$~{\cmsq} \citep{anand07}. The $N(\MgII)$ is
useful to place limits on the metallicity of the low ionization gas
phase. Figure~\ref{fig:7} presents how the column densities of the various ions
change with respect to the ionization parameter, log $U$, for
different values of $N(\HI)$ and metallicity. For a given metallicity
and ionization parameter, i.e., a certain density, an increase in
$N(\HI)$ would correspond to an increase in the size of the
absorber. Also, with increasing ionization parameter, the neutral
fraction of hydrogen declines such that to converge on the same value
of $N(\HI)$, the size of the absorber has to further increase. It
is evident from Figure~\ref{fig:7} that for $N(\HI) = 10^{15}$~{\cmsq}, at
sub-solar metallicity (e.g. $0.1$Z$_\odot$), the model column density of {\MgII} is
inadequate to explain the observed $N(\MgII)$ even for the weakest
{\MgII} lines in our sample.  For a given log $U$, the column
densities of the ionization stages of various elements scale almost
linearly with both $N(\HI)$ and metallicity. Thus, higher $N(\MgII)$ can be
recovered by raising either $N(\HI)$ or metallicity. For example, at $N(\HI) =
10^{15}$~{\cmsq}, by raising the metallicity by 1 dex (to Z$_\odot$),
we find that the ionization models reproduce the observed column
densities in the weaker {\MgII} systems ($N(\MgII) < 10^{12}$~{\cmsq})
in our sample. For the same $N(\HI)$ value, at 10Z$_\odot$, a
substantial fraction of the range of observed $N(\MgII)$ is covered by
the ionization models, except for those systems with $N(\MgII) >
10^{13}$~{\cmsq}. Alternatively, with a 1 dex increase in $N(\HI)$,
the curves shift correspondingly such that systems with $N(\MgII) <
10^{12}$~{\cmsq} can be produced in 0.1Z$_\odot$ gas. However, for
$N(\HI) > 10^{17}$~{\cmsq}, the low ionization gas cloud is an
optically thick, Lyman-limit absorber (i.e. able to produce a break in
the spectrum of the background quasar at $\lambda = 912$~{\AA} in the
rest-frame of the absorber).

It can be concluded that the column density of {\MgII} is a suitable
parameter for estimating limits on the metallicity of the
absorber. Our sample of weak {\MgII} systems spans a range of 2 dex in
{\MgII} column density. Assuming solar abundance pattern, the
metallicity in many of their low ionization gas clouds is constrained
to be at least 0.1 Z$_\odot$ if the gas is optically thin in neutral
hydrogen ($N(\HI) < 10^{17}$~{\cmsq}, see \S~\ref{sec:8}).  
Moreover, the strongest {\MgII} lines ($N(\MgII) > 10^{13}$~{\cmsq}) 
among the weak systems require supersolar metallicity.

\subsection {Constraints from {\FeII}}
\label{sec:4.2}

In our sample of weak absorbers, 32\% (66/205) of {\MgII} clouds have
{\FeII} detected at the $> 3\sigma$ level, out of which 81\% are firm
detections (i.e. detections unaffected by blending with other
absorption features).  The column density ratio, N(\FeII)/N(\MgII),
falls between 0.02 and 4.0. The range of values for the ratio remains
unchanged even when we exclude upper limit measurements. Among the
clouds with {\FeII} detected, 13 are single cloud systems and the
remaining 53 are part of multiple cloud systems.

Constraints for metallicity, similar to the ones derived using
observed $N(\MgII)$, can also be derived based on $N(\FeII)$.  
Figure~\ref{fig:7} illustrates how the column density of {\FeII} changes with
ionization parameter for different values of $N(\HI)$ and metallicity.
At $N(\HI) \leq 10^{16}$~{\cmsq} and Z$ = 0.1$Z$_{\odot}$, $N(\FeII) <
10^{11.2}$~{\cmsq} which is inadequate to explain the observed column
density in systems with {\FeII} detected. By raising $N(\HI)$ by one
dex, we find a corresponding increase in the column density of {\FeII}
in the models, such that a column density of $N(\FeII) <
10^{12.2}$~{\cmsq} is possible at sub-solar metallicity. This still
does not account for some fraction ($\sim 10$\%) of the observed {\FeII} lines. With
metallicity increased to solar and super-solar values the models begin
to produce enough {\FeII} to explain the full range of observed values. In
general, we can infer that for the systems in which {\FeII} is
detected in our sample, the metallicity is constrained to values of Z
$\geq$ Z$_\odot$ if $N(\HI) < 10^{17}$~{\cmsq}.

It is evident from the column density comparison in Figure~\ref{fig:6} that, for
a given $N(\MgII)$, the observed $N(\FeII)$ has a spread of $\sim
1$~dex between the various systems. This spread is also evident in
Figure~\ref{fig:5} which compares the rest-frame equivalent widths.  For gas
that is optically thin, the ratio of column density between various
ionization stages does not depend on metallicity since all individual
column densities scale linearly. An exception to this can occur
(discussed in \S~\ref{sec:4.4}) for certain ions at supersolar
metallicities where cooling leads to much lower gas temperatures. In
the optically thin regime, for a given abundance pattern, the ratio of
column densities between different elements varies primarily with
ionization parameter. The relative strength of {\FeII} compared to
{\MgII} in a system is particularly sensitive to ionization parameter
for log $U > -4.0$. Thus we over-plot, in Figure~\ref{fig:8}, the observed column density
ratios of {\FeII} to {\MgII} on a Cloudy grid of photoionization
models. The Cloudy grid is for a range of log
$U$ and $N(\HI)$ at sub-solar, solar and super-solar metallicities.
The censored data points that occupy the left of Figure~\ref{fig:8} 
are systems in which {\MgII} is very weak. The {\FeII}, being even weaker, is not
detected at the $3 \sigma$ significance threshold. The $S/N$ of the
best of our sample of quasar spectra are comparable and therefore the
envelope of the ratio $N(\FeII)/N(\MgII)$ for censored data points is
seen as increasing with decreasing $N(\MgII)$.

We can place constraints of log $U$ assuming that the {\FeII} and
{\MgII} arise in the same phase.  To begin with, we notice that the
column density ratios of all clouds in our sample confine the
ionization parameter to log $U < -2.0$, corresponding to $n_H >
0.002$~{\cc} (for log $n_\gamma = -4.70$ at $z = 2$).  
At 0.1Z$_{\odot}$, the systems with {\FeII} detected require {\HI}
column densities greater than $10^{16}$~{\cmsq}. By increasing the
metallicity, the grids shift to the right proportionately such that
the same {\FeII} to {\MgII} ratios can now be recovered from weaker
{\HI} lines ($N(\HI) \sim 10^{14} - 10^{15}$~{\cmsq}). Thus, if the
low ionization gas is thin in neutral hydrogen, the metallicity in
systems where {\FeII} is detected is constrained to solar or
super-solar values. 

In the sample of 17 single cloud weak absorbers studied by Rigby
{\etal} (2002), a subset of systems with log $[N(\FeII)/N(\MgII)] > -0.3$
were classified as {\it iron-rich}. These systems were found to
have high metallicity ($ > 0.1 Z_{\odot}$), particularly strong
constraints on density (log $U < -4.0$, $n_{H} > 0.09$~{\cc}), and small
sizes [$N(\HI) + N(\HII) < 10^{18}$~{\cmsq}, R $ < 10 $~pc]. The
relatively high {\FeII} to {\MgII} ratio indicated that the high
density, low ionization gas in the {\it iron-rich} systems is not
$\alpha$-enhanced. Following the definition of \citet{weak2}, we find that 30
clouds in our sample are {\it iron-rich}, excluding censored data
points.  Comparing the data to the Cloudy grid, we find that the
ionization parameter in these systems is constrained to an upper limit
on the ionizing parameter between -3.2 and -3.7 depending on the
difference in ionizing photon number density between $z = 2$ and $z
=1$ respectively. A limit of log $U < -3.7$ translates to a density of
$n_H > 0.05$~{\cc} in the low ionization gas for log $n_\gamma =
-5.04$~{\cc} (the number density of ionizing photons with $h\nu \geq
13.6$~eV at $z = 1$).  The density constraint for the {\FeII} rich
systems translates into a small upper limit for the thickness ($R <
10$~pc) of the absorber.  In systems where {\FeII} is weak
compared to {\MgII}, the constraint on density is much lower
(log $U < -2.0$).

In this analysis, we have assumed a Solar abundance pattern.  Changing
the abundance of any element from this pattern would result in a
corresponding change in all ionization stages of that
element. Thus, the {\it iron-rich} systems can have a lower constraint on
density if the abundance of iron in the low ionization cloud is
enhanced relative to the solar abundance pattern, since the Cloudy
grids would be shifted upwards. Such an abundance pattern is not
physically well motivated.  On the other hand, an $\alpha-$enhanced
abundance pattern is ruled out for these {\it iron-rich} systems as its
effect would be to shift the Cloudy grids further down such that the
ionization models will not be able to reproduce the observed {\FeII}
to {\MgII} ratio. The $\alpha-$enhancement is, however, conceivable 
for the clouds in which {\FeII} is low compared to {\MgII}. The ionization
model, in that case, would infer higher densities for the low ionization
gas.

Finally, \citet{weak2} found that the $N(\FeII)/N(\MgII)$ in
their HIRES sample had a bimodal distribution with an apparent gap of
$\sim 0.5$~dex at $-0.8 <$ log $[N(\FeII)/N(\MgII)] < -0.3$. It was
therefore used as basis for defining the {\it iron-rich} systems, and
to suggest that there may be a separate class where {\FeII} is weak
relative to {\MgII}. Figure~\ref{fig:9}, shows the histogram distribution of
the {\FeII} to {\MgII} ratio for our sample of weak {\MgII} single and
multiple clouds. The bin size is equivalent to the gap in the
distribution that \citet{weak2} found for their sample. The
distribution from our sample does not suggest a bimodality, either for
single or multiple clouds. This remains true for smaller bin sizes as well.
Hence the apparent gap that was suggested
by the \citet{weak2} data can be attributed to inadequate
sample size. Additionally, we also note that there is no difference in the
observed {\FeII} to {\MgII} ratio between single and multiple cloud systems.
The individual clouds in the multiple cloud systems have the similar
log $U$ constraints as single clouds, with {\it iron-rich} systems detected
in both category. 

\subsection{Constraints from {\MgI}}
\label{sec:4.3}

In this section, we explain the constraints that are available from
the observed {\MgI} to {\MgII} ratio in weak systems.  In the past,
single phase photoionization models (using CLOUDY 90; \citet{ferland98})
have failed to reproduce the observed {\MgI} to {\MgII} ratio in some
strong {\MgII} systems \citep{rauch02, cwc03, ding03}. 
The {\MgI}/{\MgII} ratio derived from the models was
lower than the observed neutral to singly-ionized ratio.  To
circumvent this, a separate phase was proposed in which the {\MgI}
ionization fraction is higher \citep{cwc03, ding03}. 
This separate phase would have a higher density ($n_H >
1$~{\cc}) and lower temperature ($T < 600$~K) than the gas phase
associated with the {\MgII} absorption. The {\MgI} lines corresponding
to such low temperatures are very narrow ($b \sim 2$~{\kms}) and are
therefore unresolved at the $R=45,000$ of the earlier HIRES and UVES
observations. However, through superhigh spectral resolution
observations, at $R = 120,000$ ($\Delta v = 2.5$~{\kms}), it has been
demonstrated that the {\MgI} lines are not narrower than what is
derived for $R=45,000$ (Narayanan {\etal} 2007). 

Compared to {\FeII}, only a few weak {\MgII} systems in our sample
have {\MgI} detected at the $3\sigma$ level. Out of the 200 weak
{\MgII} clouds for which there is coverage, {\MgI} is detected in only
7 single cloud systems and in 20 clouds in multiple cloud
systems. Both single and multiple clouds span roughly the same range
of values for the {\MgI} to {\MgII} ratio, between -2.2 and -0.5,
considering only firm detections. The neutral magnesium fraction (Mg$^0$/Mg$_{total}$) 
is thus small, compared to the {\MgII} fraction ({\MgII}/Mg$_{total}$), in these systems. 
This most likely explains the large
scatter in the range of limits, evident in Figures~\ref{fig:5} and ~\ref{fig:6} , 
since we are
sampling a large number of quasar spectra with differences in
sensitivity. The spectra with the highest
$S/N$ ratio in our sample, however, constrain the {\MgI} column
density to values as low as $10^{9.5}$~{\cmsq}, $\sim 2$~dex smaller
than $N(\MgII)$, indicating that the neutral fraction in the {\MgII}
phase is indeed not very high.

Figure~\ref{fig:10} shows the observations compared to the grid of Cloudy
(version 07.02.01) ionization models. To begin with, the ionization
models are able to recover the observed {\MgI} to {\MgII} ratio from a
single phase. Compared to the CLOUDY (version 90) grid of ionization
models presented in Churchill {\etal} (2003), the models displayed in
Figure~\ref{fig:10}, have the {\MgI}/{\MgII} fraction higher by $\sim 0.5$~dex for a
given log $U$. The difference in the ionization fraction of magnesium
is a result of improvements in the rate coefficients for charge
transfer reactions, incorporated into the more modern versions of
Cloudy \citep{kingdon96}. The relevance of the charge transfer reaction  
(H + Mg$^+$ $\rightarrow$ H$^+$ + Mg) in controlling the ionization
fractions of  {\MgI} and {\MgII} has also been noted by \citet{tappe04}. For our sample of
weak {\MgII} systems, the ionization models derived from the revised
version of the photionization code suggests that a single phase
solution is possible. In fact, the observed ratio of {\MgI} to {\MgII}
in strong {\MgII} absorbers can also now be explained without invoking
a separate cold phase for {\MgI}.

We find that, in a large majority of the systems for which information on
{\MgI} is available, the ionization parameter is confined to log $U < -2.5$, for
solar and super-solar metallicity. At Z $<$ Z$_\odot$, the constraint
on ionization parameter is higher by $\sim 1$~dex. The fraction
{\MgI}/{\MgII} is expected to decrease with an increase in the ionization
conditions in the gas. This is evident in Figure~\ref{fig:7}. Therefore, the
systems with higher {\MgI} to {\MgII} ratios will have lower
constraints on ionization parameter (log $U < -3.0$) and
correspondingly higher constraints on density ($n_H > 0.02$~{\cc}),
identical to iron-rich systems. Moreover, if the {\HI} lines are
weaker ($N(\HI) < 10^{16}$~{\cmsq}), solar or supersolar metallicities
will be necessary to reproduce the observed {\MgI} and {\MgII} column
densities (see Figure~\ref{fig:10}) .

\subsection{Constraints from {\CII}}
\label{sec:4.4}

In our sample, all the 15 weak {\MgII} systems for which there is
coverage of {\CII}~$\lambda 1335$~{\AA} show prominent {\CII}
lines. Among these, eight are single cloud and the remaining are
multiple cloud systems. The ratio of $N({\CII})$ to $N({\MgII})$ is
always significantly greater than 1, with a range of values between 4
and 60.   A larger oscillator strength for the {\MgII}$\lambda$ 2796
line, and a longer wavelength compared to {\CII}$\lambda$ 1335, leads to
them having comparable equivalent widths.

The Cloudy grid of single phase photoionization models, comparing the
column density of {\CII} to {\MgII}, is shown in Figure~\ref{fig:11}. We
find that for most of the systems, the ionization parameter would be
constrained to log $U > -2.5$, implying a gas phase density of $n_H <
0.006$~{\cc}.  Such large values of log $U$ would be inconsistent
with that inferred for the low ionization phase for the clouds in
which {\FeII} is detected.  However, 
in Figure~\ref{fig:15}, we compare the observed {\CII}
to {\MgII} ratio against {\FeII} to {\MgII} in systems with
simultaneous coverage of both lines.  Only for two systems ($z =
1.585464$, and $z = 1.988656$), do we have firm (i.e. measurements that
are not limits) detections for both {\CII} and {\FeII}. In these two
systems we find $N({\FeII})$ to be $\sim 1$~dex smaller than
$N(\MgII)$, and therefore a constraint of log $U
> -2.5$ (see the $z \geq 1.5$ iron grid).  This is consistent with the log
$U$ derived using the {\CII} to {\MgII} ratio in the corresponding
systems allowing for a single low ionization
phase solution.  The density of this low ionization phase would
be smaller than what is estimated for the {\it iron-rich}
systems.  Unfortunately, such a definite statement cannot be extended
for all the other systems plotted in Figure~\ref{fig:15}, as their {\FeII}
measurements are upper limits. Nonetheless, even these limits are consistent
with a low density.

Many of the detected clouds in Figure~\ref{fig:8} did not appear in
Figure~\ref{fig:15} because {\CII} could not be measured for them due
to contamination or lack of coverage.  For these clouds the inferred
ionization parameters range from log $U \sim -2.2$ to $\sim -3.5$.  If
the {\CII} and {\MgII} arise in same phase, we would expect the
{\CII} in the clouds in which {\FeII} is detected to have
$N({\CII})/N({\MgII}) > 10$, such that consistent log $U$ values
would be derived from {\CII}. Such a direct comparison between the
measured {\CII} and {\FeII} may not apply because {\CII} may arise
partly from a higher ionization phase.  In addition to the low
ionization phase, most weak {\MgII} systems also have an associated
high ionization phase where the density is low ($n_{\H} \leq
10^{-3}$~{\cc}).  Although dominated by higher ionization states of
carbon ({\CIII} and {\CIV}), the {\CII} ionization fraction
(i.e. {\CII}/C$_{total}$) can be non-negligible in this phase. For
example, at log $U = -1.5$, $N({\HI}) = 10^{15}$~{\cmsq}, and 
Z = 0.3Z$_\odot$ (typical values derived from photoionization models,
e.g. see Table 5 of \citet{misawa07}), $N(\CII) =
10^{13.3}$~{\cmsq}, which is comparable to the detected {\CII} in many
of our clouds.  The $N(\MgII)$ contribution from this high ionization
phase is negligible.  In summary, even though {\CII} is detected in
all weak {\MgII} systems, since it does not arise exclusively in the low
ionization phase, it may not provide as robust a constraint on the
ionization parameter as does {\FeII}.

We note, also, that the grid with a metallicity of 10Z$_\odot$ does
not cover many of the {\CII} data points.  At this metallicity,
temperatures fall to $T < 1000$~K because of metal cooling, even
at log $U > -1.5$.  The gas, including both magnesium and carbon,
is less heavily ionized at the low temperatures, but the
effect is stronger for {\MgII} so that the density of {\MgII} 
is larger relative to {\CII}.  Clouds that have supersolar metallicity constraints,
based on other transitions, would then need to have a large
contribution to the {\CII} from a separate phase.

\subsection {Constraints from {\SiII}}
\label{sec:4.5}

In our sample, we find that the {\SiII} column density
is comparable to the column density of {\MgII}, as shown in Figure~\ref{fig:6}.
The ratio of column densities has values between 0.2 and 3.2, with majority of them
at $\sim 1$.  Most of the clouds have {\SiII} to {\MgII} ratios that
fall on the grid of Cloudy models in Figure~\ref{fig:12}.  However, the grids do not provide
much leverage in determining log $U$ since {\SiII} and {\MgII} are
similar over most of the parameter space, particularly for solar and
higher metallicities.

The single phase ionization models suggest that systems in which
$N(\SiII) < N(\MgII)$ require log $U > -1.5$ and high metallicity,
assuming a Solar abundance pattern. This is evident in the Cloudy
grids where data points corresponding to low {\SiII} to {\MgII} column
density ratio are below the model expectations for Z $<$
Z$_\odot$. Only for supersolar metallicities, do any of the models
reproduce the low {\SiII} to {\MgII} ratio. This is also evident in
Figure~\ref{fig:7} where the {\SiII} and {\MgII} ionization curves cross only at
super-solar metallicity for log $U > -1.5$, corresponding to $n_H <
10^{-3}$~{\cc}. The low {\SiII} to {\MgII} ratio can also be obtained
from single phase models by lowering the abundance of silicon compared
to other $\alpha-$process elements, in which case the metallicity could
be lower, and the density higher.

In general, the {\SiII} does not provide a robust constraint on the
ionization parameter for the low ionization gas.  In the small number
of clouds for which both {\SiII} and {\CII} are covered they usually
provide consistent constraints on log $U$, taking into account that
some fraction of the {\CII} can arise in a separate phase.

\subsection {Constraints from {\AlII}}
\label{sec:4.6}
  
Figure~\ref{fig:13} shows the Cloudy grid of photoionization models, with
measurements of {\AlII} to {\MgII} overplotted.  Within the sample of
systems, we find that $N({\AlII)}$ is always smaller than
$N({\MgII})$, sometimes by as much as 0.5 to 1.5 dex.  For a solar or
smaller metallicity, many systems (those with $N({\AlII})\geq 0.1
N({\MgII})$) are covered by the photoionization grids.  Particularly
at solar metallicity, the ratio of {\AlII} to {\MgII} is not very
sensitive to ionization parameter, so it cannot be used to effectively
measure this quantity.  Of special note are a number of systems, both
at high and low redshift, with $N({\AlII})\leq 0.1 N({\MgII})$.  These
points are not covered by the grid for all metallicities, based on a
solar abundance pattern.  The 10Z$_\odot$ grid does extend (for $\log
U > -1.5$) to somewhat lower values of {\AlII} to {\MgII}.  A
supersolar metallicity could help to explain some systems with a low
{\AlII} to {\MgII} ratio, such as the $z=1.68079$ system towards Q~$0429-4091$
described in Misawa {\etal} (2007).  However, for many of these systems, the most likely
explanation would be a reduction of the aluminum abundance relative to
magnesium by up to $\sim$0.7 dex.  A reduction would effectively
shift the grids down by the same amount.  Such an abundance pattern is
feasible, since it is consistent with $\alpha$-enhancement.  We note
that the same shift should apply for these systems in the {\AlIII}
grids as well.

\subsection {Constraints from {\AlIII}}
\label{sec:4.7}
 
Among the 74 systems with simultaneous coverage of {\MgIIdblt}~{\AA}
and {\AlIIIdblt}~{\AA} lines, {\AlIII} is detected in 12 single cloud
systems and in 44 clouds in 20 multiple cloud systems. The ratio of
{\AlIII} to {\MgII} column density falls within the range 0.04 to
0.98, excluding upper limits. The Cloudy grids of single phase models
are shown in Figure~\ref{fig:14}.  Assuming that the {\AlIII} and {\MgII} are
produced in the same phase, for the {\AlIII} detections, the ionization
parameter ranges from $-3.5 <$ log $U < -1.0$.

In the previous section, in order to reconcile the {\AlII} to {\MgII}
grid with the points below that grid, we proposed a reduction of the
aluminum abundance relative to magnesium.  If this is applied to
the {\AlIII}, it shifts this grid downwards so that it does not
cover some of the data points.  If the systems with small {\AlII}
to {\MgII} also have small {\AlIII} to {\MgII} this is not a problem,
though it does require large log $U$ for even these systems.
This is found to be the case in Figure~\ref{fig:15}, where we have plotted
the ratio of {\AlIII} to {\MgII} versus {\AlII} to {\MgII}.  For all
of the systems below the {\AlII} grid (see Figure~\ref{fig:13}),
$N$({\AlIII})~$< 0.3 N({\MgII})$.  

Although we have not identified specific clouds for which {\AlIII} and
{\MgII} cannot arise in the same phase, we note that some of our {\AlIII}
detections imply large ionization parameters, log $U > -2.0$.  This is
even more the case if we rely on a decrease of the aluminum abundance.
So either there is a sub-population of weak {\MgII} clouds which are
of a higher ionization state, or some of the {\AlIII} is produced in a
higher ionization phase, such as the one giving rise to the bulk of the
{\CIV} absorption.

\subsection {{\AlIII} to {\AlII} ratio}
\label{sec:4.8}

In damped Ly-$\alpha$ (DLA) systems, the chemical abundance
estimations are often carried out under the assumption that the ionization
corrections are not significant, since the gas is expected
to be predominantly in the low ionization phase. However, the
detection of {\AlIII} lines at the same velocity as the low ionization
lines in several DLAs lead \citet{vladilo01} to investigate the
relevance of ionization corrections for these systems.  In their
analysis, \citet{vladilo01} observed that the
$N(${\AlIII}$)/N(\AlII)$ ratio in DLA systems exhibits an
anti-correlation with $N(\HI)$. This relationship was described as
intrinsic to DLAs, and was used to suggest that the {\AlIII} to
{\AlII} ratio in these systems could be a sensitive probe of the
ionization conditions in the gas. Using a sample of sub-DLA systems,
\citet{dz03} examined if this anti-correlation
extends to lower $N(\HI)$. They found the {\AlIII} to {\AlII}
ratio in sub-DLAs to be in the same range as for DLA systems. In other
words, the anti-correlation trend did not seem to extend to sub-DLAs
($10^{19} < N(\HI) < 10^{20.3}$~{\cmsq}). However, more recently
\citet{meiring07} found that the anti-correlation could apply
even to sub-DLAs, based on a different sample of systems.

In our sample, 28 weak {\MgII} clouds have measurements of both
{\AlIII} and {\AlII}, of which 15 are firm detections in both
(i.e. measurements that are not limits). In Figure~\ref{fig:16}, we plot their
ratio with respect to the corresponding $N(\MgII)$. We also plot, in
an adjacent panel, the {\AlIII} to {\AlII} ratio in DLA and sub-DLA
systems as a function of $N(\HI)$, based on information extracted from
the literature \citep{vladilo01, dz03, meiring07}. 
We find the {\AlIII} to {\AlII} ratio in weak
{\MgII} systems to be considerably higher than in DLAs and
sub-DLAs. On average, the ratio is $\sim 0.5 - 1$~dex higher than what
has been measured for the other two classes of systems. This indicates
that the ionization conditions are higher in weak {\MgII}
systems than in DLA or sub-DLA systems. 

Based on photoionization modeling (see \S~\ref{sec:4.1}) we have
concluded that a large fraction of weak {\MgII} clouds have a
metallicity of solar or higher if the clouds are optically thin in
neutral hydrogen.  The observed redshift number density of weak
{\MgII} absorbers is too large for all of them to be Lyman limit
systems (as explained in
\S~\ref{sec:8}).  If $N(\HI) < 10^{17}$~{\cmsq} for the weak systems
plotted in Figure~\ref{fig:16}, then we conclude that
the anti-correlation trend discovered by Vladilo {\etal} (2001) for DLA
systems, and supported by \citet{meiring07} for sub-DLA systems,
continues to lower $N(\HI)$ values.

\section {Evolution of the low ionization phase structure}
\label{sec:5}

One of our objectives in carrying out the chemical and ionization
analysis on a large sample of weak {\MgII} systems is to find out if
there are any evolutionary trends observable in the absorber
population.  Our VLT/UVES sample of weak {\MgII} systems span the
redshift interval $0.4 < z < 2.4$. Photoionization constraints have
already suggested that a range of ionization properties and
metallicities can be expected for the low ionization phase. It would be
unusual to assume that the entire population of weak {\MgII} systems
are tracing some unique type of physical process/structure, given these
variations and the large redshift interval surveyed.

To investigate, we compared the observed {\FeII} to {\MgII} ratio
between the various systems, as it is a reliable constraint on
density and chemical enrichment history. Figure~\ref{fig:17} shows the measured
$N(\FeII)/N(\MgII)$ as a function of redshift.
Because of the many non-restrictive limits at small $N({\MgII})$,
particularly for low redshift clouds, it is hard to evaluate whether
there is a significant relationship between $N(\FeII)/N(\MgII)$
and $z$.  In order to consider this issue, we separated clouds
with log $N({\MgII}) < 12.2$, those that were
likely to have only limits on {\FeII} (based on inspection of
Figure~\ref{fig:8}), and considered only the stronger
of the weak {\MgII} clouds, plotted in {\it red} color in the top panel of
Figure~\ref{fig:17}.  It appears
that there is an anti-correlation between $N(\FeII)/N(\MgII)$ and
$z$.  At high redshift, there is an absence of detections with
larger $N(\FeII)/N(\MgII)$ values, while at low redshift there
are many detections with large $N(\FeII)/N(\MgII)$ and few limits
that could even be consistent with small $N(\FeII)/N(\MgII)$
values.  We applied a Kolmogorov-Smirnov (K-S) test to compare the
distributions of $N(\FeII)/N(\MgII)$ at $z \geq 1.5$ and $z < 1.5$
for the clouds with log $N({\MgII}) > 12.2$.
In the cases where only upper limits are available, we conservatively
include these as values when performing the K-S test.  
The distributions are shown in histogram form in the
lower panel of Figure~\ref{fig:17}.  We find that
there is a probability of only P(KS) = 0.006 (KS statistic D = 0.505) that
the two samples are drawn from the same distribution.  The probability
is likely to decrease if upper limits could be replaced with actual
detections.  Thus we find that the
observed anti-correlation of $N(\FeII)/N(\MgII)$ with $z$ is
statistically significant for log $N({\MgII}) > 12.2$ clouds.

We have found an absence of log $N({\MgII}) > 12.2$ clouds at high
redshift with large values of $N(\FeII)/N(\MgII)$ and an apparent absence of
log $N({\MgII}) > 12.2$ clouds at low redshift with small values of
$N(\FeII)/N(\MgII)$.  There are a number of low redshift clouds with
limits that could be consistent with small values.  Furthermore,
for weaker clouds (with log $N({\MgII}) < 12.2$), there are some examples of
low $N(\FeII)/N(\MgII)$ values at low redshifts.  We conclude
that large $N(\FeII)/N(\MgII)$ clouds are present only at $z<1.5$ and
not at higher redshifts.  The other population with small
$N(\FeII)/N(\MgII)$ exists both at low and high redshifts.

As demonstrated earlier, systems in which log $N(\FeII)/N(\MgII) >
-0.3$ are constrained to have a high density (log $U < -3.7$, $n_H >
0.05$~{\cc}).  Those with a lower {\FeII} to {\MgII} ratio have lower
densities, ranging down to log $U < -2.0$, which corresponds to $n_H <
0.001$~{\cc}.  The observed trend in the {\FeII} to {\MgII} ratio with
redshift, therefore, could imply the absence of high density clouds in
the low ionization phase in weak absorbers at high-$z$.  Such
variations in the phase structure are plausible if the weak systems are probing a different
combination of astrophysical systems/processes at $z \sim 2$ and $z
\sim 1$. Furthermore, if the absorbers are optically thin {\HI}
clouds, then we are also seeing a change in the thicknesses
of the low ionization gas clouds, from kiloparsec-scale at $z \sim 2$
to a range of values including both parsec-scale and kiloparsec-scale
clouds at $z \sim 1$.

Alternatively, gas clouds that are enriched primarily by Type II SNe
events will have [$\alpha$/Fe] $> 0$, in which case the observed
{\FeII} to {\MgII} column density ratios will be low.  Thus the
observed trend could also indicate that the weak {\MgII} clouds are
predominantly $\alpha$-enhanced at high redshift, with an increasing
contribution to the population at lower redshift from clouds with a
higher iron-abundance.  Increasing the [$\alpha$/Fe] in the Cloudy
models, would then lead to low {\FeII} to {\MgII} clouds having high
densities (log $U < -3.0$, $n_H > 0.01$~{\cc}), similar to the {\it
iron-rich} clouds.  The relevance of abundance pattern variations is
discussed in detail in \S~\ref{sec:8}, where we speculate on the
physical origin of these absorbers at the two redshift epochs.

\section{Weak {\MgII} Absorbers \& Satellites of Strong {\MgII} Systems}
\label{sec:6}

Strong {\MgII} systems are understood to be absorption arising in the
disk and extended halos of normal galaxies \citep{bergeron91, steidel94, steidel02}.  
Their broad ($\Delta$v
$\sim 150$~{\kms}) and kinematically complex {\MgII} line profiles are
found to be consistent with this picture \citep{charlton98}. A characteristic 
feature in many $z \sim 1$ strong {\MgII} systems is
weak, kinematic subsystems separated in velocity from the dominant
absorption component \citep{cwcvogt01}. Such kinematic subsystems are
likely to be gas clouds in the extended halo of the absorber in an
arrangement analogous to the Galactic high velocity cloud (HVC) and 
intermediate velocity cloud (IVC) populations.  In
earlier work, we hypothesized that a non-negligible fraction of weak
{\MgII} systems could be the extragalactic analogs of Milky Way HVCs,
in which a random line of sight intercepts the surrounding halo
cloud(s), but misses the optically-thick absorber. The possibility of
such an event is favored strongly by some recent observations which
find a patchy distribution (less than unity covering factor) for the
gas in the extended halos of galaxies \citep{tripp05, cwc07}. 
For a patchy halo, a sight line that passes only
through the clouds in the halo is more likely to produce a weak
{\MgII} system than a strong one \citep{cwc05}.  Our
hypothesis was primarily based on the observed evolution in the
redshift number density ($dN/dz$) of weak {\MgII} systems and the
evolution in the gas kinematics of strong {\MgII} absorbers over the
same redshift interval of $0.4 < z < 2.4$ \citep{anand07, mshar07}.

To extend this postulate further, and also to test its validity, we
compared the {\FeII} to {\MgII} ratio for the low ionization gas in weak
{\MgII} systems to that for the satellite clouds of strong {\MgII} systems
presented in \citet{cwcvogt01}. Based on an observed break in
the velocity distribution of Voigt Profile components in their sample
of strong {\MgII} systems, \citet{cwcvogt01} specified clouds
at $|\Delta v| > 40$~{\kms} as intermediate velocity or high velocity
subsystems (i.e. satellite clouds). The satellite clouds in that
sample were separated in velocity by as much as $|\Delta v| \sim
350$~{\kms} from the system center, with a median value of $|\Delta v|
= 165$~{\kms}. In Figure~\ref{fig:18}, we plot the {\FeII} to {\MgII} column
density ratio of these subsystems and compare it to the same in our
sample of weak {\MgII} clouds. We have included only those weak
absorbers that are within $0.4 \leq z \leq 1.2$, equivalent to the
redshift interval of the \citet{cwcvogt01} sample.

The comparison shows that the weak {\MgII} clouds closely resemble the
satellite clouds of strong {\MgII} systems. To begin with, the column
density of {\MgII} in the satellite clouds spans roughly the same
range of values as that of weak {\MgII} systems. The {\FeII} to
{\MgII} in the satellite clouds have a scatter which is also
comparable to the scatter in weak absorbers at the same redshift. A
Kolmogorov-Smirnov test estimates that the two
samples are consistent with being drawn from the same distribution [P(KS) =
0.633, D=0.196]. Comparable to the subset of {\it iron-rich} weak absorbers are
several satellite clouds with $N(\FeII) \sim N(\MgII)$, which
consequently constrains their density to $n_{\H} > 0.05$~{\cc}. In
addition, a significant subset of the satellite clouds also have much
lower densities ($n_{\H} < 0.001$~{\cc}), which make them analogous
to the $\alpha$-enhanced weak {\MgII} clouds.

\section {Summary}
\label{sec:7}

Using a large sample of recently discovered weak {\MgII} systems
\citep{anand07}, we have derived constraints on the chemical
and ionization conditions in their low ionization gas. In addition to
{\MgII}, we have measured the equivalent widths and column densities of
a number of other prominent metal lines associated with these
absorbers. The significant results reported in this paper can be
summarized as follows:

\noindent 1. In our sample of 100 weak {\MgII} systems, we find that
only 48\% are single cloud absorbers. This fraction is smaller than
the past results of \citet{weak2} where the majority (67\%) of
weak {\MgII} absorbers were found to be single cloud systems, but is
consistent within errors. The VLT/UVES sample that we consider in this
paper is a factor of $\sim 5$ larger than the Keck/HIRES sample used
by \citet{weak2}. We find no evidence for an evolution in the
ratio of single to multiple cloud absorbers over $0.4 < z < 2.4$.

\noindent 2. We find the equivalent widths and column densities of {\CII}
and {\SiII} are well correlated with the equivalent widths of {\MgII}, with minimal
scatter in the respective relationships. The column densities of
{\CII} and {\SiII} yield the following relationships with
{\MgII}; log $N(\CII) = (0.82 \pm 0.10)$ log $N(\MgII) + 3.42$, and
log $N(\SiII) = (1.02 \pm 0.25)$ log $N(\MgII) - 0.26$. The presence
of a significant correlation in the equivalent widths, extends the possibility of using {\CII}
and {\SiII} as proxy doublets for detecting analogs of weak {\MgII}
systems at $z > 2.5$ in the optical spectra of quasars.

\noindent 3. If a large fraction of weak {\MgII} clouds are sub-Lyman limit 
systems (i.e. optically thin in {\HI} with $N(\HI) <
10^{17}$~{\cmsq}), then the observed column density of {\MgII}
constrains the metallicity in the low ionization gas to Z $\geq 0.1$
Z$_\odot$. We also find the neutral fraction of magnesium to be very
low in almost all weak {\MgII} clouds, approximately $\sim 2$~dex
smaller than the corresponding $N(\MgII)$.

\noindent 4. From assuming a solar abundance pattern, we find
that the clouds for which $N(\FeII) \sim N(\MgII)$ have their
ionization parameters constrained to log $U < -3.7$, corresponding to
$n_H > 0.05$~{\cc}. If the low ionization gas is optically thin in
neutral hydrogen, then this places an upper limit of $R < 10$~pc on
the thickness of these gas clouds.  Similarly, clouds with
$N(\FeII) << N(\MgII)$ are constrained to have higher ionization
parameters (log $U \sim -2$ in some cases) and lower densities.
If the weak {\MgII} clouds, in which {\FeII} is observed to be weak 
relative to {\MgII}, are $\alpha$-enhanced, then that would yield 
higher constraints on density similar to the $N(\FeII) \sim N(\MgII)$ 
absorbers.

\noindent 5. In the past, ionization models using CLOUDY (version 90) have often 
not succeeded in recovering the observed {\MgI} to {\MgII} ratio in
both strong and weak {\MgII} systems. The ionization fraction of
{\MgI}, compared to {\MgII}, predicted by the models was not
sufficiently large to explain the observed
$N(\MgI)/N(\MgII)$. Therefore, a separate, cold (T $\sim 500$~K), high
density (n$_{\H} > 1$~{\cc}) phase, centered at the same velocity as
the {\MgII} phase was proposed in order to recover the observed {\MgI} in
the models. However, in the current version of Cloudy (ver 07.02.01),
with improvements in the rate coefficients of charge transfer
reactions, the model {\MgI} to {\MgII} fraction is higher by $\sim
0.5$~dex for a given ionization parameter log $U$. Such an increase
makes it consistent for {\MgI} to be in the same low ionization phase
as {\MgII}, in both weak as well as strong {\MgII} systems.

\noindent 6. Most of our {\CII} and {\SiII} measurements are for systems
at $z > 1.5$. In single phase models, the constraints from {\CII} and 
{\SiII} are typically high for the ionization parameter (log $U > -2.5$),
which is inconsistent with the constraints derived for clouds in which
$N(\FeII) \sim N(\MgII)$. However, we also find an evolution in the
relative strength of {\FeII}, compared to {\MgII}, such that towards
higher redshift ($z > 1.5$) there might be a paucity of {\it iron-rich}
systems (see Figure~\ref{fig:17}) . The absorbers in our sample, for 
which there is simultaneous coverage
of {\CII}, {\MgII} and {\FeII}, suggest that the $N(\FeII)$ could be sufficiently
small compared to $N(\MgII)$ in the high redshift clouds. Moreover, a non-negligible
fraction of {\CII} can arise in the high ionization gas, traced by {\CIII} and {\CIV},
such that {\CII},  in itself, cannot be used to determine the physical
conditions in the low ionization gas in weak absorbers. 

\noindent 7. We find that deviations from a solar abundance pattern
is required to explain the observed column density of {\AlII} in many
weak {\MgII} clouds. In particular, systems in which $N(\AlII) < 0.1
N(\MgII)$ require the abundance of aluminum in the low ionization gas
to be lowered by $\sim 0.7$~dex, consistent with $\alpha$-enhacement.
Models with super-solar metallicity generally produce less {\AlII}
relative to {\MgII}, but some reduction of the aluminum abundance is
still required for many clouds.  When the abundance of aluminum is
reduced, models underpredict {\AlIII} absorption unless the ionization
parameter is high, which is sometimes inconsistent with that derived from
other ions.  This suggests that {\AlIII}, like {\CII}, sometimes arises
partly in a separate, higher ionization phase.

\noindent 8. In our sample, we find a relative absence of weak {\MgII}
clouds with $N(\FeII) \sim N(\MgII)$ at high redshift ($z \geq 1.5$)
compared to many detections of $N(\FeII) \sim N(\MgII)$ towards
low-$z$. This observed trend can be interpreted in two ways : (1) an
absence of high density, low ionization gas at high-$z$ and/or (2) the presence
of  [$\alpha$/Fe] $ > 0$ in weak {\MgII} clouds at high-$z$. The other
population of weak absorbers, in which $N(\FeII) << N(\MgII)$, are 
detected at all intervals within $0.4 < z < 2.4$.

\noindent 9. We find similarities between the observed column density
of {\MgII} as well as the {\FeII} to {\MgII} column density ratio in weak {\MgII} clouds
and the high velocity subsystems (i.e. satellite clouds) of strong {\MgII} absorbers. 
The range of $N(\MgII)$ and $N(\FeII)/N(\MgII)$ for the two groups are comparable. 
This could be suggestive of the fact that some fraction of weak absorbers could be
probing a similar type of physical structure as the satellites of strong {\MgII} systems.

\section {Discussion}
\label{sec:8}

Weak {\MgII} absorbers have been identified over a large redshift
interval $0 < z < 2.4$ \citep{weak1, anand05, anand07}, corresponding 
to a great majority 
of the history of the universe. Within this interval, their redshift
number density ($dN/dz$) is found to be evolving, with a peak value of
$dN/dz = 1.76$ at $<z> = 1.2$ \citep{anand07}. Towards lower
redshift, the decrease in number density follows the expected curve
for a non-evolving population (for a $\Lambda$CDM concordance model)
\citep{anand05}. At $z > 1.2$, the $dN/dz$ has been found to
decrease rapidly, such that an extrapolation to $z > 3$ would yield a
value of zero. In other words, the observed redshift number density
does not suggest that a significant population of weak {\MgII} systems
exists at $z > 3$. In contrast, the number density of Lyman limit
systems (LLSs) has been found to increase towards high redshift. At $z
\sim 0.7, 1.5$ and 3, the $dN/dz$ of LLSs is estimated to be 0.7, 1.1
and 1.9, respectively \citep{strengler95, sargent89}. These values are 
in turn closely matched by the redshift number
density of strong {\MgII} systems (with $W_r(2796) > 0.3$~{\AA}) at
those same redshifts \citep[e.g.][]{nestor05}. Therefore, a
substantial fraction of the observed weak {\MgII} clouds at $z\sim0.7$
and $z\sim 1.5$ ought to be gaseous structures that are optically thin
in {\HI} (i.e. sub-Lyman limit systems with $N(\HI) <
10^{17}$~{\cmsq}). This, consequently, would constrain the metallicity
in the low ionization gas of many weak {\MgII} absorbers to Z $\geq$
Z$_\odot$, in order to reproduce the observed column density of
{\MgII}. Detailed photoionization models, where information on the
{\HI} column density has been available, further support this
inference \citep{charlton03, masiero05, misawa07}.

Bearing in mind these observed number statistics and constraints on
the chemical and ionization conditions described in this work, we now
proceed to discuss the plausible hosts of these low ionization, high metallicity 
weak absorbers.  Given the range of ionization properties 
and chemical abundances, it would be unusual to assume that the entire 
population of weak {\MgII} systems would
correspond to one unique type of astrophysical process/structure at 
all redshifts.

\citet{schaye07} have recently suggested that weak {\MgII} clouds
are likely to arise in gas ejected from starburst supernova-driven
winds during an intermediate stage in free expansion, before settling
in pressure equilibrium with the surrounding IGM.  The galaxy
populations detected at high redshift ($z \sim 2 - 3$) are found to be
rapidly star-forming, with a star-formation rate of $10 - 100$~M$_\odot$ yr$^{-1}$
\citep[e.g.][]{pettini01, choi06}. The starburst events
associated with these could give rise to galactic scale outflows that can
displace large amounts of chemically enriched gas from the ambient ISM
into the extended halo \citep[e.g.][]{heckman01, pettini01}. The strong 
clustering of {\CIV} systems with Lyman break
galaxies, which dominate the star formation rate at high-$z$, is possibly
a signature of such outflows \citep{adelberger03, adelberger05a}. Such 
supernova driven winds are observed to have a multiphase
structure, with a non-negligible fraction of the interstellar gas in a
warm neutral phase ($T < 10,000$~K) traced by such lines as {\CII},
{\SiII}, {\FeII} and {\AlII} \citep{schwartz06} and a cold neutral 
component ($T \sim 100$~K)
detected in {\NaI} \citep{heckman00, rupke02}.  A
sight line that directly intercepts the outflow close to the
starburst region is likely to produce a very strong, saturated, and
kinematically broad absorption feature \citep{bond01}. However,
as described in \citet{schaye07}, as the wind material moves
farther into the outskirts of the extended halo of the galaxy, the
column densities would decrease in response to a decreasing density in
the ambient medium.  At this stage, fragments in the wind, generated
through hydrodynamical instabilities, would manifest as weak {\MgII}
clouds, and later as weaker {\CIV} absorption associated with {\HI}
lines in the Ly-$\alpha$ forest \citep{zonak04, schaye07}.

The interstellar clouds, ejected from correlated supernova events, are
likely to be highly chemically enriched because of the close
association with the feedback from star formation. \citet{simcoe06}
discovered evidence for such chemically enriched gas (Z $>
0.1$Z$_\odot$) at $z \sim 2.3$, at distances of $\sim 100 - 200$~kpc
from luminous star-forming galaxies, which they interpret as feedback
from supernova winds or perhaps tidally stripped gas. The low
ionization lines such as {\MgII}, {\FeII}, {\AlII}, {\CII}, and {\SiII} in the absorbers
presented in that study have column densities similar to 
those of weak {\MgII} clouds in our sample. Material that is directly related
to star-forming events is likely to have [$\alpha$/Fe] $> 0$.  Weak
{\MgII} clouds associated with such events would therefore have
$N(\FeII) << N(\MgII)$.  This is consistent with the dominant
population of high redshift ($z \sim 2$) weak {\MgII} clouds, and with
some fraction of the clouds towards low ($z \sim 1$) redshift.  The
high metallicity weak {\CIV} absorption clouds presented in \citet{schaye07} were estimated to have sizes that are small (R $\sim100$~pc),
less than the Jean's length for self-gravitating clouds, implying that
they are likely to be short lived.  Such a transient physical nature 
is also a feature of weak {\MgII} clouds (Narayanan {\etal} 2005), and
is anticipated for relics of winds.

The metal enriched interstellar gas expelled from the disk would
resemble the high velocity gaseous structures surrounding the Milky Way, 
as they move through the galaxy's halo. 
\citet{ellison04}, from estimating the coherence scales of low and high ionization gas
associated with weak absorbers, have suggested a scenario in which weak
{\MgII} absorption could arise in the outskirts of ordinary galaxies,
where the filling factor of the low ionization clouds (i.e. number of
clouds per cubic parsec) is small compared to that in the center.  For
low ionization gas, the coherence scale is $\sim 2$~kpc, i.e.,
there is a high probability of seeing weak {\MgII} absorption along
two lines of sight separated by this distance \citep{ellison04}.
However, the separate low ionization phases must not fully cover this $\sim 2$~kpc
region, since individual absorbing clouds are not seen along both
lines of sight separated by tens to hundreds of parsecs \citep{rauch99}.  
Also, photoionization models have shown that cloud
line-of-sight thicknesses are often $< 100$~pc.  This suggests a
clustering of separate clouds on a $\sim 2$~kpc scale, as well
as implying a flattened geometry for the coherent structure, consistent
with the findings of \citet{milni06}.  This scale could
be consistent with dwarf galaxies \citep{ellison04} or with 
tidal streams. Sight lines through gas stripped in tidal interactions 
of galaxies can also produce sub-Lyman limit systems, and related weak {\MgII}
absorbers. Gas that is tidally stripped in merger or accretion events
could also form stars and provide a source of enriched gas clouds
to the halos of high redshift galaxies.  These would be analogous to
the Milky Way circumgalactic gaseous streams, related to accretion of
interstellar gas from satellite galaxies.

In this context, we emphasize that the Milky Way analogs of weak {\MgII} absorbers are not likely to be the HVC complexes 
detected in 21cm and/or H$\alpha$ emission, since those have $N(\HI) > 10^{18}$~{\cmsq} \citep{wakker97, putman03}. The weak absorbers must instead correspond to a population of halo clouds with  lower {\HI} column densities. Spectroscopic observations in the ultraviolet along various sight lines through the Milky Way halo have detected such high velocity gas in which the {\HI} column densities are sub-Lyman limit \citep[$N(\HI) \sim 10^{16.5}$~{\cmsq}; ][]{collins04, fox05, ganguly05}. These clouds, which exhibit multiple gaseous phases, have the column densities of {\CII}, {\SiII} and {\FeII} constrained to values similar to what we find for these ions in our weak {\MgII} sample. Using the HST/STIS archival spectra of quasars, \citet{richter08} have identified a population of high velocity clouds in the Milky Way halo in which $N(\HI) < 10^{18}$~{\cmsq}, with a few having $N(\HI) \lesssim 10^{17}$~{\cmsq}. The low ionization metal lines associated with these halo clouds are kinematically narrow and weak, identical to the high-$z$ weak {\MgII} systems. Two of the sight lines that cover {\MgII} measure $W_r(2796) < 0.2$~{\AA}. These observations lend further support to the proposition that at least some fraction of the weak {\MgII} absorption systems are likely to have their physical origin in gas clouds residing in the halos of high-$z$ galaxies.

The star formation per co-moving volume is known to be roughly
constant between $z \sim 1$ and $z \sim 3$ \citep[e.g.][]{bouwens03, wang06}. 
In this same interval however, the number
density of weak {\MgII} clouds has been found to decline from a value
of $dN/dz = 1.76$ at $z = 1.2$ to $dN/dz = 0.65$ at $z=2.2$ \citep{anand07}. 
If a significant fraction of weak {\MgII} clouds,
especially those in which $N(\FeII) << N(\MgII)$, form in supernova
driven outflows from star-forming galaxies, then it would seem
perplexing that a declining trend is observed for $dN/dz$ from $z \sim
1$ to $z \sim 2.4$.  We would expect $dN/dz$ of weak {\MgII} absorbers to be 
not decreasing so drastically if they were {\it all}
directly connected to interactions and outflows.  However, we find a
spread in the physical properties of weak {\MgII} absorbers, and an
evolution in these properties from $z \sim 1$ to $z \sim 2$.
We would expect that those weak {\MgII} absorbers with
$N({\FeII}) << N({\MgII})$ could be consistent with an origin
in superwind condensations, since that process would lead to
$\alpha$-enhancement and could produce high metallicities.  We
have found such absorbers both at $z \sim 1$ and $z \sim 2$, and
their numbers are roughly consistent with a constant $dN/dz$ for
the sub-population, and thus with a constant star formation rate over the 
same interval.

We then must also consider the sub-population of weak {\MgII}
absorbers with $N({\FeII}) \sim N({\MgII})$.  These objects have a
$dN/dz$ that peaks at $z \sim 1$, and they are apparently rare at $z
\sim 2$.  When we combine the two populations, the $N({\FeII}) \sim
N({\MgII})$ and the $N({\FeII}) << N({\MgII})$ clouds, the result is a
$dN/dz$ that declines from $z \sim 1$ to $z \sim 2.4$, as is observed.
The evolution and the physical properties of the $N({\FeII}) \sim
N({\MgII})$ absorbers provide clues to their origins.  It has been shown that, to reproduce the observed solar composition of iron,  
nucleosynthetic yields from Type Ia events have to included \citep{timmes95}. 
Thus, the weak {\MgII} clouds in which the {\FeII} to {\MgII} column density ratio approximately reflects Solar abundance (i.e. $N(\FeII) \sim N(\MgII)$, see Figure~\ref{fig:6}) should also be Type Ia-enriched. Enrichment by Type Ia supernovae requires that there is a $\sim 1$ billion year delay from
the onset of star formation until the elements produced in the
Type Ia enter the interstellar medium.  What is needed to explain
these clouds is a process that peaks at $z \sim 1.5$ ($\sim 1$~Gyr
before the peak of $dN/dz$ at $z =1.2$) to produce
the stars that subsequently give rise to the iron enrichment.
It has been noted in \citet{lynch07} and \citet{misawa07} that this peak 
closely compares to the peak in the global star
formation rate in dwarf galaxies \citep{kauffmann04, bauer05}.
More generally, for Type Ia SNe to contribute to enrichment, the
absorbing structure must persist for more than $1$ billion
years.  This may exclude superwinds from starbursts and tidal
debris, that are likely to be relatively short-lived.  However,
it would be consistent with Type Ia enriched gas trapped in
the potential wells of dwarf galaxies, or with intergalactic
star-forming structures in the cosmic web \citep{weak2, milni06}.  
Either of these sites could
also be consistent with the coherence lengths of low ionization gas
as estimated in \citet{ellison04}. 

In the local universe, \citet{stocke04} and \citet{keeney06} have 
associated weak {\MgII} absorbers with unbound winds from dwarf starburst 
galaxies. In their respective
examples, a post-starburst galaxy is identified at impact parameters of $71h^{-1}$~kpc
and $33h^{-1}$~kpc from the line of sight. Compared to
massive star-forming galaxies, the halo escape velocities are smaller
for dwarfs and hence, they are more efficient in transporting metal
enriched gas clouds into intergalactic environments. Yet, outflows
associated with starbursts from dwarfs are also likely to be $\alpha$-enhanced,
such that the $N(\FeII) \sim N(\MgII)$ weak absorbers would remain largely
unexplained. Thus the clouds that have large $N({\FeII})$ require 
a site that has been previously enriched by local star formation, such as
the dwarf galaxies themselves.

We conclude, based on the range of physical properties derived for the weak
{\MgII} absorbers, that they arise from at least two types of astrophysical
processes.  The $N(\FeII) << N(\MgII)$ clouds, some of which are $\alpha$--enhanced,
are produced in processes related to the concerted action of massive stars, such as superwinds.
They thus evolve in number along with the global star formation rate in massive
galaxies, so we would expect roughly a constant $dN/dz$ from $z\sim 1$ to $z\sim 5$
from this population.  The $N(\FeII) \sim N(\MgII)$ clouds, which would be enriched
by local Type Ia supernova, could be housed in relatively dense pockets within
the potential wells of dwarf galaxies or intergalactic structures.  Because the
star formation rate in dwarfs peaks at a later epoch ($z \sim 1.5$), we expect
these clouds to emerge from $z\sim 2$ to $z\sim 1$.  The two populations combined
would lead to a gradual increase in the total weak {\MgII} absorber $dN/dz$
from $z\sim 2.4$ to $z\sim 1$.  This picture would predict that the $dN/dz$
of weak {\MgII} absorbers would remain constant from $z\sim 2.4$ up to higher
redshifts, because the star formation rate in massive galaxies was constant.

However, the redshift interval $1 < z < 3$ also corresponds to the
epoch over which  hierarchical structure growth and mergers are most
active. The fraction of interacting galaxies, and proto-galaxies
with irregular luminosity profiles,
is observed to be higher towards high redshift \citep[$z > 1$, $\sim 40$\%, ][]{abraham96, vandenbergh96, conselice03, elmegreen05}. Indirect evidence of this
dynamical evolution of galaxies was also noticed by \citet{mshar07} in the evolution of the kinematic profiles of strong {\MgII} systems. Towards high redshift ($z \sim 2$), the
kinematic profiles in a number of strong {\MgII} lines were found to
be particularly complex, with absorption in multiple clouds
that were linked with each other continuously in velocity.  This
is suggestive of ongoing accretion events.  In contrast, $z \sim 0.5$
strong {\MgII} systems typically have a distinct region of strong
absorption due to several blended clouds, surrounded by one or more
weaker high velocity components (i.e. satellite clouds).  This
is the expected absorption signature for quiescent disk/halo galaxies.
It was proposed by Narayanan {\etal} (2007) that the superpositions
of the numerous halo clouds present in high redshift proto-galactic
structures lead to a rarity of isolated weak {\MgII} absorbers during
that epoch.  Instead, the weak {\MgII} absorbers would be consolidated
into a stronger {\MgII} absorber.  The result would be a deficit of
weak {\MgII} absorbers at $z>2.5$.  Near-IR high resolution spectroscopic observations 
are needed to determine the $dN/dz$ of weak {\MgII} absorbers at $z > 2.5$
and to determine their metallicities, ionization conditions, and
chemical abundances.

\acknowledgements
We express our gratitude to the ESO for the public data archive and for providing the MIDAS UVES pipeline. The authors are grateful to Gary Ferland for answering questions related to the charge transfer reaction rates and also for making the Cloudy photoionization code openly available. We also wish to thank an anonymous referee for providing valuable suggestions which improved the scope of the paper significantly. AN is thankful to Eric Feigelson for several useful discussions on the application of survival analysis in statistical calculations. This research was funded by NASA under grants NAG5-6399 and NNG04GE73G and by the National Science Foundation (NSF) under grant AST-04-07138.


\clearpage
\LongTables
\begin{landscape}
\begin{deluxetable}{lllcrrrrrrr}
\tabletypesize{\scriptsize} 
\tablewidth{0pt}
\tablecaption{EQUIVALENT WIDTH OF METAL LINES ASSOCIATED WITH WEAK {\MgII} ABSORBERS}
\setlength{\tabcolsep}{0.06in}
\tablehead{
\colhead{QSO}&
\colhead{$\lambda~{\AA}$}&
\colhead{z$_{abs}$}&
\colhead{type}&
\colhead{{\MgII} 2796}&
\colhead{{\MgI} 2853}&
\colhead{{\FeII} 2383}&
\colhead{{\AlII} 1671}&
\colhead{{\CII} 1335}&
\colhead{{\SiII} 1260}&
\colhead{{\AlIII} 1855}
}
\startdata
3c336 & 3530 - 6650 & 0.702901 & s & $	0.028 {\pm} 0.004	$ & $	< 0.010	$ & $	< 0.009 	$ & $	...	$ & $	... 	$ & $	...	$ & $	...	$	\\

CTQ 0298 & 3520 - 8550 & 1.256069 & s & $	0.057 {\pm} 0.004	$ & $	< 0.004	$ & $	...	$ & $	...	$ & $	... 	$ & $	...	$ & $	...	$	\\

Q 0001-2340 & 3060 - 10070 & 0.452394 & m & $	0.138 {\pm} 0.003	$ & $	< 0.019 ^*	$ & $	0.019 {\pm} 0.001	$ & $	...	$ & $	...	$ & $	...	$ & $	...	$	\\

 &  & 0.685957 & s & $	0.035 {\pm} 0.001	$ & $	< 0.002 	$ & $	...	$ & $	...	$ & $	...	$ & $	...	$ & $	< 0.008 	$	\\
	
 & & 1.651484 & s & $	0.077 {\pm} 0.001	$ & $	0.010 {\pm} 0.001	$ & $	< 0.004 ^* 	$ & $	< 0.003	$ & $	0.075 {\pm} 0.001	$ & $	< 0.081 ^* 	$ & $	0.004 {\pm} 0.001	$	\\

Q 0002-4220 & 3160 - 10070 & 1.446496 & s & $	0.042 {\pm} 0.000	$ & $	0.002 {\pm} 0.000	$ & $	...	$ & $	...	$ & $	...	$ & $	...	$ & $	< 0.001	$	\\

 &  & 1.988656 & m & $	0.276 {\pm} 0.001	$ & $	... 	$ & $	0.016 {\pm} 0.003	$ & $	0.081 {\pm} 0.001	$ & $	0.171 {\pm} 0.001	$ & $	0.180 {\pm} 0.001	$ & $	0.045 {\pm} 0.002	$	\\
	
Q 0011+0055 & 3770 - 10000 & 0.487264 & s & $	0.244 {\pm} 0.019	$ & $	< 0.040 	$ & $	   ...	$ & $	...	$ & $	...	$ & $	...	$ & $	...	$	\\
	
&  & 1.395635 & s & $	0.186 {\pm} 0.004	$ & $	< 0.013	$ & $	...	$ & $	...	$ & $	...	$ & $	...	$ & $	0.043 {\pm} 0.008 	$	\\
	
 &  & 1.777926 & s & $	0.127 {\pm} 0.003	$ & $	< 0.051	$ & $	...	$ & $	0.059 {\pm} 0.006	$ & $	...	$ & $	...	$ & $	...	$	\\
	
Q 0013-0029 & 3060 - 9890 & 0.635069 & m & $	0.205 {\pm} 0.014	$ & $	< 0.005 	$ & $	< 0.053 ^*	$ & $	...	$ & $	...	$ & $	...	$ & $	...	$	\\
	
&   & 0.857469 & m & $	0.150 {\pm} 0.005	$ & $	< 0.005 	$ & $	0.014 {\pm} 0.004	$ & $	...	$ & $	...	$ & $	...	$ & $	0.023 {\pm} 0.003	$	\\
	
 &   & 1.146770 & m & $	0.047 {\pm} 0.001	$ & $	< 0.009	$ & $	0.004 {\pm} 0.001	$ & $	...	$ & $	...	$ & $	...	$ & $	< 0.034	$	\\
	
Q 0042-2930 & 3530 - 6800 & 0.798665 & m & $	0.239 {\pm} 0.007	$ & $	< 0.007	$ & $	0.063 {\pm} 0.004	$ & $	...	$ & $	...	$ & $	...	$ & $	...	$	\\
	
&   & 1.091866 & s & $	0.162 {\pm} 0.005	$ & $	< 0.012	$ & $ 0.69 {\pm} 0.001	$ & $	...	$ & $	...	$ & $	...	$ & $< 0.011	$	\\
	
Q 0100+130 & 3520 - 10000 &1.758694 & m & $	0.108 {\pm}  0.008	$ & $	< 0.013	$ & $	...	$ & $	...	$ & $	0.125 {\pm} 0.004	$ & $	...	$ & $	...	$	\\

 &		 & 2.298494 & m & $	0.232 {\pm}  0.004	$ & $	 ...	$ & $	< 0.003	$ & $	...	$ & $	< 0.291 $ & $	0.081 {\pm} 0.002	$ & $	...	$	\\
	
Q 0109-3518 & 3060 - 10070 & 0.769646 & s & $	0.044 {\pm} 0.001	$ & $	0.002 {\pm} 0.000	$ & $	< 0.001	$ & $	...	$ & $	...	$ & $	...	$ &	$	< 0.005 ^*	$	\\
	
 & 	& 0.896004 & m & $	0.020 {\pm} 0.001	$ & $	< 0.001 	$ & $	< 0.003 	$ & $	< 0.009 	$ & $	...	$ & $	...	$ & $	...	$	\\
	
 &	& 1.182696 & m & $	0.135 {\pm} 0.001	$ & $	0.009 {\pm} 0.001	$ & $	0.016 {\pm} 0.001	$ & $	< 0.020 ^*	$ & $	...	$ & $	... 	$ &	$	< 0.030 ^*	$	\\
	
Q 0122-380 & 3060 - 10190 &0.822606 & m & $	0.269 {\pm} 0.003	$ & $	0.020 {\pm} 0.002	$ & $	0.034 {\pm} 0.003	$ & $	...	$ & $	...	$ & $	...	$ & $	< 0.020 ^*	$	\\
	
 & 	& 0.910117 & m & $	0.060 {\pm} 0.004	$ & $	< 0.006 	$ & $	< 0.003 	$ & $	...	$ & $	...	$ & $	...	$ & $	...	$	\\
	
 & 	& 1.174224 & s & $	0.020 {\pm} 0.001	$ & $	< 0.007	$ & $	< 0.006	$ & $	< 0.005	$ & $	... 	$ & $	...	$ & $	< 0.005	$	\\
	
 & 	& 1.450076 & m & $	0.061 {\pm} 0.003	$ & $	< 0.009	$ & $	...	$ & $	< 0.005	$ & $	...	$ & $	...	$ & $	< 0.004	$	\\
	
 & 	& 1.911015 & m & $	0.199 {\pm} 0.006	$ & $	< 0.015	$ & $	0.020 {\pm} 0.002	$ & $	0.037 {\pm} 0.001	$ & $	0.104 {\pm} 0.001	$ & $	0.084 {\pm} 0.001	$ & $	0.015 {\pm} 0.001	$	\\
	
 &	& 1.974182 & m & $	0.271 {\pm}  0.020	$ & $	< 0.019	$ & $	< 0.008	$ & $	0.020 {\pm} 0.001	$ & $	0.251 {\pm} 0.006	$ & $	< 0.323 ^*	$ & $	0.057 {\pm} 0.007	$	\\
	
Q 0128-2150 & 3050 - 6800 &1.398315 & s & $	0.018 {\pm} 0.001	$ & $	...	$ & $	< 0.016	$ & $	...	$ & $	< 0.050	$ & $	...	$ & $	...	$	\\
	
 & 	& 1.422086 & s & $	0.042 {\pm} 0.001	$ & $	...	$ & $	...	$ & $	...	$ & $	< 0.060	$ & $	...	$ & $	...	$	\\
	
Q 0130-4021 & 3550 - 6800 & 0.962487 & m & $	0.089 {\pm} 0.004	$ & $	< 0.011 	$ & $	...	$ & $	...	$ & $	...	$ & $	...	$ & $	...	$	\\
	
Q 0136-231 & 3500 - 6640 & 1.261761 & s & $	0.102 {\pm} 0.003	$ & $	< 0.013	$  & $	< 0.009	$ & $	0.026 {\pm} 0.003	$ & $	...	$ & $	...	$ & $	0.021 {\pm} 0.003	$	\\
	
 & 	 & 1.285796 & s & $	0.021 {\pm} 0.003	$ & $	< 0.013	$ & $	 < 0.009	$ & $	< 0.011	$ & $	...	$ & $	...	$ & $	< 0.010	$	\\
	
 & 	& 1.353662 & m & $	0.170 {\pm} 0.004	$ & $	... 	$ & $	< 0.019	$ & $	...	$ & $	...	$ & $	...	$ & $	0.029 {\pm} 0.003	$	\\
	
Q 0141-3932 & 3060 - 10000 & 1.781686 & s & $	0.039 {\pm} 0.001	$ & $	< 0.011	$ & $	< 0.008	$ & $	0.006 {\pm} 0.001	$ & $	0.025 {\pm} 0.001	$ & $	0.020 {\pm} 0.001	$ & $	0.004 {\pm} 0.000	$	\\
	
Q 0151-4326 & 3060 - 10070 & 0.737248 & s & $	0.022 {\pm} 0.001	$ & $	< 0.001	$ & $	< 0.002	$ & $	...	$ & $	...	$ & $	...	$ & $	< 0.003	$	\\
	
 & 	& 1.708492 & s & $	0.026 {\pm} 0.001	$ & $	< 0.002	$ & $	0.005 {\pm} 0.001	$ & $	...	$ & $	...	$ & $	...	$ & $	0.003 {\pm} 0.001	$	\\
	
Q 0237-23 & 3060 - 10070 & 1.184624 & m & $	0.140 {\pm} 0.002	$ & $	0.004 {\pm} 0.000	$ & $	0.009 {\pm} 0.001	$ & $	0.033 {\pm} 0.002	$ & $	...	$ & $	...	$ & $	< 0.016	$	\\
	
Q 0328-272 & 3500 - 6630 & 0.570827 & s & $	0.168 {\pm} 0.008	$ & $	< 0.013	$ & $	< 0.044 ^* 	$ & $	...	$ & $	...	$ & $	...	$ & $	...	$	\\
	
 &	& 1.269042 & m & $	0.105 {\pm} 0.007	$ & $	< 0.016	$ & $	< 0.015	$ & $	< 0.031	$ & $	...	$ & $	...	$ & $	< 0.017	$	\\
	
Q 0329-2550 & 3060 - 10070 &0.992899 & m & $	0.279 {\pm} 0.002	$ & $	< 0.001	$ & $	0.027 {\pm} 0.002	$ & $	...	$ & $	...	$ & $	...	$ & $	0.005 {\pm} 0.001	$	\\
	
&	 & 1.398230 & m & $	0.025 {\pm} 0.001	$ & $	< 0.006	$ & $	< 0.003	$ & $	...	$ & $	...	$ & $	...	$ & $	...	$	\\
	
Q 0329-3850 & 3070 - 8500 &0.929608 & s & $	0.073 {\pm} 0.002	$ & $	< 0.003	$ & $	< 0.003	$ & $	< 0.007 	$ & $	...	$ & $	...	$ & $	< 0.004 	$	\\
	
 &	& 0.970957 & s & $	0.055 {\pm} 0.001	$ & $	< 0.003 	$ & $	< 0.004	$ & $	< 0.011 	$ & $	...	$ & $	...	$ & $	< 0.031 ^*	$	\\
	
Q 0429-4901 & 3050 - 10080 & 0.584249 & s & $	0.016 {\pm} 0.002	$ & $	< 0.004 	$ & $	< 0.004 	$ & $	...	$ & $	...	$ & $	...	$ & $	...	$	\\
	
 &	& 1.680766 & s & $	0.015 {\pm} 0.001	$ & $	...	$ & $	< 0.012	$ & $	< 0.009	$ & $	0.012 {\pm} 0.001	$ & $	< 0.024	$ & $	0.005 {\pm} 0.001	$	\\
	
Q 0453-4230 & 3060 - 10070 &0.895865 & s & $	0.034 {\pm} 0.001	$ & $	< 0.001	$ & $	< 0.001	$ & $	< 0.003 ^*	$ & $	...	$ & $	...	$ & $	...	$	\\
	
 & 	& 1.039517 & m & $	0.189 {\pm} 0.003	$ & $	...	$ & $	0.105 {\pm} 0.001	$ & $	...	$ & $	... 	$ & $	... 	$ & $	...	$	\\
	
Q 0549-213 & 3500 - 6640 & 1.343495 & s & $	0.181 {\pm} 0.010	$ & $	...	$ & $	< 0.012	$ & $	< 0.12	$ & $	... 	$ & $	...	$ & $	< 0.013	$	\\
	
Q 0551-3637 & 3060 - 9370 &0.505437 & m & $	0.052 {\pm} 0.014	$ & $	< 0.013	$ & $	...	$ & $	...	$ & $	...	$ & $	...	$ & $	...	$	\\
	
 & 	& 1.491972 & s & $	0.176 {\pm} 0.004 	$ & $	< 0.018	$ & $	< 0.010	$ & $	0.043 {\pm} 0.005	$ & $	0.108 {\pm} 0.010	$ & $	...	$ & $	< 0.019	$	\\
	
Q 0810+2554 & 3050 - 6640 & 0.821727 & m & $	0.252 {\pm} 0.001	$ & $	0.019 {\pm} 0.001	$ & $	...	$ & $	...	$ & $	...	$ & $	...	$ & $	< 0.003	$	\\
	
 &	& 0.831511 & m & $	0.158 {\pm} 0.002	$ & $	0.004 {\pm} 0.001	$ & $	...	$ & $	< 0.008 	$ & $	...	$ & $	...	$ & $	< 0.031 ^*	$	\\
	
Q 0926-0201 & 3060 - 10000 &1.096336 & s & $	0.020 {\pm} 0.001	$ & $	< 0.001	$ & $	< 0.006	$ & $	< 0.013	$ & $	...	$ & $	...	$ & $	< 0.023	$	\\
	
 &	 & 1.232206 & m & $	0.069 {\pm} 0.004	$ & $	< 0.003	$ & $	0.010 {\pm} 0.001	$ & $	< 0.002	$ & $	...	$ & $	...	$ & $	< 0.005	$	\\
	
Q 0940-1050 & 3110 - 10070 & 2.174535 & s & $	0.035 {\pm} 0.001	$ & $	< 0.008	$ & $	< 0.005	$ & $	0.004 {\pm} 0.000	$ & $	< 0.029 ^*	$ & $	< 0.033 ^*	$ & $	0.011 {\pm} 0.000	$	\\
	
Q 1122-1648 & 3060 - 10070 & 0.806215 & m & $	0.249 {\pm} 0.001	$ & $	< 0.003	$ & $	0.032 {\pm} 0.001	$ & $	...	$ & $	...	$ & $	...	$ & $	0.045 {\pm} 0.001	$	\\
	
 &	 & 1.234140 & m & $	0.200 {\pm} 0.000	$ & $	0.010 {\pm} 0.000 	$ & $	0.017 {\pm} 0.000	$ & $	< 0.133	$ & $	...	$ & $	...	$ & $	0.028 {\pm} 0.000	$	\\
	
Q 1140+2711 & 3775 - 10000 & 2.196632 & m & $	0.193 {\pm} 0.002	$ & $	< 0.061 ^* 	$ & $	...	$ & $	...	$ & $	0.173 {\pm} 0.001	$ & $	< 0.295 ^*	$ & $	...	$	\\
	
Q 1151+068 & 3705 - 10000 & 1.153704 & s & $	0.108 {\pm} 0.003	$ & $	< 0.008	$ & $	< 0.004	$ & $	...	$ & $	...	$ & $	...	$ & $	< 0.016	$	\\
	
Q 1157+014 & 3520 - 7400 & 1.330502 & s & $	0.120 {\pm} 0.002	$ & $	< 0.007	$ & $	0.020 {\pm} 0.004	$ & $	0.024 {\pm} 0.006	$ & $	...	$ & $	...	$ & $	0.013 {\pm} 0.002	$	\\
	
Q 1158-1843 & 3070 - 10070 & 0.506041 & s & $	0.022 {\pm} 0.001	$ & $	< 0.003	$ & $	< 0.021 ^*	$ & $	...	$ & $	...	$ & $	...	$ & $	...	$	\\
	
 &	& 0.818119 & m & $	0.063 {\pm} 0.001	$ & $	< 0.001	$ & $	0.009 {\pm} 0.001	$ & $	...	$ & $	...	$ & $	...	$ & $	< 0.004 	$	\\
	
Q 1209+0919 & 3520 - 7770 & 1.264983 & s & $	0.083 {\pm} 0.007	$ & $	< 0.019	$ & $	< 0.015	$ & $	...	$ & $	... 	$ & $	...	$ & $	...	$	\\
	
Q 1229-021 & 3530 - 6650 & 0.700377 & s  & $	0.010 {\pm} 0.001	$ & $	< 0.004	$ & $	< 0.006	$ & $	...	$ & $	...	$ & $	...	$ & $	...	$	\\
	
 &	& 0.768862 & s & $	0.033 {\pm} 0.002	$ & $	< 0.005 	$ & $	< 0.005	$ & $	...	$ & $	...	$ & $	...	$ & $	...	$	\\
	
 &	& 0.830858 & m & $	0.134 {\pm} 0.000	$ & $	0.011 {\pm} 0.001	$ & $	0.032 {\pm} 0.002	$ & $	...	$ & $	... 	$ & $	...	$ & $	...	$	\\
	
Q 1418-064 & 3765 - 9945 & 1.516673 & s & $	0.075 {\pm} 0.003	$ & $	< 0.019 ^*	$ & $	...	$ & $	...	$ & $	...	$ & $	...	$ & $	< 0.047 ^*	$	\\
	
 &	& 2.174224 & s & $	0.178 {\pm} 0.004	$ & $	< 0.023 ^*	$ & $	0.018 {\pm} 0.003	$ & $	...	$ & $	...	$ & $	...	$ & $	...	$	\\
	
Q 1444+014 & 3520 - 5830 & 0.509719 & m & $	0.193 {\pm} 0.007	$ & $	0.022 {\pm} 0.003	$ & $	< 0.129 ^*	$ & $	...	$ & $	...	$ & $	...	$ & $	...	$	\\
	
 &	& 1.101989 & m & $	0.033 {\pm} 0.005	$ & $	< 0.007	$ & $	0.005 {\pm} 0.000	$ & $	...	$ & $	...	$ & $	...	$ & $	0.032 {\pm} 0.005	$	\\
	
Q 1448-232 & 3060 - 10070 & 1.019191 & m & $	0.033 {\pm} 0.005	$ & $	...	$ & $	< 0.002	$ & $	< 0.070	$ & $	... 	$ & $	...	$ & $	< 0.002	$	\\
	
 &	& 1.473201 & m & $	0.269 {\pm} 0.009	$ & $	0.011 {\pm} 0.000	$ & $	0.007 {\pm} 0.001	$ & $	0.024 {\pm} 0.001	$ & $	...	$ & $	< 0.116 ^* 	$ & $	< 0.051	$	\\
	
 & 	& 1.585464 & m & $	0.088 {\pm} 0.002	$ & $	0.008 {\pm} 0.001	$ & $	0.005 {\pm} 0.000	$ & $	0.005 {\pm} 0.001	$ & $	0.069 {\pm} 0.001	$ & $	0.044 {\pm} 0.001	$ & $	0.022 {\pm} 0.002	$	\\
	
Q 1621-0042 & 3530 - 6800 & 1.174521 & m & $	0.237 {\pm} 0.012	$ & $	0.004 {\pm} 0.001	$ & $	0.003 {\pm} 0.001	$ & $	...	$ & $	...	$ & $	...	$ & $	...	$	\\
	
Q 1629+120 & 3050 - 6800 & 1.379330 & m & $	0.142 {\pm} 0.007	$ & $	< 0.018	$ & $	< 0.017	$ & $	...	$ & $	...	$ & $	...	$ & $	...	$	\\
	
Q 2000-330 & 3495 - 9945 & 1.249864 & s & $	0.032 {\pm} 0.001	$ & $	< 0.003	$ & $	< 0.004	$ & $	...	$ & $	... 	$ & $	...	$ & $	...	$	\\
	
Q 2044-168 & 3520 - 9900 & 1.342525 & m & $	0.057 {\pm} 0.004	$ & $	< 0.013	$ & $	...	$ & $	< 0.014	$ & $	...	$ & $	...	$ & $	< 0.013	$	\\
	
Q 2059-360 & 3750 - 9280 & 1.242973 & s & $	0.015 {\pm} 0.001	$ & $	< 0.006	$ & $	...	$ & $	...	$ & $	...	$ & $	...	$ & $	< 0.009	$	\\
	
 &	& 1.399947 & m & $	0.109 {\pm} 0.002	$ & $	< 0.006	$ & $	0.006 {\pm} 0.002	$ & $	...	$ & $	...	$ & $	...	$ & $	< 0.011	$	\\
	
Q 2116-358 & 3530 - 6640 & 0.539154 & s & $	0.115 {\pm} 0.004	$ & $	< 0.018	$ & $	...	$ & $	...	$ & $	...	$ & $	...	$ & $	...	$	\\
 &	 & 0.775358 & m & $	0.183 {\pm} 0.010	$ & $	... 	$ & $	< 0.013	$ & $	...	$ & $	...	$ & $	...	$ & $	...	$	\\
	
Q 2132-433 & 3500 - 6640 & 0.793600 & m & $	0.184 {\pm} 0.007	$ & $	0.012 {\pm} 0.005	$ & $	0.030 {\pm} 0.007	$ & $	...	$ & $	...	$ & $	...	$ & $	...	$	\\
	
Q 2204-408 & 3520 - 6800 & 1.335279 & m & $	0.052 {\pm} 0.004	$ & $	< 0.009	$ & $	< 0.008	$ & $	...	$ & $	...	$ & $	...	$ & $	...	$	\\
	
Q 2206-199 & 3420 - 6640 & 0.948363 & m & $	0.265 {\pm} 0.007	$ & $	< 0.005	$ & $	0.016 {\pm} 0.002	$ & $	...	$ & $	...	$ & $	...	$ & $	< 0.007	$	\\
	
 & 	& 1.297044 & s & $	0.148 {\pm} 0.001	$ & $	0.008 {\pm} 0.001	$ & $	0.005 {\pm} 0.001	$ & $	< 0.064 ^*	$ & $	... 	$ & $	...	$ & $	< 0.005	$	\\
	
Q 2217-2818 & 3060 - 9890 & 0.599512 & s & $	0.115 {\pm} 0.003	$ & $	 0.005 {\pm} 0.001 	$ & $	< 0.016 ^*	$ & $	...	$ & $	...	$ & $	...	$ & $	...	$	\\
	
 & 	& 0.786515 & m & $	0.204 {\pm} 0.001	$ & $	< 0.001 	$ & $	0.025 {\pm} 0.005	$ & $	...	$ & $	...	$ & $	...	$ & $	< 0.028 ^*	$	\\
	
 &	& 1.054310 & m & $	0.046 {\pm} 0.002	$ & $	< 0.002	$ & $	< 0.002	$ & $	< 0.021	$ & $	...	$ & $	...	$ & $	...	$	\\
	
 &	& 1.082780 & m & $	0.125 {\pm} 0.001	$ & $	< 0.001	$ & $	< 0.001	$ & $	< 0.003	$ & $	...	$ & $	...	$ & $	< 0.003 ^*	$	\\
	
 &	& 1.200162 & m & $ 	0.099 {\pm} 0.002	$ & $	< 0.001 ^*	$ & $	< 0.002 ^* 	$ & $	...	$ & $	...	$ & $	...	$ & $	< 0.022 ^*	$	\\
	
 &	& 1.555849 & m & $	0.268 {\pm} 0.001	$ & $	< 0.076	$ & $	0.043 {\pm} 0.000	$ & $	0.036 {\pm} 0.000	$ & $	< 0.271 ^*	$ & $	0.125 {\pm} 0.002	$ & $	< 0.006  ^*	$	\\
	
Q 2222-3939 & 3530 - 6640 & 1.227553 & s & $	0.114 {\pm} 0.005	$ & $	< 0.019	$ & $	< 0.013	$ & $	< 0.013	$ & $	...	$ & $	...	$ & $	< 0.012	$	\\
	
Q 2225-2258 & 3050 - 10000 & 0.831374 & m & $	0.033 {\pm} 0.002	$ & $	< 0.003	$ & $	< 0.005 	$ & $	< 0.03 	$ & $	...	$ & $	...	$ & $	< 0.005 	$	\\
	
 &	 & 1.433018 & m & $	0.167 {\pm} 0.002	$ & $	< 0.011	$ & $	...	$ & $	0.036 {\pm} 0.003	$ & $	0.162 {\pm} 0.004	$ & $	0.010 {\pm} 0.010	$ & $	0.009 {\pm} 0.002	$	\\
	
Q 2243-6031 & 3140 - 10000 & 0.828081 & m & $	0.263 {\pm} 0.005	$ & $	< 0.003 	$ & $	< 0.007 	$ & $	...	$ & $	...	$ & $	...	$ & $	< 0.009	$	\\
	
 &	& 1.389707 & m & $	0.106 {\pm} 0.022	$ & $	< 0.003	$ & $	...	$ & $	< 0.007	$ & $	...	$ & $	...	$ & $	...	$	\\
	
 &	& 1.755704 & s & $	0.108 {\pm} 0.001	$ & $	0.007 {\pm} 0.001	$ & $	< 0.004 	$ & $	< 0.008	$ & $	< 0.106 ^*	$ & $	< 0.058 ^*	$ & $	...	$	\\
	
Q 2314-409 & 3520 - 6640 & 0.843114 & s & $	0.043 {\pm} 0.003	$ & $	< 0.008	$ & $	< 0.010 	$ & $	...	$ & $	...	$ & $	...	$ & $	...	$	\\
	
Q 2347-4342 & 3100 - 10070 & 1.109640 & s & $	0.040 {\pm} 0.004	$ & $	< 0.003	$ & $	< 0.003	$ & $	...	$ & $	...	$ & $	...	$ & $	...	$	\\
	
 &	& 1.405367 & s & $	0.074 {\pm} 0.001	$ & $	< 0.003	$ & $	< 0.004	$ & $	...	$ & $	< 0.095 ^*	$ & $	...	$ & $	< 0.030 ^*	$	\\
	
 &	& 1.796233 & s & $	0.160 {\pm} 0.001	$ & $	< 0.009	$ & $	0.010 {\pm} 0.001	$ & $	...	$ & $	< 0.163 ^*	$ & $	0.066 {\pm} 0.001	$ & $	0.004 {\pm} 0.000	$

\enddata
\tablecomments{$*$ indicates lines that were contaminated by absorption features at other redshifts. In most cases, the contamination was from {\HI} lines of the Ly-$\alpha$ forest. The second column gives the wavelength coverage of the UVES spectrum of each quasar, the third column gives the redshift of each absorber, and the fourth column indicates whether the weak {\MgII} absorption was in a single cloud (s) or in multiple clouds (m). Columns 5 - 11 are the total rest-frame equivalent widths of the respective lines.}
\end{deluxetable}
\clearpage
\end{landscape}
\clearpage

\medskip
\medskip
\begin{center}
\vspace{2 in}
\textsf{\large NOTE: Table 2 lists the velocity, column density and Doppler parameter for {\MgI}, {\MgII}, {\FeII}, {\CII}, {\SiII}, {\AlII} and {\AlIII} associated with the 100 weak {\MgII} systems in our sample. The table (18 pages) can be downloaded from \bf{http://www.astro.wisc.edu/$\sim$anand/vptable.pdf}}
\end{center}
\clearpage


\clearpage

\begin{landscape}
\begin{figure*}
\begin{center}
\epsscale{0.4}
\rotatebox{270}{\plotone{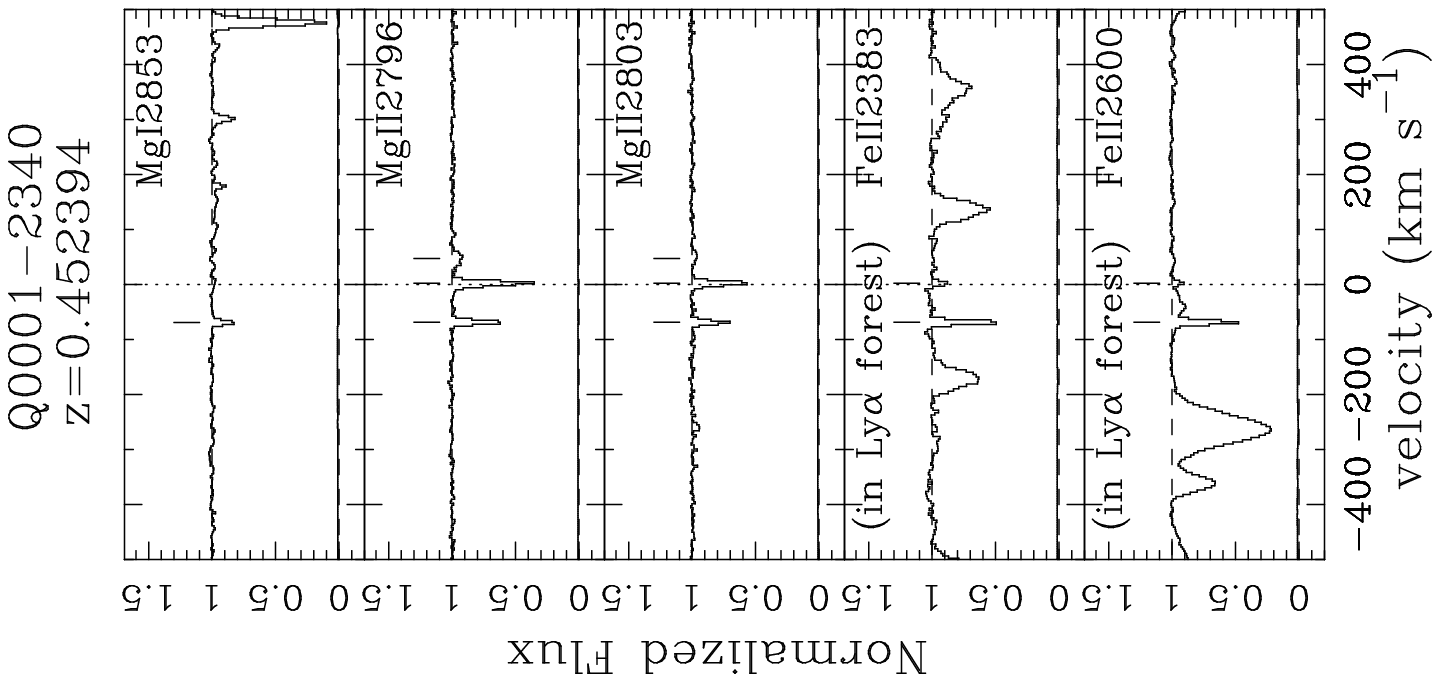}}
\hspace{8 mm}
\epsscale{0.4}
\rotatebox{270}{\plotone{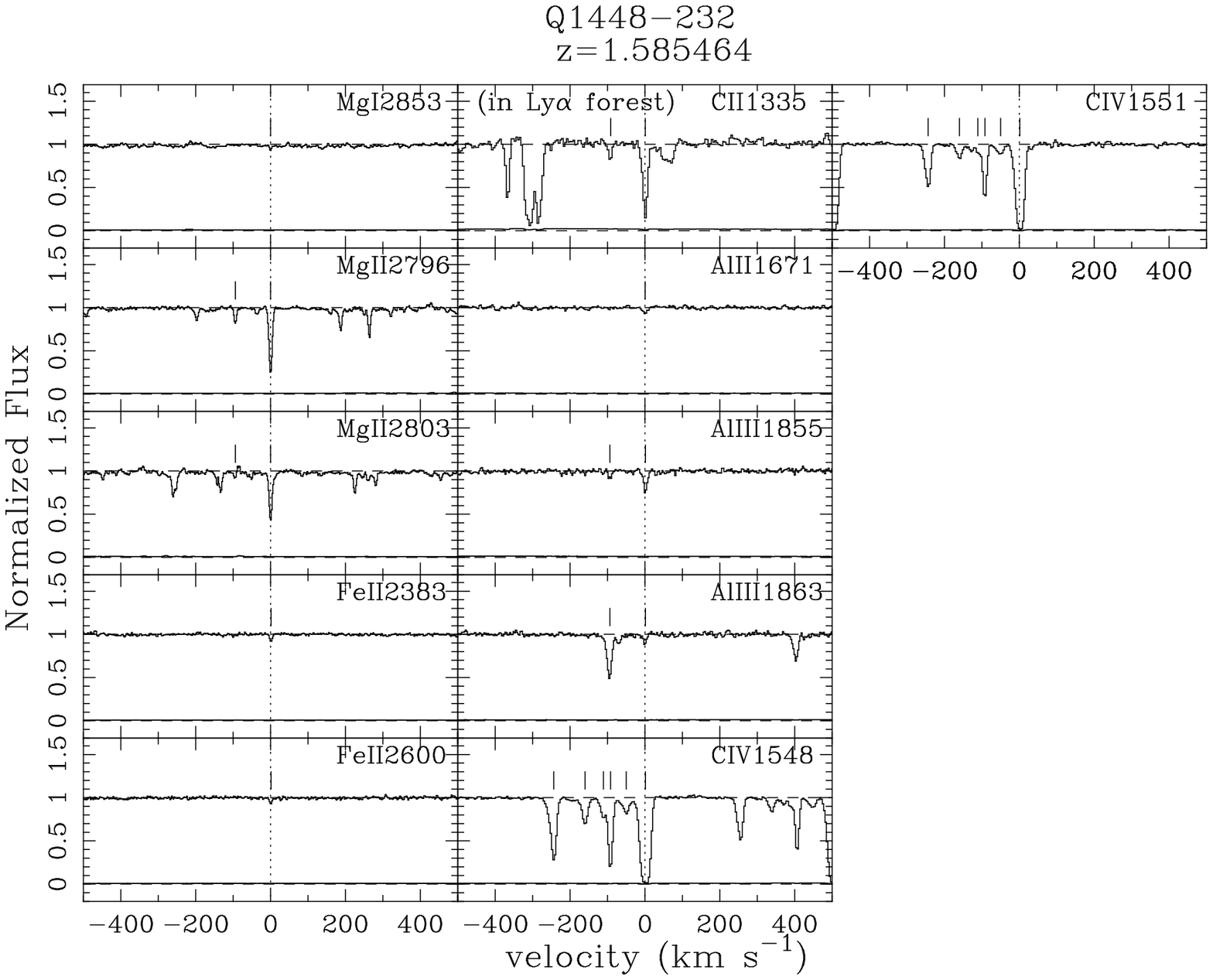}} \\
\epsscale{0.4}
\rotatebox{270}{\plotone{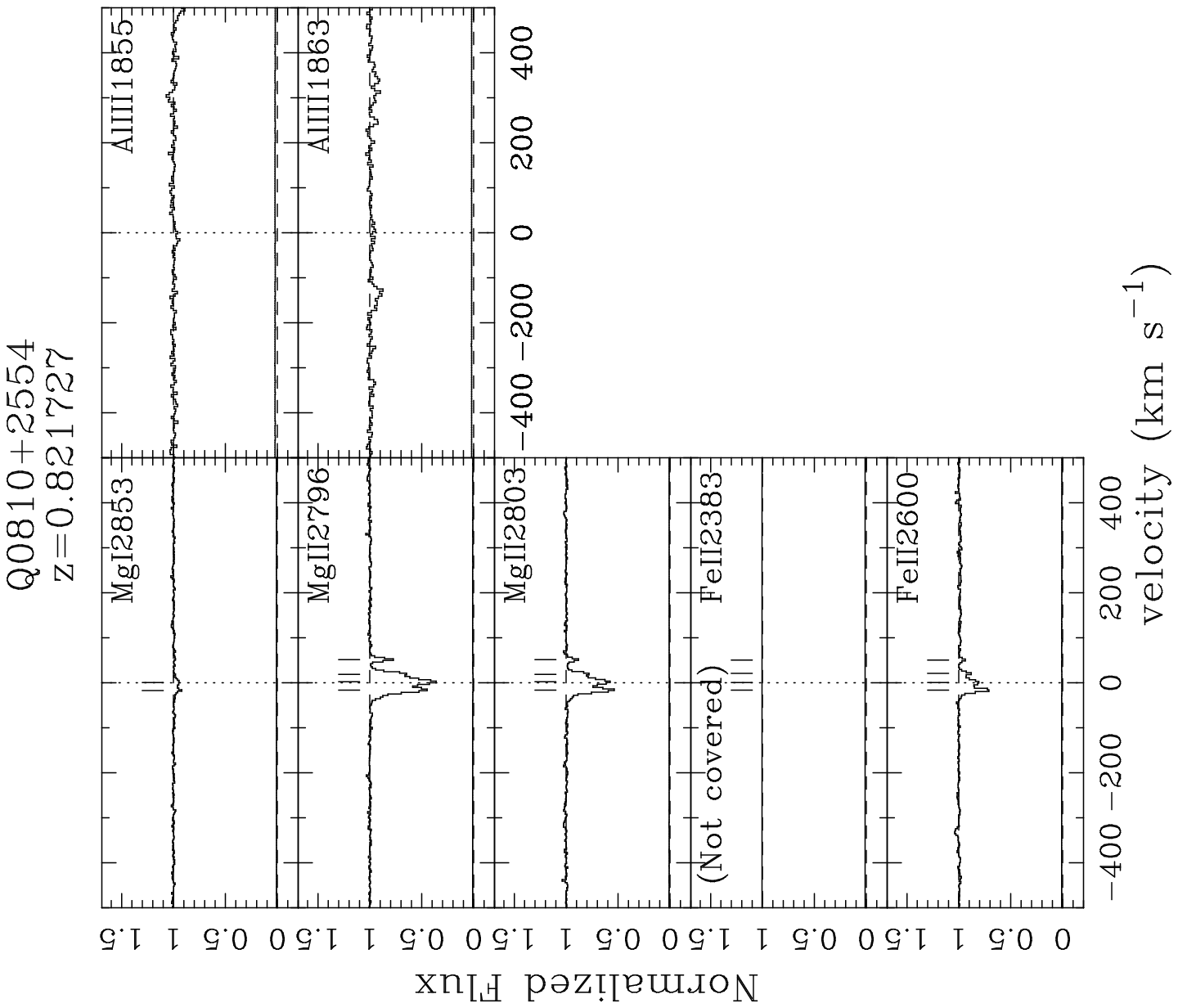}}
\hspace{1 mm}
\epsscale{0.4}
\rotatebox{270}{\plotone{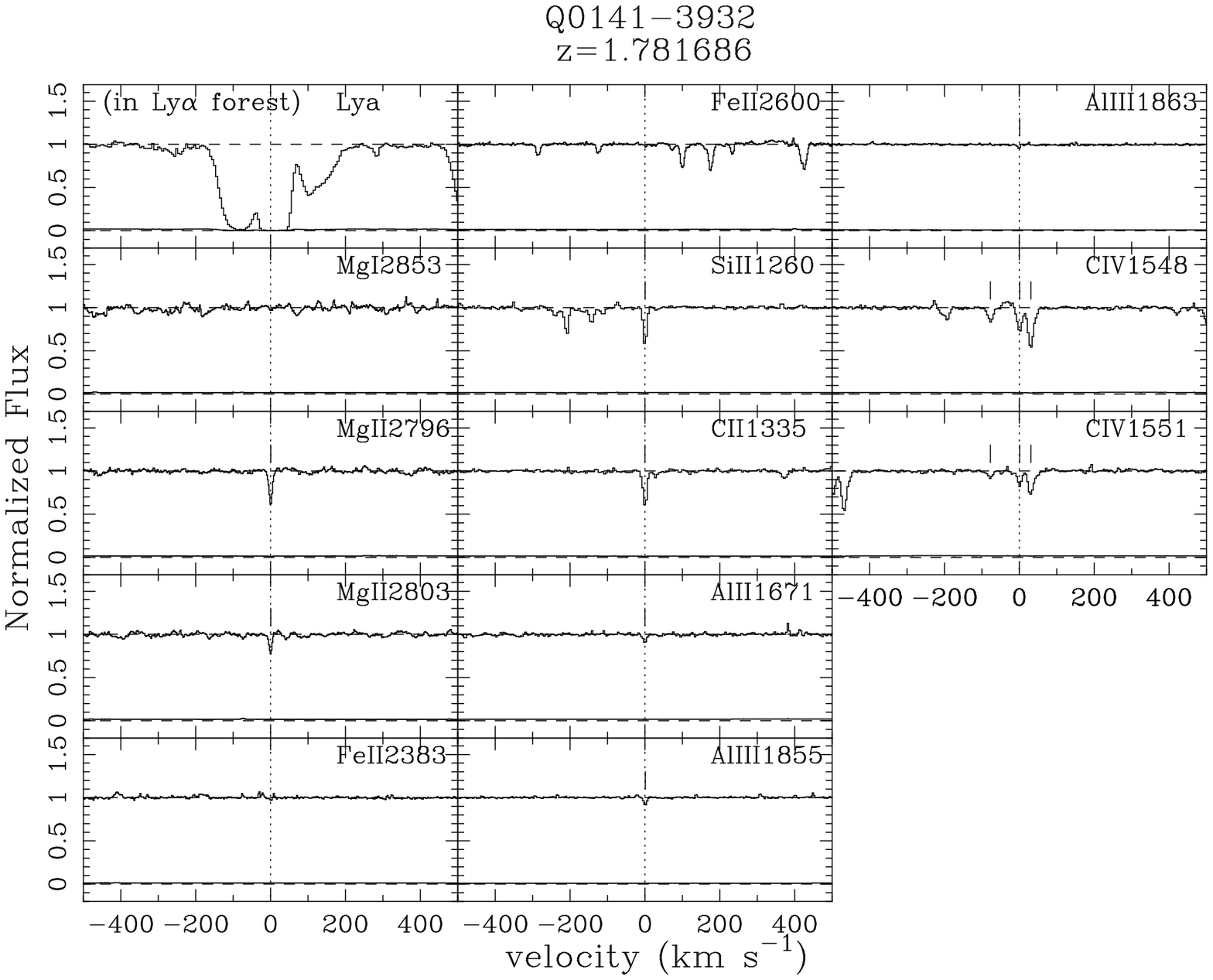}}
\end{center}
\protect
\caption{A few example systems plots from the $100$ weak {\MgII} absorbers in our sample represented on a velocity scale with the system redshift corresponding to $0$~{\kms}. The vertical axis is continuum normalized scale for the spectra. The vertical tick marks correspond to the centroid of the absorption components in each spectral feature, derived from Voigt profile fitting. Regions that are blueward of the Ly-$\alpha$ emission of the quasar are labeled in the respective panels as {\it in  Ly-$\alpha$ forest}. The quasar name and redshift of the system are also labeled in each system plot. ({\it Note: The full set of system plots will be presented in the online version of the journal.})}
\label{fig:1}
\end{figure*}
\clearpage
\end{landscape}

\begin{figure*}
\begin{center}
\epsscale{1.0}
\vspace{1.5 in}
\rotatebox{90}{\plotone{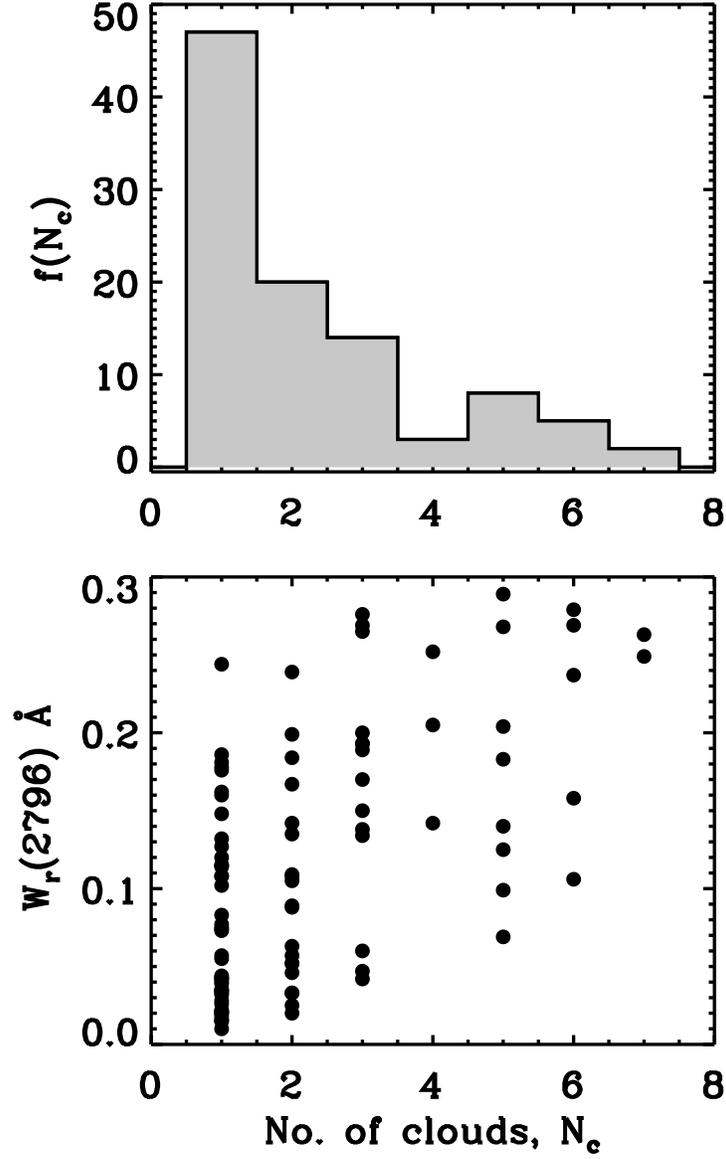}}
\end{center}
\protect
\caption{The {\it top panel} shows the frequency distribution of the number of weak {\MgII} 
clouds (i.e. number of Voigt profile components) in our sample. Single cloud systems account 
for 48\% of the total number of systems.The {\it bottom panel} compares the rest-frame integrated equivalent width of $\MgII \lambda 2796$ line as a function of the number of clouds. Most of the weaker systems [$W_r(2796) < 0.1$~{\AA}] have absorption only in one or two clouds.}
\label{fig:2}
\end{figure*}
\clearpage

\begin{figure*}
\epsscale{1.0}
\begin{center}
\vspace{2 in}
\plotone{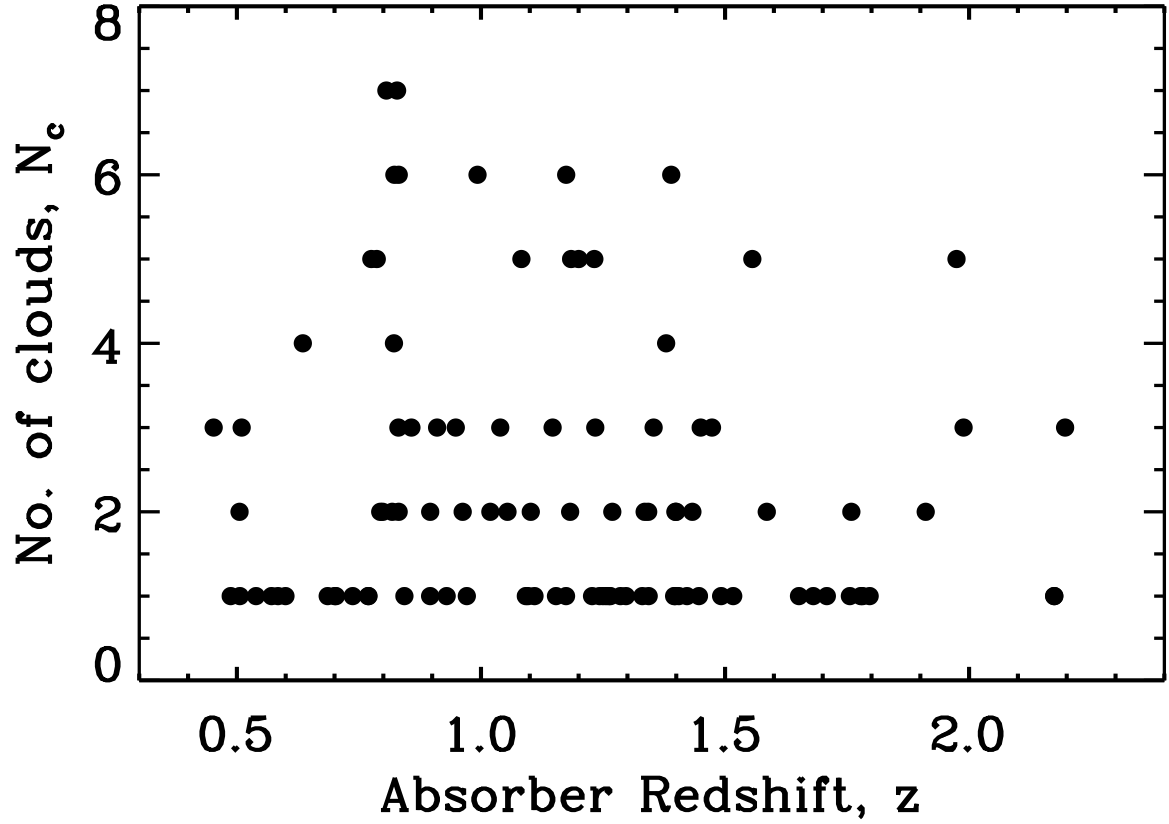}
\end{center}
\protect
\caption{Distribution of single and multiple cloud weak {\MgII} systems 
as a function of redshift. The redshift distribution is comparable between the two types of 
systems (N$_c = 1$ and N$_c \geq 2$; i.e., single and multiple cloud) and they do not exhibit preference for either 
low ($z \sim 1$) or high ($z \sim 2$) redshift. }
\label{fig:3}
\end{figure*}
\clearpage

\begin{figure*}
\epsscale{1.0}
\begin{center}
\vspace{2 in}
\plotone{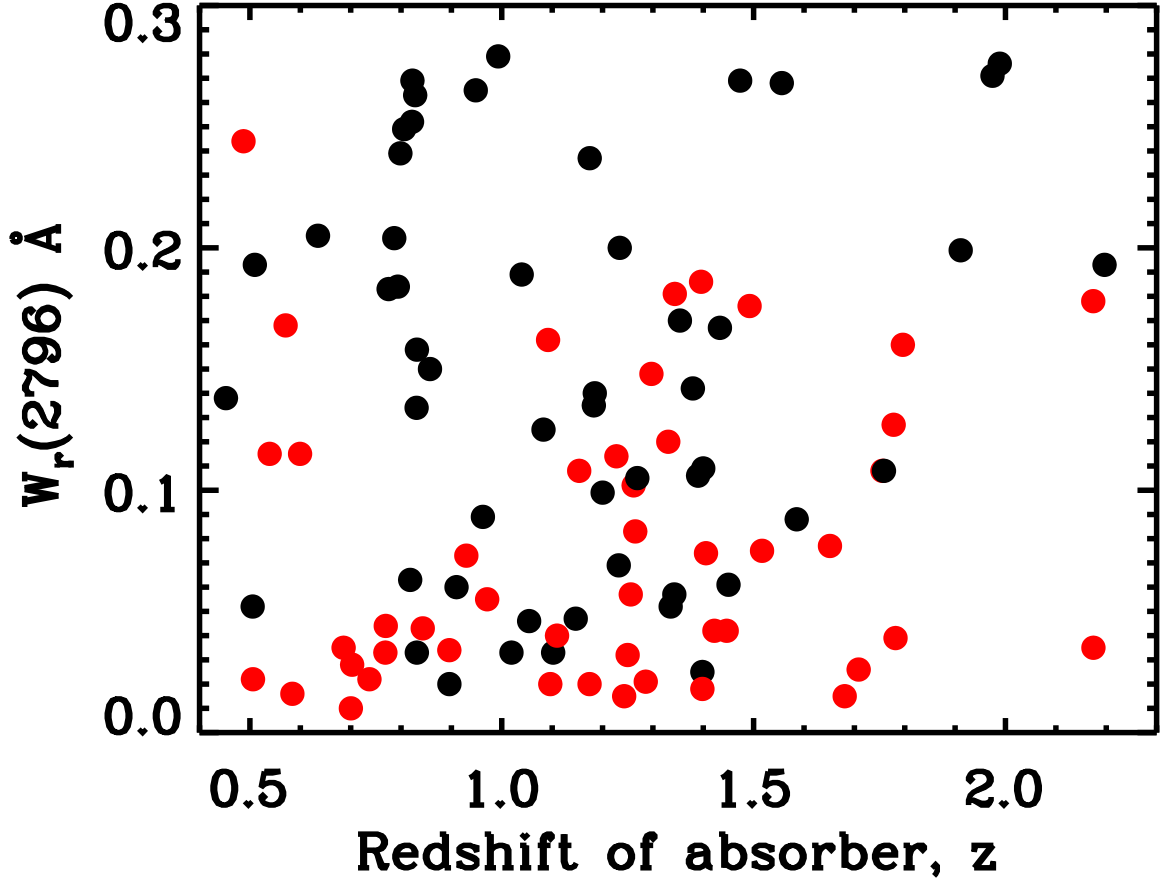}
\end{center}
\protect
\caption{The total (i.e. integrated) rest-frame equivalent width of $\MgII \lambda 2796$~{\AA} in each absorber, as a function of the absorber's redshift. The {\it red} and {\it black} data points correspond to single and multiple cloud systems respectively. It is evident that there is no significant correlation between the strength of weak absorber and its redshift.}
\label{fig:4}
\end{figure*}
\clearpage

\begin{landscape}
\begin{figure*}
\epsscale{0.6}
\begin{center}
\vspace{0.8 in}
\rotatebox{90}{\plotone{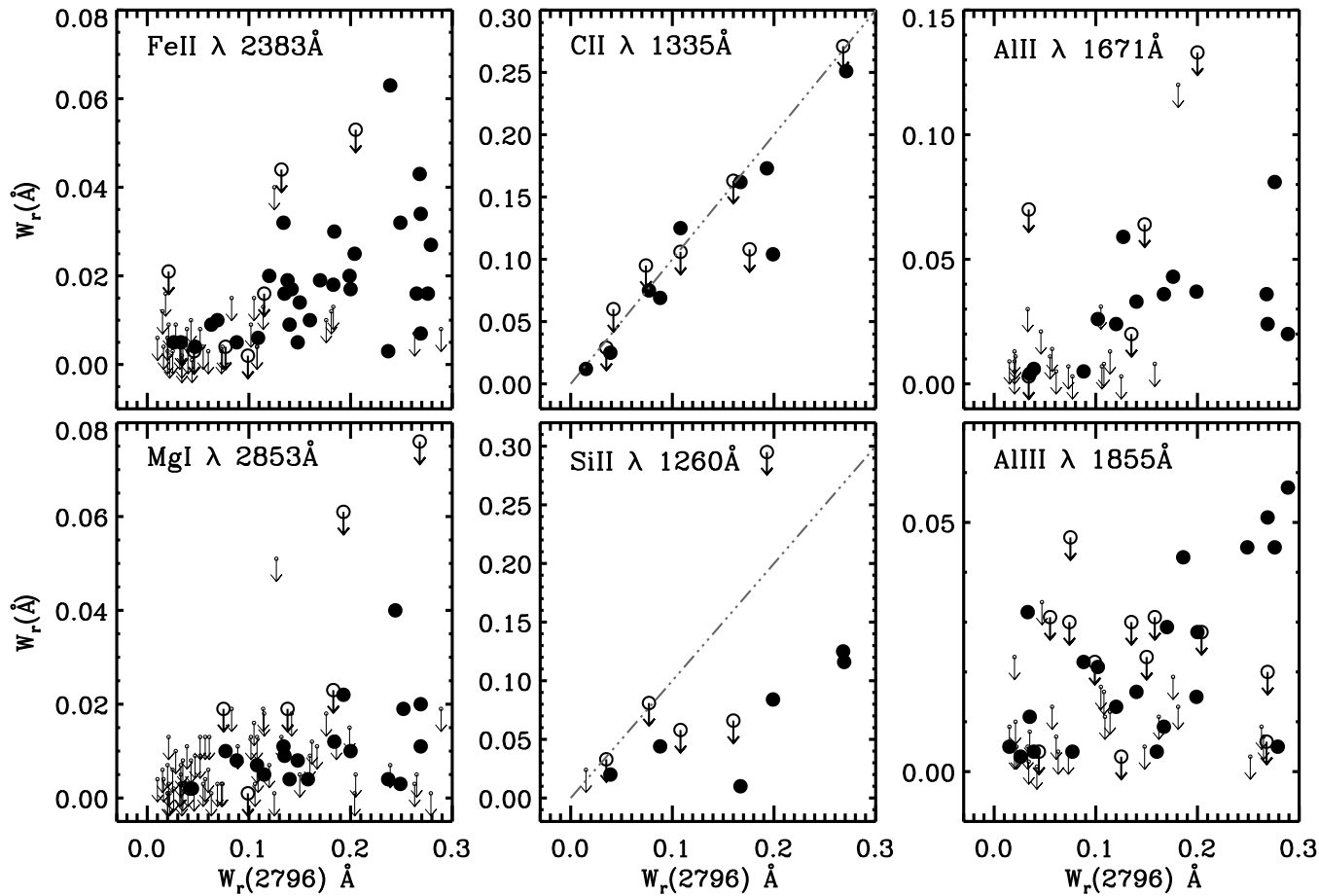}}
\end{center}
\protect
\caption{The comparison of the total rest-frame equivalent widths of {\MgII} ($W_r(2796)$) in each absorber, with other corresponding low ionization transitions and {\AlIII}. The large solid dots are 3$\sigma$ detections, the large open circles are detections that are affected by blending with an absorption feature at some other redshift, in which case the measurement is considered as an upper limit, indicated by a downward pointing arrow from the open circle. Non-detections at the 3$\sigma$ level are plotted using just the downward pointing arrow. In most cases, the blending was from {\HI} lines of the Ly-$\alpha$ forest. The equivalent width corresponding to non-detections at the 3$\sigma$ level are plotted using downward pointing arrows. The {\it dash-dot} line in the {\CII} and {\SiII} panels are the $y = x$ line for comparison with {\MgII}. The correlation coefficients (Spearman's $\rho$) between $W_r(\MgII)$ and $W_r(\FeII)$, $W_r(\MgI)$, $W_r(\CII)$, $W_r(\SiII)$, $W_r(\AlII)$, $W_r(\AlIII)$ are  0.64(0.00), 0.50(0.00), 0.75(0.00), 0.38(0.20), 0.67(0.00), and 0.38(0.00) respectively. The value in parentheses represents the significance level, i.e., the probability that the observed value of $\rho$ would be greater than or equal to the actual value by chance.}
\label{fig:5}
\end{figure*}
\clearpage
\end{landscape}

\begin{landscape}
\begin{figure*}
\epsscale{0.7}
\begin{center}
\rotatebox{90}{\plotone{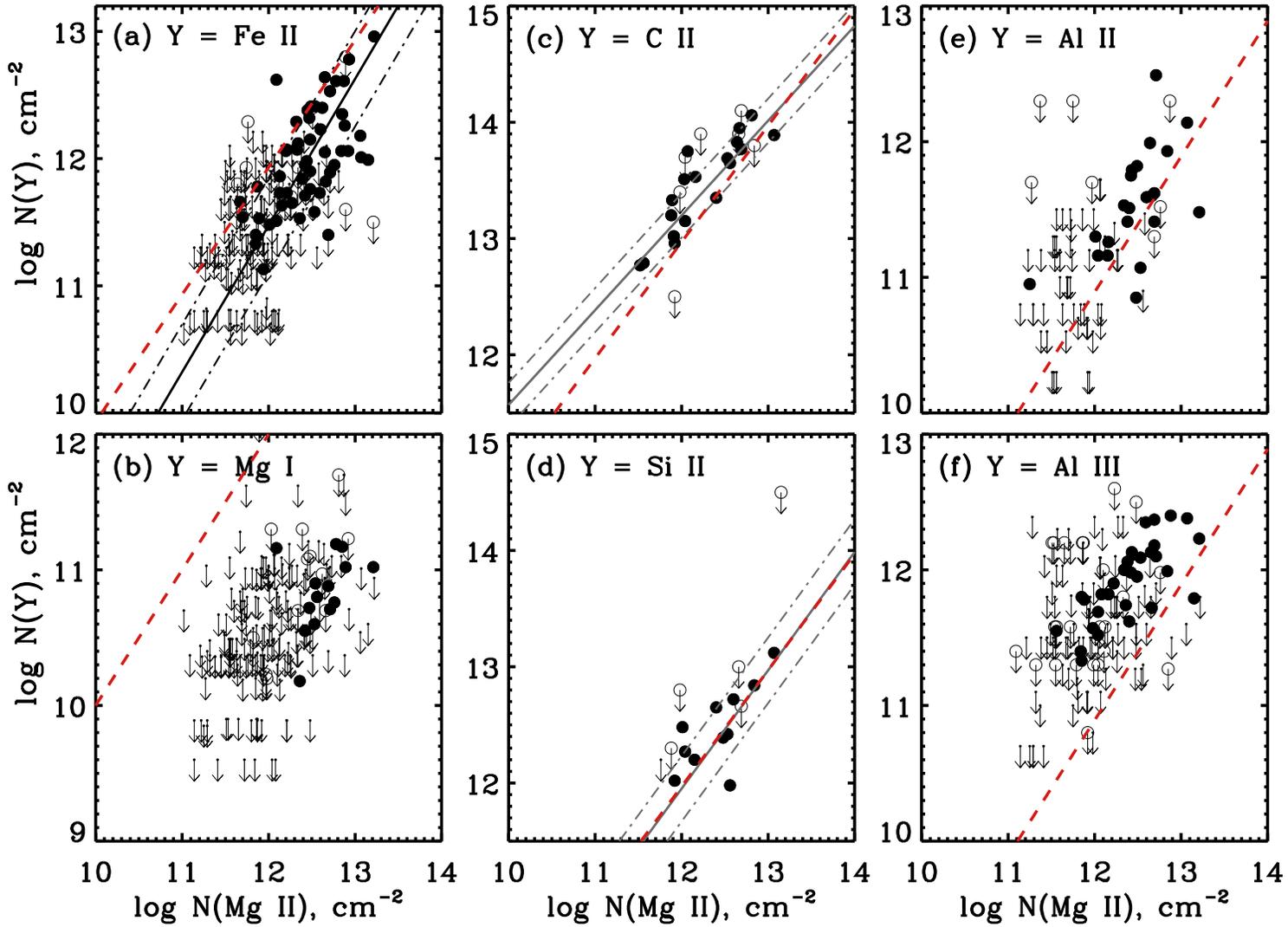}}
\end{center}
\protect
\vspace{0.2in}
\caption{The comparison of {\MgII} column densities with the column density of other low 
ionization transitions and {\AlIII}. The large solid dots are 3$\sigma$ detections, the large open circles are detections that are affected by blending with an absorption feature at some other redshift, 
in which case the measurement is considered as an upper limit, indicated by a downward 
pointing arrow from the open circle. The column density of non-detections, estimated from the $3\sigma$ 
equivalent width limit, are plotted using just the downward pointing arrows. The solid line is a linear 
regression fit formalizing the relationship between the two ions. The standard deviation of the 
corresponding fits are indicated by the {\it dashed-dot} line. The regression lines are drawn for the three ions ({\FeII}, {\CII} and {\SiII}) that display the least scatter with N(\MgII). The {\it red dashed} line in each panel indicates the solar abundance pattern based on values given in \citet{grevesse98,allende01,allende02} and \citet{holweger01}. In the lower left panel, the {\it red dashed} line therefore corresponds to $ y = x$.} 
\label{fig:6}
\end{figure*}
\clearpage
\end{landscape}

\begin{landscape}
\begin{figure*}
\epsscale{0.9}
\plotone{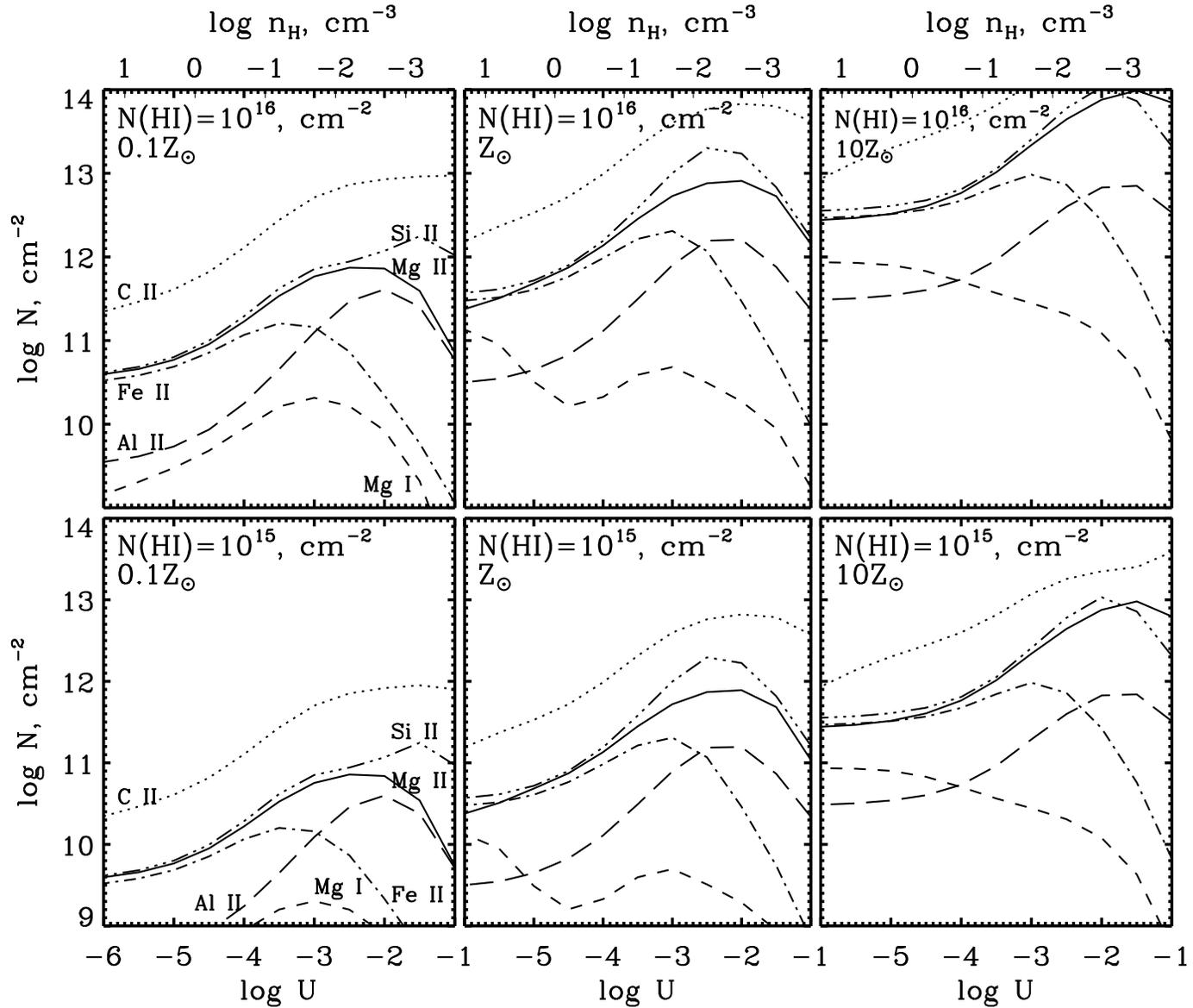}
\protect
\vspace{0.3in}
\caption{Cloudy photoionization curves indicating how the column density of {\SiII}, {\CII}, {\MgII}, {\FeII}, {\AlII} and {\MgI} change with ionization parameter (log $U$), metallicity ($Z$)
and neutral hydrogen column density [$N(\HI)$]. The density ($n_{\H}$) was calculated using the expression log $n_{\H}$ = log $n_\gamma$ - log $U$, for log $n_\gamma$ = -4.70, corresponding to the number density of ionizing photons ($h\nu \geq 13.6$~eV) at $z = 2$.}
\label{fig:7}
\end{figure*}
\clearpage
\end{landscape}

\begin{landscape}
\begin{figure*}
\epsscale{0.7}
\begin{center}
\rotatebox{270}{\plotone{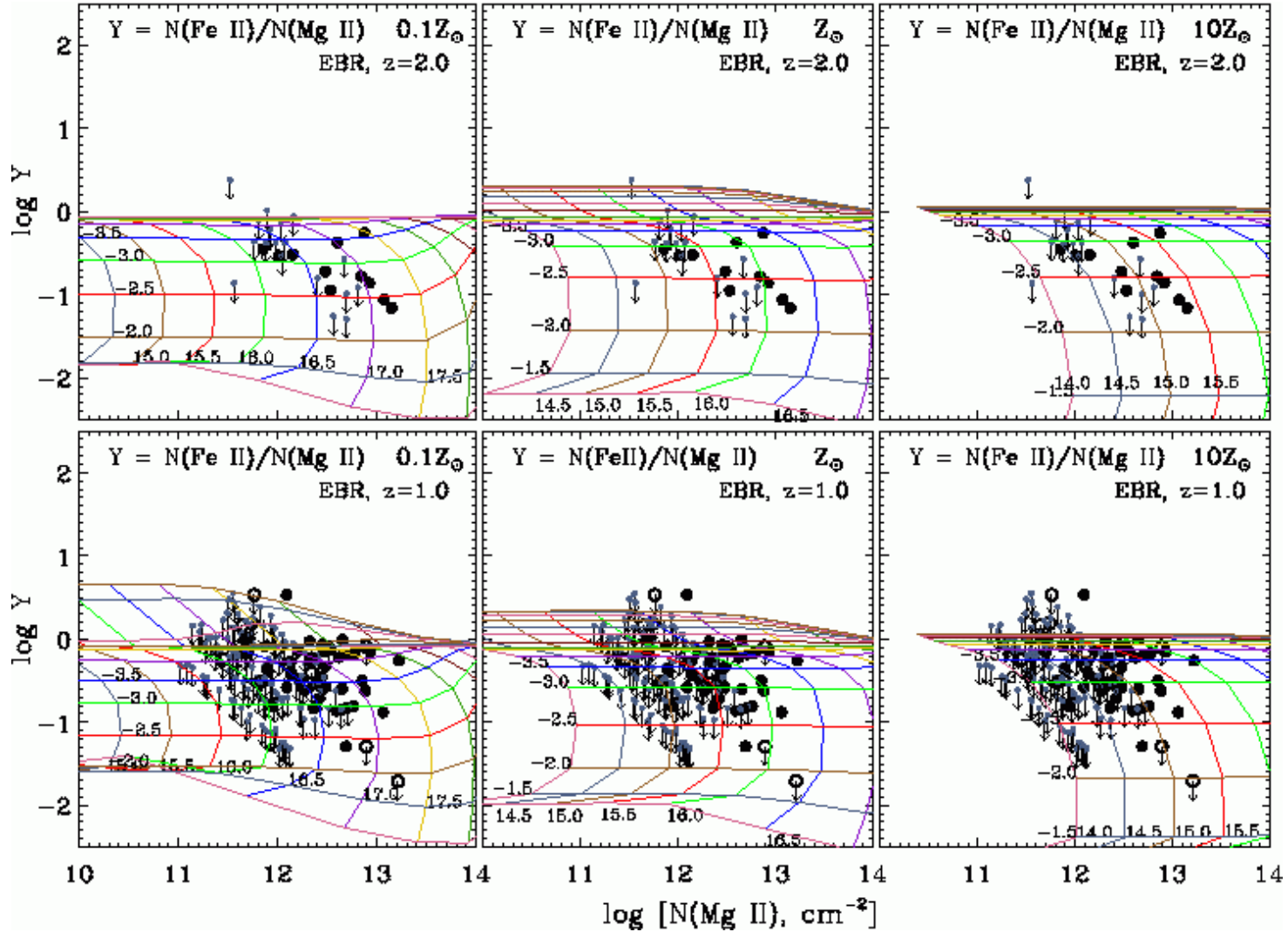}}
\end{center}
\protect
\caption{Cloudy grid of photoionization models for 0.1Z$_\odot$, Z$_\odot$ and 
10Z$_\odot$ metallicity with measurements of {\FeII} and {\MgII} column density over-plotted. 
Vertical curves correspond to lines of constant $N(\HI)$ and the horizontal curves lines of 
constant ionization parameter (log $U$). The Cloudy models were computed for the intensity of 
the extragalactic ionizing background radiation (EBR) at $z = 2$ and $z=1$. Weak {\MgII} 
clouds at $z \geq 1.5$ are plotted in the EBR at $z = 2$ panel and clouds at $z < 1.5$ are 
plotted in the EBR at $z =1$ panel. The large solid dots are 3$\sigma$ detections, the large open circles 
are detections that are affected by blending with an absorption feature at some other redshift, 
in which case the measurement is considered as an upper limit, indicated by a downward 
pointing arrow from the open circle. The column density of non-detections, estimated from the $3\sigma$ 
equivalent width limit, are plotted using just the downward pointing arrows.}
\label{fig:8}
\end{figure*}
\clearpage
\end{landscape}

\begin{figure*}
\epsscale{0.55}
\begin{center}
\vspace{1.5 in}
\plotone{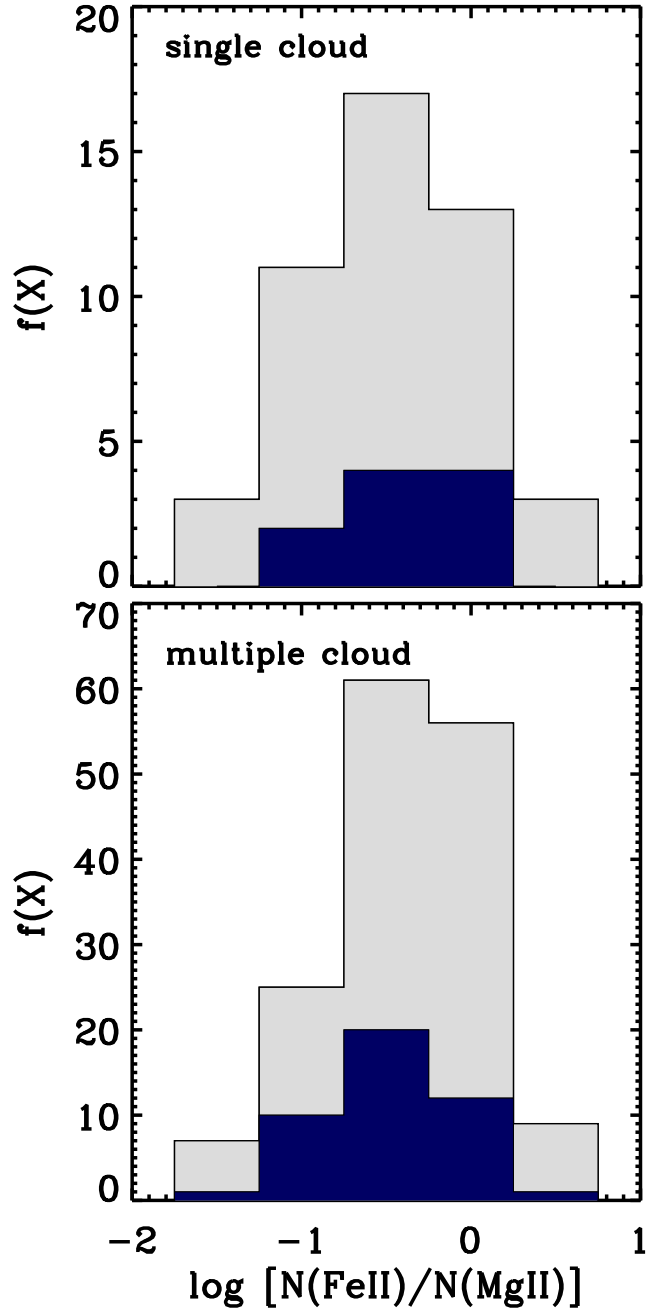}
\end{center}
\protect
\caption{The distribution of {\FeII} to {\MgII} column density ratio for single and multiple cloud 
systems. The histogram filled with the light color includes measurements that are upper limits in {\FeII}, whereas the 
histogram filled with the dark color excludes those. For multiple clouds, we have included the {\FeII} to {\MgII} ratio in individual clouds. It is evident from the figure that both single and 
multiple cloud systems span roughly the same range of {\FeII} to {\MgII} column density ratios.}
\label{fig:9}
\end{figure*}
\clearpage

\begin{landscape}
\begin{figure*}
\epsscale{0.75}
\rotatebox{270}{\plotone{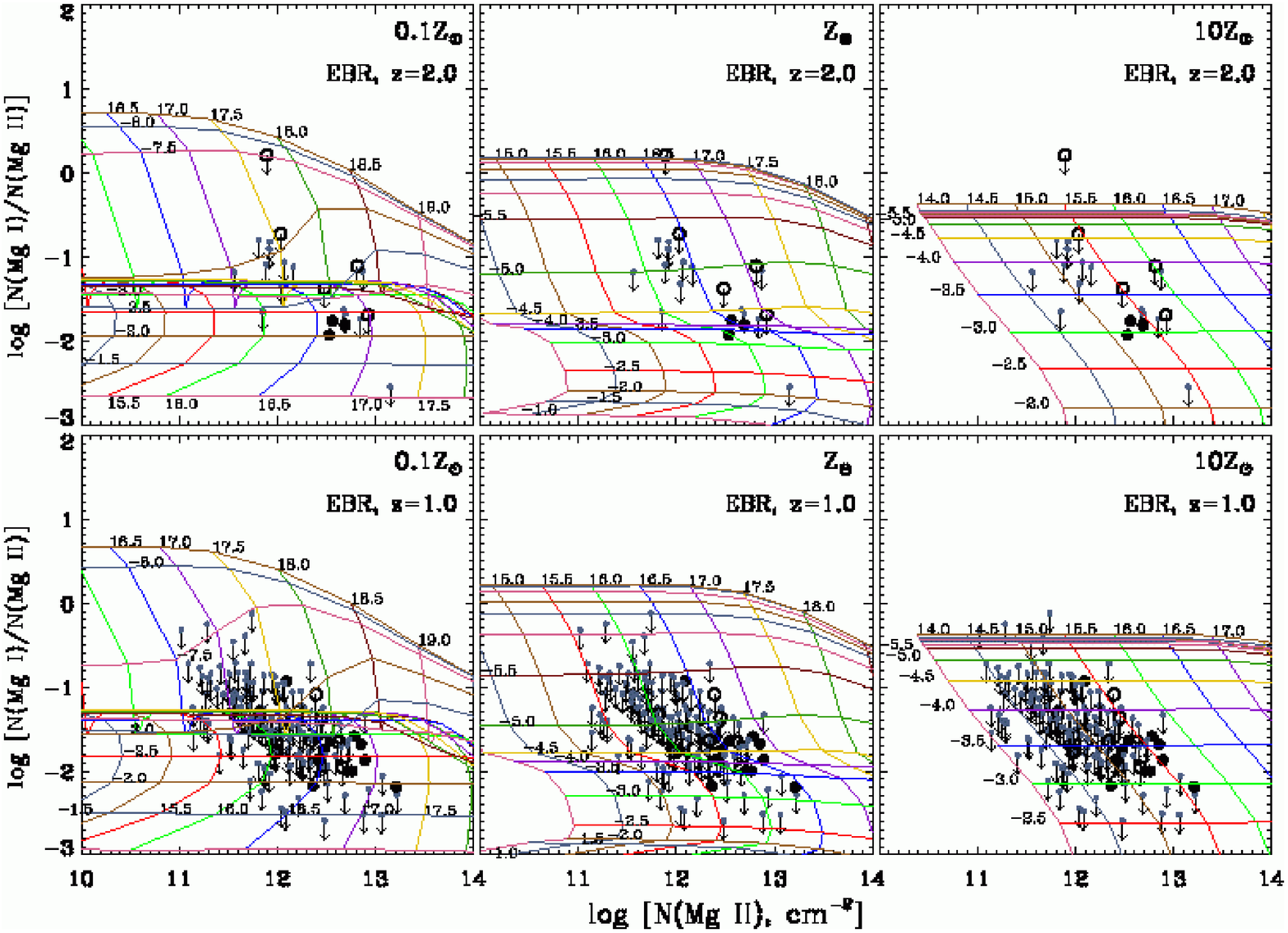}}
\protect
\caption{Cloudy grid of photoionization models with measurements of {\MgI} and {\MgII} 
column density over-plotted. The description of the Cloudy curves and the data points are the 
same as in Figure~\ref{fig:8}.}
\label{fig:10}
\end{figure*}
\clearpage
\end{landscape}

\begin{landscape}
\begin{figure*}
\epsscale{0.3}
\begin{center}
\rotatebox{90}{\plotone{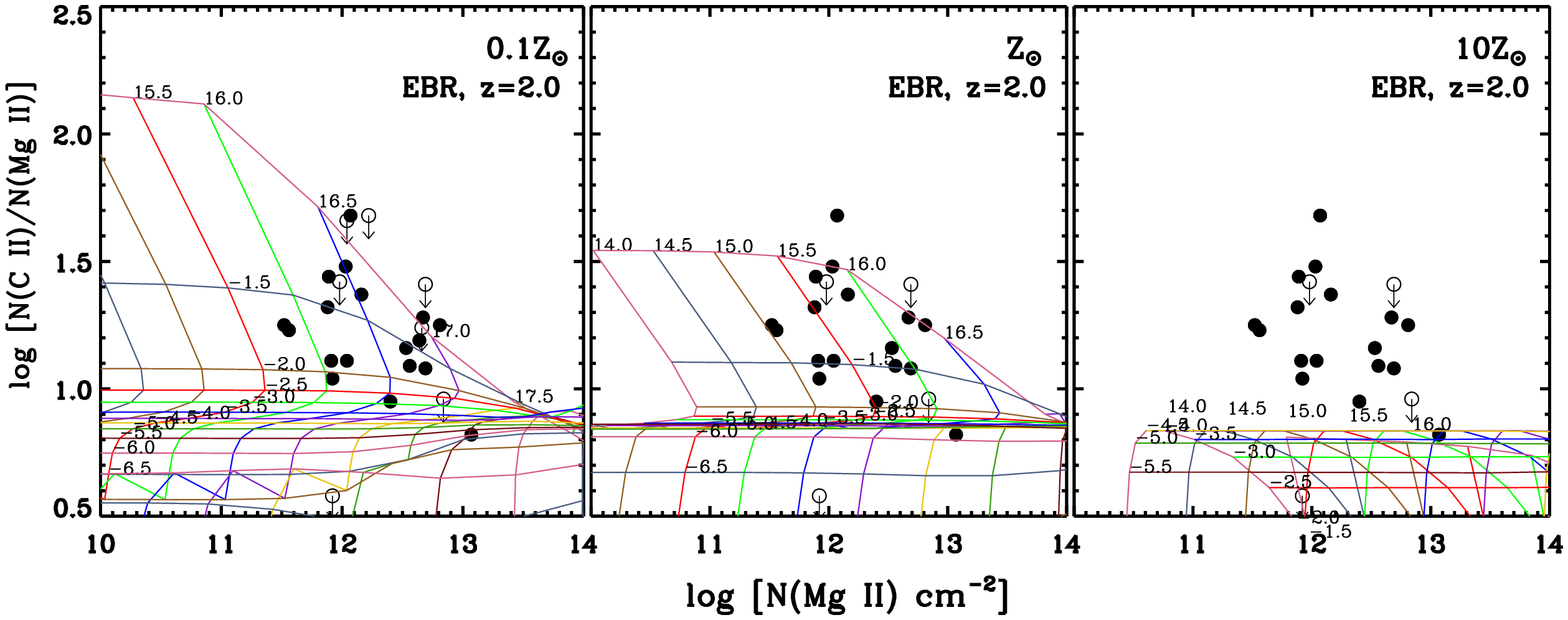}}
\end{center}
\protect
\vspace{0.1in}
\caption{Cloudy grid of photoionization models with measurements of {\CII} and {\MgII} column 
density over-plotted. The description of the Cloudy curves and the data points are the same as 
in Figure~\ref{fig:8}. Almost all weak {\MgII} absorbers with coverage of $\CII \lambda 1335$ are at $z \geq 1.5$ and therefore we plot all of them on the $z = 2$ EBR plot.}
\label{fig:11}
\vspace{0.2in}
\epsscale{0.30}
\begin{center}
\rotatebox{90}{\plotone{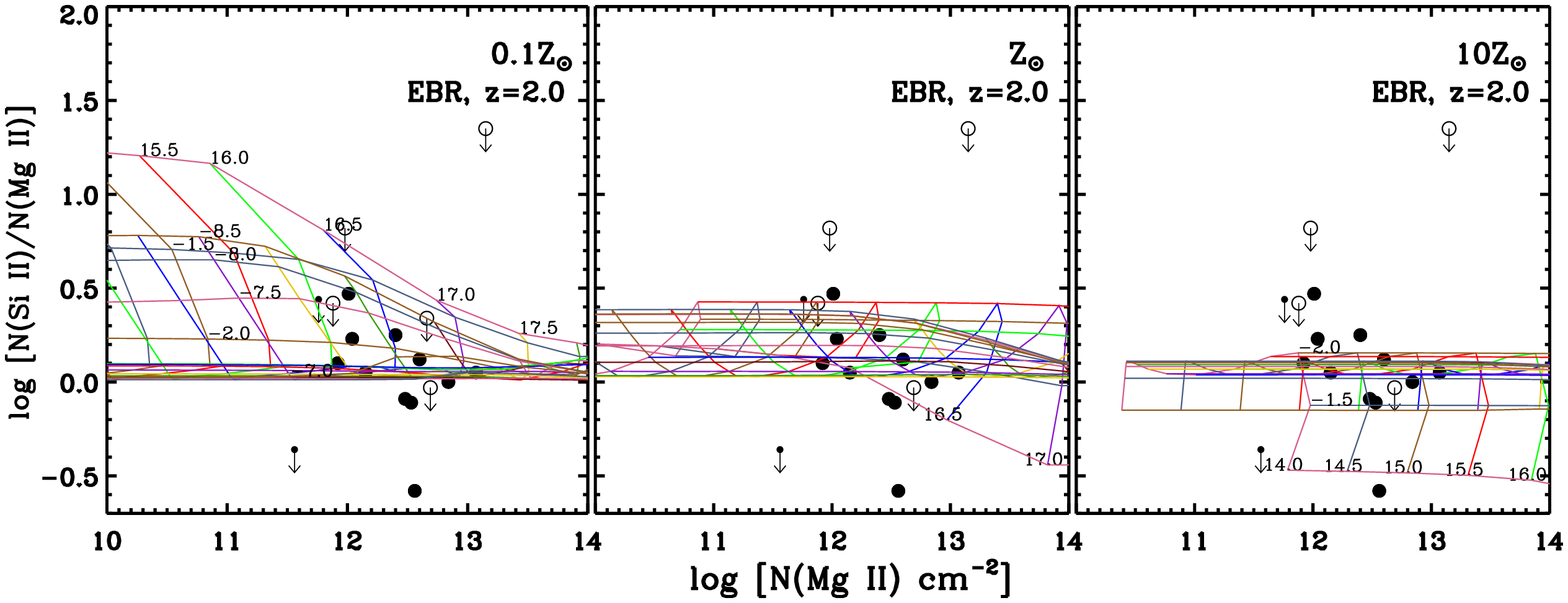}}
\end{center}
\protect
\vspace{0.1in}
\caption{Cloudy grid of photoionization models with measurements of {\SiII} and {\MgII} column 
density over-plotted. The description of the Cloudy curves and the data points are the same as 
in Figure~\ref{fig:8}. Almost all weak {\MgII} clouds with coverage of $\SiII \lambda 1260$ are at $z \geq 1.5$ and therefore we plot all of them on the $z = 2$ EBR plot.}
\label{fig:12}
\end{figure*}
\clearpage
\end{landscape}

\begin{landscape}
\begin{figure*}
\epsscale{0.65}
\rotatebox{270}{\plotone{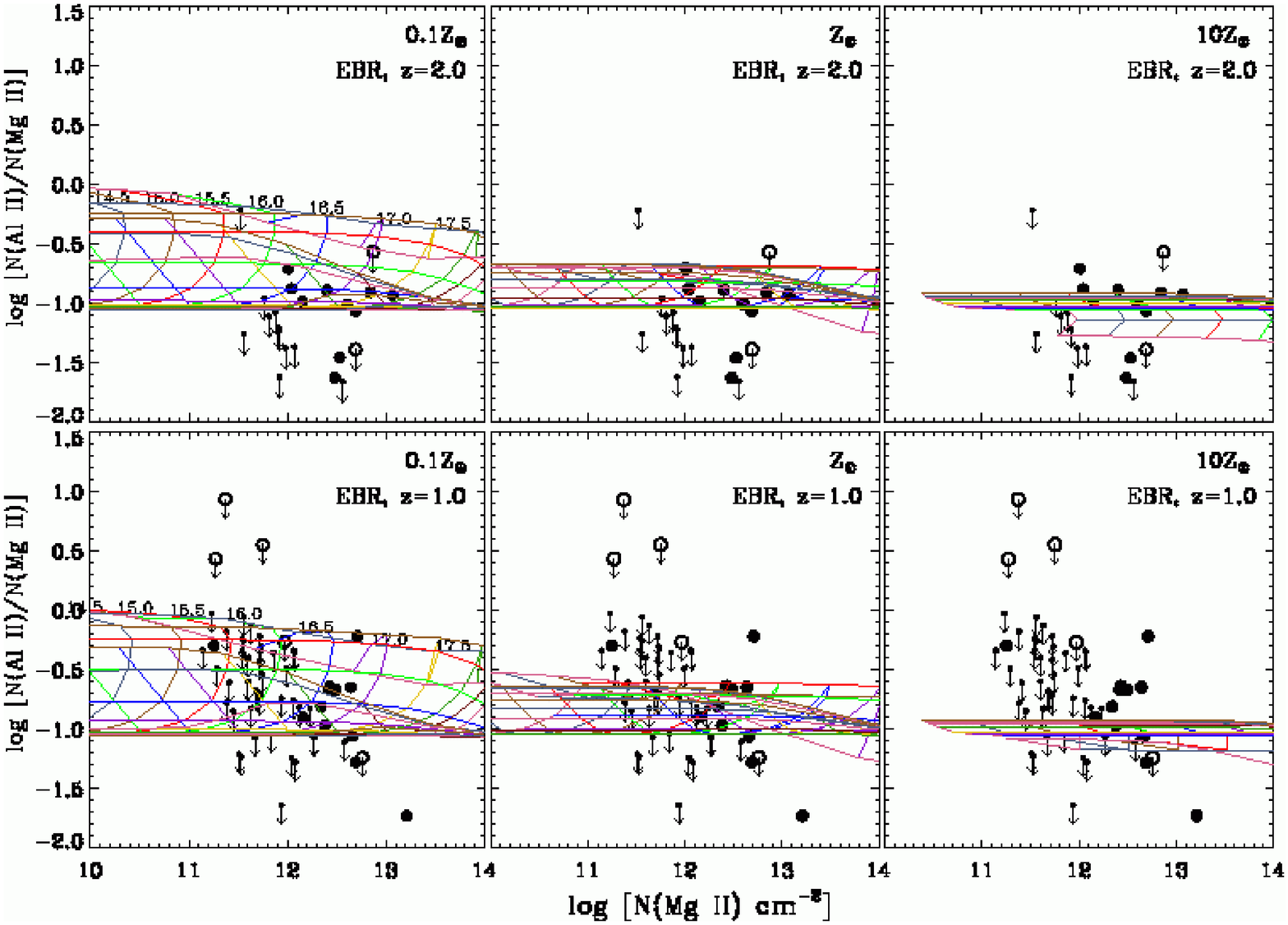}}
\protect
\vspace{0.3in}
\caption{Cloudy grid of photoionization models with measurements of {\AlII} and {\MgII} column 
density over-plotted. The description of the Cloudy curves and the data points are the same as 
in Figure~\ref{fig:8}.}
\label{fig:13}
\end{figure*}
\clearpage
\end{landscape}

\begin{landscape}
\begin{figure*}
\epsscale{1.0}
\vspace{0.6 in}
\plotone{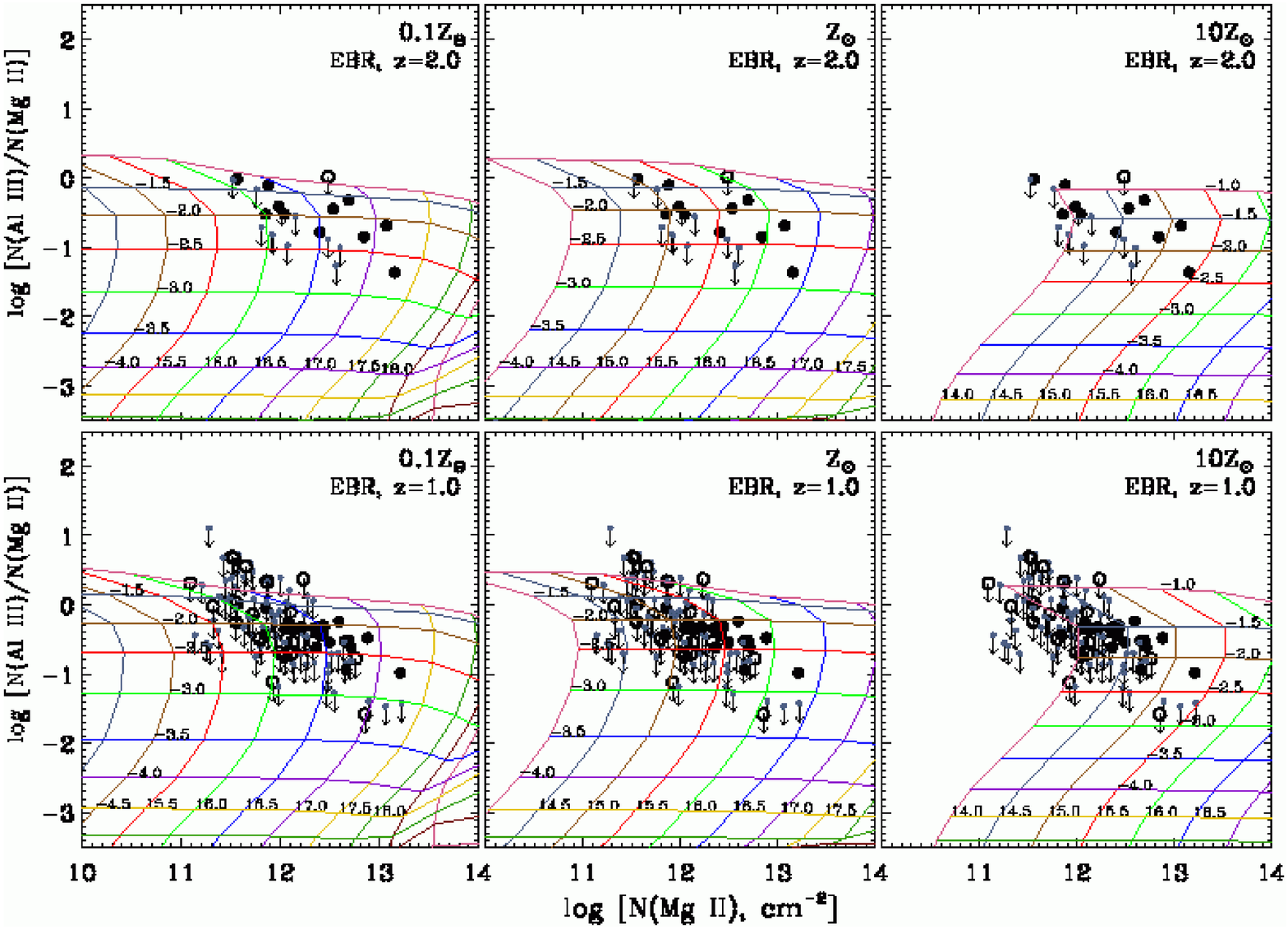}
\protect
\caption{Cloudy grid of photoionization models with measurements of {\AlIII} and {\MgII} 
column density over-plotted. The description of the Cloudy curves and the data points are the 
same as in Figure~\ref{fig:8}.}
\label{fig:14}
\end{figure*}
\clearpage
\end{landscape}

\begin{landscape}
\begin{figure*}
\epsscale{0.40}
\begin{center}
\vspace{2 in}
\rotatebox{90}{\plotone{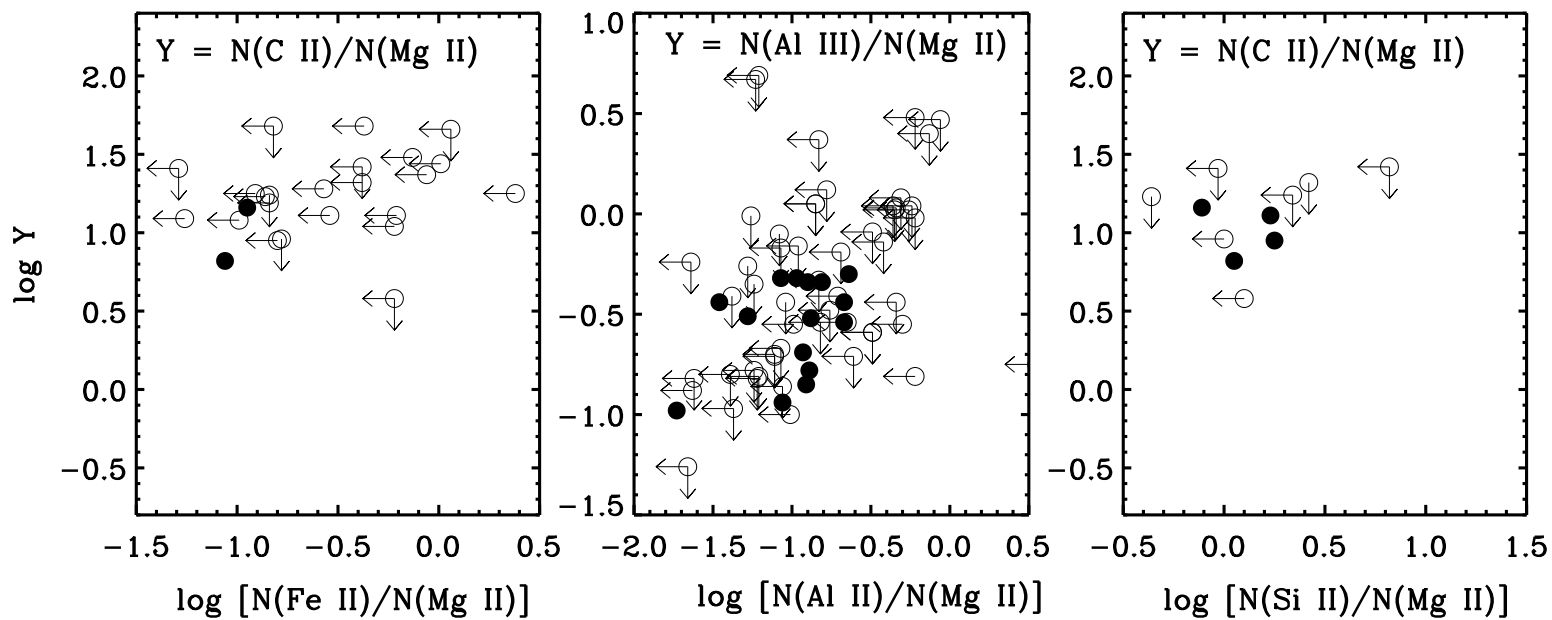}}
\end{center}
\protect
\vspace{0.2in}
\caption{{\it Left Panel :} The figure compares the {\CII} to {\MgII} ratio with {\FeII} to {\MgII} in weak {\MgII} clouds that have measurements for both {\CII} and {\FeII}. The filled circles correspond to weak {\MgII} clouds that had firm measurements for both {\CII} and {\FeII}. The open circles correspond to data points in which either {\CII} or {\FeII} or both have measurements that are upper limits. {\it Middle Panel :} compares the {\AlIII} to {\MgII} ratio with {\AlII} to {\MgII} in weak {\MgII} clouds that have measurements for both {\AlIII} and {\AlII}. {\it Right Panel :} compares the {\CII} to {\MgII} ratio with {\SiII} to {\MgII} in weak {\MgII} clouds that have simultaneous measurements for both {\CII} and {\SiII}.}
\label{fig:15}
\end{figure*}
\clearpage
\end{landscape}

\begin{landscape}
\begin{figure*}
\begin{center}
\epsscale{0.8}
\vspace{1.3 in}
\plotone{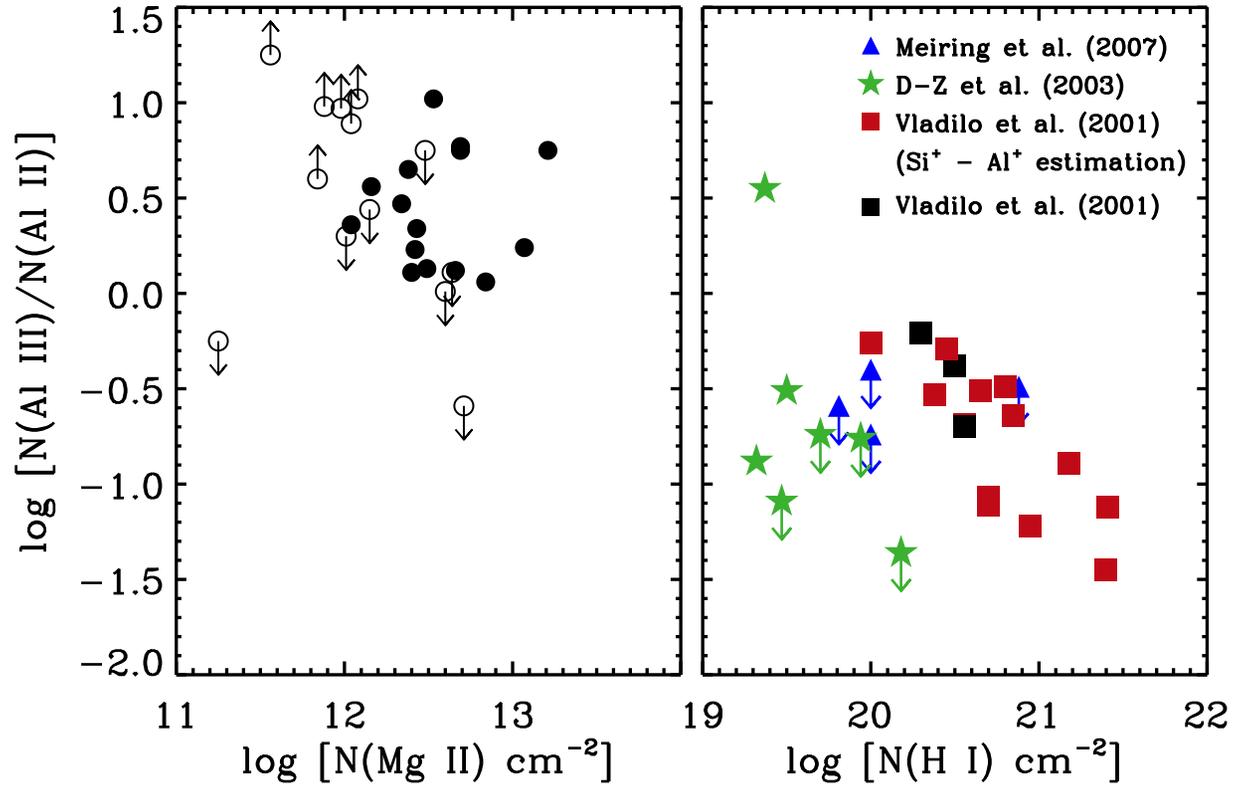}
\end{center}
\protect
\caption{On the {\it left panel} is the {\AlIII} to {\AlII} ratio in weak {\MgII} clouds discussed in this paper. Measurements that are upper limits in {\AlIII} and {\AlII} are indicated using downward and upward pointing arrows respectively. The {\it right panel} shows the {\AlIII} to {\AlII} ratio in sub-DLA and DLA systems taken from the literature. If weak {\MgII} clouds are optically thin in {\HI} (i.e. $N(\HI) < 10^{17}$~{\cmsq}), then the measurements indicate that the anti-correlation between $N(\HI)$ and {\AlIII} to {\AlII} ratio that is observed for DLA and sub-DLA systems, also extends to lower $N(\HI)$ values.}
\label{fig:16}
\end{figure*}
\clearpage
\end{landscape}

\begin{figure*}
\epsscale{0.62}
\vspace{2 in}
\plotone{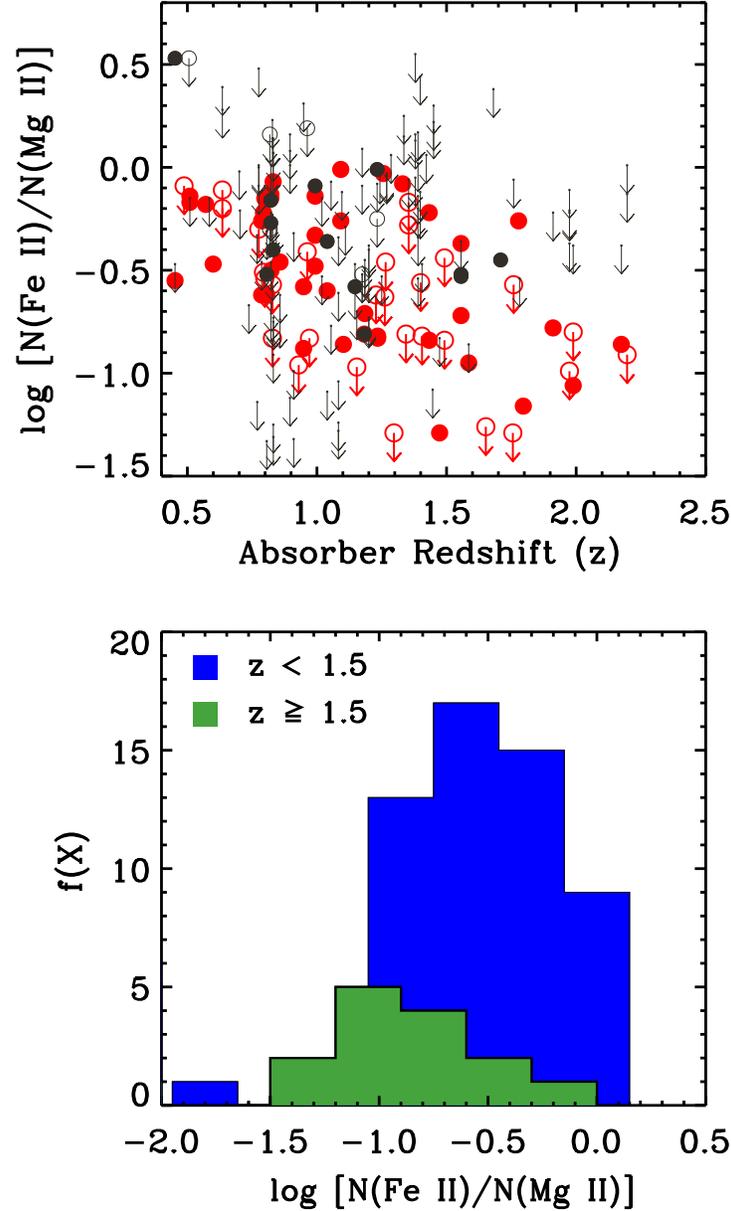}
\protect
\caption{{\it Top panel : } shows the {\FeII} to {\MgII} ratio in all the weak {\MgII} clouds as a function of redshift of the system. The filled circles are firm detections (i.e. detections that are not limits), the large open circles with arrow pointing downward 
corresponds to {\FeII} lines that are affected by blending with an absorption feature at some 
other redshift, in which case the measurement is considered as an upper limit. The non-
detections at the $3\sigma$ level are plotted using downward pointing arrows. Those weak {\MgII} clouds in which $N(\MgII) > 10^{12.2}$~{\cmsq} are plotted in {\it red}. {\it Bottom 
panel :} shows the distribution of {\FeII} to {\MgII} in absorbers at $z \geq 1.5$ and $z < 1.5$. The width of each bin in the distribution is 0.3. The data used to create the frequency distribution includes only those weak {\MgII} clouds in which $N(\MgII) > 10^{12.2}$~{\cmsq}, below which a large fraction of systems only have upper limits for {\FeII}.}
\label{fig:17}
\end{figure*}
\clearpage

\begin{figure*}
\epsscale{0.7}
\vspace{2 in}
\rotatebox{90}{\plotone{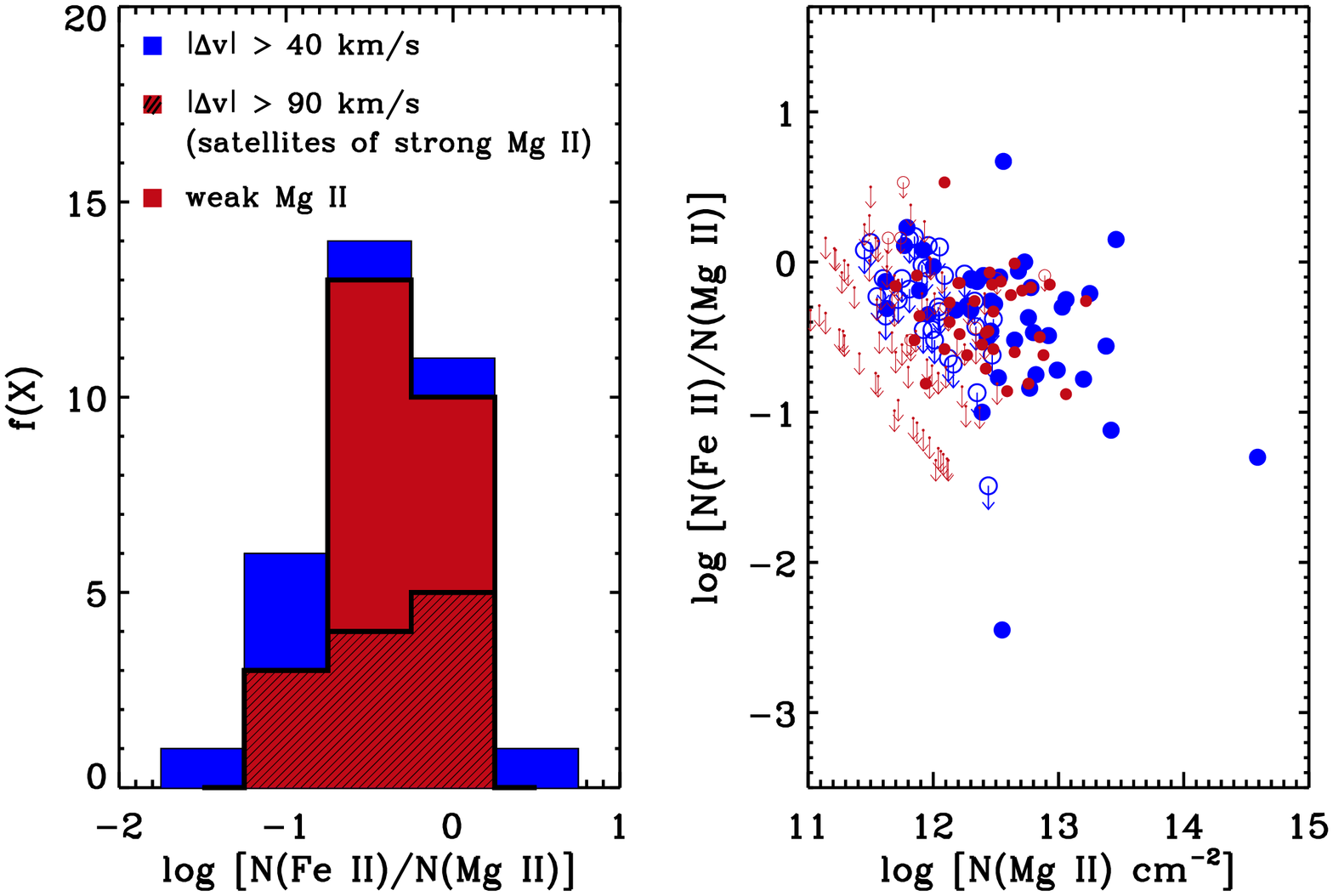}}
\protect
\vspace{0.3in}
\caption{The figure compares the column density ratio between {\FeII} and {\MgII} for weak {\MgII} clouds and the high velocity components (i.e. satellite clouds) of strong {\MgII} systems. In the {\it right panel} the downward pointing arrows indicate measurements that are upper limits in $N(\FeII)$. The {\FeII} and {\MgII} column densities for the satellite clouds of strong {\MgII} systems is taken from \citet{cwcvogt01}, and are represented as {\it large blue} points. The {\it small red} points correspond to weak {\MgII} clouds presented in this paper. The {\it left panel} compares the distribution of the column density ratio between {\FeII} and {\MgII} for weak {\MgII} clouds and the satellite clouds of strong {\MgII} systems. The {\it blue} histogram is for the satellite clouds that are offset in velocity from the main absorbing component by $|\Delta v| > 40$~{\kms}. The histogram shaded with lines is for the satellite clouds that are offset in velocity by $|\Delta v| > 90$~{\kms}. The {\it red} histogram is the distribution for the sample of weak {\MgII} clouds presented in this paper.}
\label{fig:18}
\end{figure*}
\clearpage


\begin{thebibliography}{XXX}

\bibitem[Abraham {\etal}(1996)]{abraham96}
Abraham, R. G., van den Bergh, S., Glazebrook, K., Ellis, R. S., Santiago, B. X., Surma, P., \& Griffiths, R. E. 1996, ApJS, 107, 1.

\bibitem[Adelberger {\etal}(2005a)]{adelberger05a}
Adelberger, K. L., Shapley, A. E., Steidel, C. C., Pettini, M., Erb, D. K., \& Reddy, N. A. 2005, ApJ, 629, 636. 

\bibitem[Adelberger {\etal}(2003)]{adelberger03}
Adelberger, K. L., Steidel, C. C., Shapley, A. E., \& Pettini, M. 2003, ApJ, 584, 45.

\bibitem[Allende Prieto {\etal}(2001)]{allende01}
Allende Prieto, C., Lambert, D. L., \& Asplund, M. 2001, \apj, 556, L63.

\bibitem[Allende Prieto {\etal}(2002)]{allende02}
Allende Prieto, C., Lambert, D. L., \& Asplund, M. 2002, \apj, 573, L137. 

\bibitem[Aracil {\etal}(2006)]{aracil06}
Aracil, B., Tripp, T. M., Bowen, D. V., Prochaska, J. X., Chen, H. -W., \& Frye, B. L. 2006, MNRAS, 367. 139.

\bibitem[Bauer {\etal}(2005)]{bauer05}
Bauer, A. E., Drory, N., Hill, G. J., \& Feulner, G. 2005, ApJ, 621, 89.

\bibitem[Bergeron {\etal}(1991)]{bergeron91}
Bergeron, J., \& Boiss\'e, P. 1991, A\&A, 243, 344.

\bibitem[Bond {\etal}(2001)]{bond01}
Bond, N. A., Churchill, C. W. C., Charlton, J. C., \& Vogt, S. S. 2001, ApJ, 562, 641.

\bibitem[Bouwens {\etal}(2003)]{bouwens03}
Bouwens, R. J. {\etal} 2003, ApJ, 595, 589.

\bibitem[Charlton {\etal}(2003)]{charlton03}
Charlton, J. C., Ding, J., Zonak, S. G., Churchill, C. W., Bond, N. A., \& Rigby, J. R. 2003, ApJ, 589, 111.

\bibitem[Charlton \& Churchill (1998)]{charlton98}
Charlton, J. C., \& Churchill, C. W. C. 1998, ApJ, 499, 181.

\bibitem[Choi {\etal}(2006)]{choi06}
Choi, P. I. {\etal}2006, ApJ, 637, 227.

\bibitem[Churchill(1997)]{cwcthesis} 
Churchill, C.~W.\ 1997, Ph.D.~Thesis

\bibitem[Churchill {\etal}(2007)]{cwc07}
Churchill, C. W. C., Kacprzak, G. G., Steidel, C. C., \& Evans, J. L. 2007, ApJ, 661, 714.

\bibitem[Churchill {\etal}(2005)]{cwc05}
Churchill, C. W. C., Kacprzak, G. G., Steidel, C. C. 2005,  IAU Colloquium Proceedings, Cambridge University Press, pp 24-41.

\bibitem[Churchill \& Le Brun(1998)]{cwc98}
Churchill, C. W., \& Le Brun, V. 1998, ApJ, 499, 677.

\bibitem[Churchill {\etal}(1999)]{weak1}
Churchill, C. W., Rigby, J. R., Charlton, J. C., \& Vogt, S. S. 1999, \apjs, 120, 51.

\bibitem[Churchill {\etal}(1996)]{cwc96}
Churchill, C. W., Steidel, C. C., \& Vogt, S. S. 1996, \apj, 471, 164. 

\bibitem[Churchill \& Vogt(2001)]{cwcvogt01}
Churchill, C. W., \& Vogt, S. S. 2001, \aj, 122, 679.

\bibitem[Churchill {\etal}(2003)]{cwc03}
Churchill, C. W. C., Vogt, S. S., \& Charlton, J. C. 2003, ApJ, 125. 98

\bibitem[Collins et al.(2004)]{collins04} Collins, J.~A., Shull, 
J.~M., \& Giroux, M.~L.\ 2004, \apj, 605, 216 

\bibitem[Conselice {\etal}(2003)]{conselice03}
Conselice, C. J., Bershady, M. A., Dickinson, M., \& Papovich, C. 2003, AJ, 126, 1183.

\bibitem[Dav\'e {\etal}(1997)]{dave97}
Dav\'e, R., Hernquist, L., Weinberg, D. H., \& Katz, N. 1997, ApJ, 477, 21.

\bibitem[Dekker {\etal}(2000)]{dekker00}
Dekker, H., D'Odorico, S., Kaufer, A., Delabre, B., \& Kotzlowski, H. 2000, SPIE, 4008, 534.

\bibitem[Dessauges-Zavadsky {\etal}(2003)]{dz03}
Dessauges-Zavadsky, M., P\'eroux, C., Kim, T. S., D'Odorico, S., \& McMahon, R. G. 2003, MNRAS, 345, 447.

\bibitem[Ding {\etal}(2003)]{ding03}
Ding, J., Charlton, J. C., Bond, N. A., Zonak, S. G., \& Churchill, C. W. 2003, ApJ, 587, 551.

\bibitem[Ellison {\etal}(2004)]{ellison04}
Ellison, S. L., Ibata, R., Pettini, M., Lewis, G. F., Aracil, B., Petitjean, P., \& Srianand, R. 2004, A\&A, 414, 79. 

\bibitem[Elmegreen {\etal}(2005)]{elmegreen05}
Elmegreen, D. M., Elmegreen, B. G., Rubin, D. S., \& Schaffer, M. A. 2005, \apj, 631, 85.

\bibitem[Feigelson, \& Nelson(1985)]{edf85}
Feigelson, E. D., \& Nelson, P. I. 1985, ApJ, 293, 192.

\bibitem[Ferland {\etal}(1998)]{ferland98}
Ferland, G., Korista, K. T., Verner, D. A., Ferguson, J. W., Kingdon, J. B., \& Verner, E. M. 1998, PASP, 110, 76.

\bibitem[Fox et al.(2005)]{fox05} Fox, A.~J., Wakker, B.~P., 
Savage, B.~D., Tripp, T.~M., Sembach, K.~R., 
\& Bland-Hawthorn, J.\ 2005, \apj, 630, 332 

\bibitem[Ganguly et al.(2005)]{ganguly05} Ganguly, R., Sembach, 
K.~R., Tripp, T.~M., \& Savage, B.~D.\ 2005, \apjs, 157, 251 

\bibitem[Grevesse \& Sauval (1998)]{grevesse98}
Grevesse, N., \& Sauval, A. J. 1998, Space Science Reviews, 85, 161-174.

\bibitem[Haardt \& Madau(1996)]{hm96}
Haardt, F., \& Madau, P. 1996, ApJ, 461, 20.

\bibitem[Heckman {\etal}(2000)]{heckman00}
Heckman, T. M., Lehnert, M. D., Strickland, D. K., \& Armus, L. 2000, ApJS, 129, 493.

\bibitem[Heckman {\etal}(2001)]{heckman01}
Heckman, T. M., Sembach, K. R., Meurer, G. R., Strickland, D. K., Martin, C. L., Clazetti, D., \& Leitherer, C. 2001, ApJ, 554, 1021.

\bibitem[Holweger {\etal}(2001)]{holweger01}
Holweger, H. 2001, {\it Solar and Galactic Composition} Workshop, Ed. by Robert F. Wimmer-Schweingruber. Pub. American Institute of Physics Conference Proceedings. Vol 598.

\bibitem[Isobe {\etal}(1986)]{isobe86}
Isobe, T., Feigelson, E. D., \& Nelson, P. I. 1986, ApJ, 306, 490.

\bibitem[Kauffmann {\etal}(2004)]{kauffmann04}
Kauffmann, G., White, S. D. M., Heckman, T. M., M\'enard, B., Brinchmann, J., Charlot, S., Tremonti, C., \& Brinkmann, J. 2004, MNRAS, 353, 713.

\bibitem[Keeney {\etal}(2006)]{keeney06}
Keeney, B. A., Stocke, J. T., Rosenberg, J. L., Tumlinson, J., \& York, D. G. 2006, AJ, 132. 2496.

\bibitem[Khare et al.(2005)]{khare05} Khare, P., et al.\ 2005, 
IAU Colloq.~199: Probing Galaxies through Quasar Absorption Lines, 427 

\bibitem[Kingdon {\etal}(1996)]{kingdon96}
Kingdon, J. B., \& Ferland, G. F. 1996, \apjs, 106, 205.

\bibitem[Lanzetta {\etal}(1987)]{lanzetta87}
Lanzetta, K. M., Wolfe, A. M., \& Turnshek, D. A. 1987, \apj, 322, 739.

\bibitem[Lavalley {\etal}(1992)]{lavalley92}
Lavalley, M., Isobe, T., \& Feigelson, E. 1992, 25, 245.

\bibitem[Lynch {\etal}(2006)]{lynch06}
Lynch, R. S., Charlton, J. C., \& Kim, T. -S. 2006, ApJ, 640, 81. 

\bibitem[Lynch \& Charlton(2007)]{lynch07}
Lynch, R. S., \& Charlton, J. C. 2007, ApJ, 666, 64.

\bibitem[Masiero {\etal}(2005)]{masiero05}
Masiero, J. R., Charlton, J. C., Ding, J., Churchill, C. W., \& Kacprzak, G. 2005, \apj, 623, 57.

\bibitem[Meiring {\etal}(2007)]{meiring07}
Meiring, J. D., Lauroesch, J. T., Kulkarni, V. P., P\'eroux, C., Khare, P., York, D. G., \& Crotts, A. P. S. 2007, MNRAS, 376, 557.

\bibitem[Milutinovi\'c {\etal}(2006)]{milni06}
Milutinovi\'c, N., Rigby, J. R., Masiero, J. R., Lynch, R. S., Palma, C., \& Charlton, J. C. 2006, ApJ, 641, 190.

\bibitem[Misawa {\etal}(2007)]{misawa07}
Misawa, T., Charlton, J. C., \& Narayanan, A. 2007, ApJ submitted.

\bibitem[Mshar {\etal}(2007)]{mshar07}
Mshar, A. C., Charlton, J. C., Lynch, R. S., Churchill, C. W. C., \& Kim, T. S. 2007, ApJ in press, arXiv0706.0515.

\bibitem[Narayanan {\etal}(2005)]{anand05}
Narayanan, A., Charlton, J. C., Masiero, J. R., \& Lynch, R. 2005, \apj, 632, 92.

\bibitem[Narayanan {\etal}(2007)]{anand07}
Narayanan, A., Misawa, T., Charlton, J. C., \& Kim, T. S. 2007, ApJ, 660. 1093.

\bibitem[Nestor {\etal}(2005)]{nestor05}
Nestor, D. B., Turnshek, D. A., \& Rao, S. M. 2005, ApJ, 628, 637.

\bibitem[Pettini {\etal}(2001)]{pettini01}
Pettini, M., Shapley, A. E., Steidel, C. C., Cuby, J. -G., Dickinson, M., Moorwood, A. F. M., Adelberger, K. L., \& Giavalisco, M. 2001, ApJ, 554, 981.

\bibitem[Putman et al.(2003)]{putman03} Putman, M.~E., 
Bland-Hawthorn, J., Veilleux, S., Gibson, B.~K., Freeman, K.~C., 
\& Maloney, P.~R.\ 2003, \apj, 597, 948 

\bibitem[Rauch {\etal}(1999)]{rauch99}
Rauch, M., Sargent, W. L. W., \& Barlow, T. A. 1999, ApJ, 515, 500.

\bibitem[Rauch {\etal}(2002)]{rauch02}
Rauch, M., Sargent, W. L. W., Barlow, T. A., \& Simcoe, R. A. 2002, ApJ, 576, 45.

\bibitem[Richter {\etal}(2008)]{richter08}
Richter, P., Charlton, J. C., Fangano, A. P. M., Bekhti, N. B., \& Masiero, J. R. 2008, ApJ submitted.

\bibitem[Rigby {\etal}(2002)]{weak2}
Rigby, J. R., Charlton, J. C., \& Churchill, C. W. 2002, ApJ, 565, 743.

\bibitem[Rupke {\etal}(2002)]{rupke02}
Rupke, D. S., Veilleux, S., \& Sanders, D. B. 2002, ApJ, 570, 588.

\bibitem[Sargent {\etal}(1989)]{sargent89}
Sargent, W. L. W., Steidel, C. C., \& Boksenberg, A. 1989, ApJS, 69, 703.

\bibitem[Schaye {\etal}(2007)]{schaye07}
Schaye, J., Carswell, R. F., \& Kim, T. S. 2007, MNRAS, 379, 1169.

\bibitem[Schwartz {\etal}(2006)]{schwartz06}
Schwartz, C. M., Martin, C. L., Chandar, R., Leitherer, C., Heckman, T. M., \& Oey, M. S. 2006, ApJ, 646, 858.

\bibitem[Simcoe {\etal}(2004)]{simcoe04}
Simcoe, R. A., Sargent, W. L. W., \& Rauch, M. 2004, ApJ, 606, 92.

\bibitem[Simcoe {\etal}(2006)]{simcoe06}
Simcoe, R. A., Sargent, W. L. W., Rauch, M., \& Becker, G. 2006, \apj, 637, 648.

\bibitem[Steidel {\etal}(1994)]{steidel94}
Steidel, C. C., Dickinson, M., \& Persson, S. E. 1994, ApJ, 437, 75.

\bibitem[Steidel {\etal}(2002)]{steidel02}
Steidel, C. C., Kollmeier, J. A., Shapley, A. E., Churchill, C. W., Dickinson, M., \& Pettini, M. 2002, ApJ, 570, 526.

\bibitem[Steidel \& Sargent(1992)]{ss92}
Steidel, C. C., \& Sargent, W. L. W. 1992, \apjs, 80, 1.

\bibitem[Stengler-Larrea {\etal}(1995)]{strengler95}
Stengler-Larrea {\etal}. 1995, ApJ, 444, 64.

\bibitem[Stocke {\etal}(2004)]{stocke04}
Stocke, J. T., Keeney, B. A., McLin, K. M., Rosenberg, J. L., Weymann, R. J. \& Giroux, M. L. 2004, ApJ, 609, 94.

\bibitem[Tappe \& Black(2004)]{tappe04}
Tappe, A. \& Black, J. H. 2004, A\&A, 423, 943.

\bibitem[Timmes {\etal}(1995)]{timmes95}
Timmes, F. X., Woosley, S. E., \& Weaver, T. A. 1995, \apjs, 98, 617.

\bibitem[Tripp {\etal}(2006)]{tripp06}
Tripp, T. M., Aracil, B., Bowen, D. V., \& Jenkins, E. B. 2006, ApJ, 643, 77.

\bibitem[Tripp \& Bowen (2005)]{tripp05}
Tripp, T. M., \& Bowen, D. V. 2005, IAU Colloquium Proceedings, Cambridge University Press, pp 5-23.

\bibitem[van den Bergh {\etal}(1996)]{vandenbergh96}
van den Bergh, S., Abraham, R. G., Ellis, R. S., Tanvir, N. R., Santiago, B. X., \& Glazebrook, K. G. 1996, AJ, 112, 359.

\bibitem[Vladilo {\etal}(2001)]{vladilo01}
Vladilo, G., Centuri\'on, M., Bonifacio, P., \& Howk, J. C. 2001, \apj, 557, 1007.

\bibitem[Wakker 
\& van Woerden(1997)]{wakker97} Wakker, B.~P., \& van Woerden, H.\ 1997, \araa, 35, 217 

\bibitem[Wang {\etal}(2006)]{wang06}
Wang, W.-H., Cowie, L. L., \& Barger, A. J. 2006, \apj, 647, 74.

\bibitem[York et al.(2006)]{york06} York, D.~G., et al.\ 2006, 
\mnras, 367, 945 

\bibitem[Zonak {\etal}(2004)]{zonak04}
Zonak, S. G., Charlton, J. C., Ding, J., \& Churchill, C. W. C. 2004, ApJ, 606, 196.

\end{thebibliography}
\end{document}